\documentclass[sn-mathphys-num]{sn-jnl}% Math and Physical Sciences Numbered Reference Style 
\usepackage{graphicx}%
\usepackage{multirow}%
\usepackage{amsmath,amssymb,amsfonts}%
\usepackage{amsthm}%
\usepackage{mathrsfs}%
\usepackage[title]{appendix}%
\usepackage{xcolor}%
\usepackage{textcomp}%
\usepackage{manyfoot}%
\usepackage{booktabs}%
\usepackage{algorithm}%
\usepackage{algorithmicx}%
\usepackage{algpseudocode}%
\usepackage{listings}%
\usepackage{enumerate}
\usepackage[figuresright]{rotating}
\usepackage{longtable}
\usepackage{adjustbox}
\theoremstyle{thmstyleone}%
\theoremstyle{thmstyletwo}%

\theoremstyle{thmstylethree}%

\begin{document}

% \title[Article Title]{Multi-view in the Serial Sections for Histopathological Analysis}
\title[Article Title]{Evolutionary Paradigms in Histopathology Serial Sections technology}


\author[1]{\fnm{Zhenfeng} \sur{Zhuang}}\email{zhuangzhenfeng@stu.xmu.edu.cn}

\author[2]{\fnm{Min} \sur{Cen}}\email{cenmin0127@mail.ustc.edu.cn}

\author[1]{\fnm{Lei} \sur{Jiang}}\email{23020241154403@stu.xmu.edu.cn}

\author[1]{\fnm{Qiong} \sur{Peng}}\email{qpeng@stu.xmu.edu.cn}

\author[1]{\fnm{Yihuang} \sur{Hu}}\email{huyihuang@stu.xmu.edu.cn}

\author[3]{\fnm{Hong-Yu} \sur{Zhou}}\email{whuzhouhongyu@gmail.com}

\author*[1]{\fnm{Liansheng} \sur{Wang}}\email{lswang@xmu.edu.cn}

\affil*[1]{\orgdiv{Computer Science at the School of Informatics}, \orgname{Xiamen Univeristy}, \orgaddress{\city{Xiamen}, \postcode{361005}, \state{Fujian}, \country{China}}}

\affil[2]{\orgdiv{School of Artificial Intelligence and Data Science}, \orgname{University of Science and Technology of China}, \orgaddress{\city{Hefei}, \postcode{230026}, \state{Anhui}, \country{China}}}

\affil[3]{\orgdiv{Harvard Medical School}, \orgname{Harvard University}
\orgaddress{\city{Boston}, \postcode{02115}, \state{MA}, \country{USA}}}

\abstract{Histopathological analysis has been transformed by serial section-based methods, advancing beyond traditional 2D histology to enable volumetric and microstructural insights in oncology and inflammatory disease diagnostics. This review outlines key developments in specimen preparation and high-throughput imaging that support these innovations. Computational workflows are categorized into multimodal image co-registration, 3D histoarchitecture reconstruction, multiplexed immunohistochemical correlation, and cross-scale data fusion. These approaches exploit serial section-derived spatial concordance to enhance resolution in microenvironmental and molecular profiling. Despite progress, challenges remain in harmonizing heterogeneous datasets, optimizing large-scale registration, and ensuring interpretability. Future directions include AI integration, spatial transcriptomics, and applications in developmental biology and neuroscience, establishing serial section analytics as central to precision histopathology.}

% \keywords{serial sections, histopathology, registration, challenges, modality-dependent}



\maketitle

\section{Introduction}\label{section1}

Traditional tissue sectioning, while foundational in pathology, is often limited by its reliance on single two-dimensional(2D) slices, which offer only a partial and static view of complex biological structures. This limitation becomes particularly evident in the diagnosis of heterogeneous conditions such as tumors, where morphology, cellular distribution, and invasive patterns vary significantly across different tissue planes \cite{herrmann2014three, cakir2024determination}. A single section may fail to capture these variations due to sampling bias, suboptimal cutting orientation, or inadequate section thickness—either missing crucial features in overly thin slices or introducing visual ambiguity in thicker ones \cite{cuenca2020interpretation, veluponnar2024resection}. Moreover, as each section represents a single time point, traditional sectioning inherently lacks temporal or dynamic context, making it difficult to track disease progression \cite{garrison2022trends}.

To overcome these limitations, the use of consecutive or serial tissue sections has emerged as a powerful strategy. By analyzing multiple adjacent slices, researchers and clinicians can reconstruct three-dimensional tissue architecture, better capture spatial heterogeneity, and more accurately delineate lesion boundaries and invasive fronts. This multi-slice approach reduces diagnostic uncertainty, enhances the representation of complex pathological changes, and enables downstream tasks such as volumetric analysis, 3D modeling, and multimodal image registration. In recent years, the integration of serial sections with digital pathology and AI-driven analysis has further expanded the scope of this technique, facilitating high-throughput, high-precision tissue characterization \cite{bai2023deep, weissleder2020automated}.

Despite these challenges, tissue sectioning remains indispensable in modern diagnostics. Through routine methods such as Hematoxylin and Eosin (H\&E) staining and immunohistochemistry, it provides critical insights into tissue composition, cellular morphology, and biomarker expression \cite{gurcan2009histopathological, bui2019digital}. Its cost-effectiveness, ease of implementation, and long-term sample preservation make it widely applicable, even in resource-constrained settings \cite{al2023applications}. As sectioning technologies evolve—from manual microtomes to fully automated systems—the precision and reproducibility of the process continue to improve. These advancements, when coupled with strategies like serial sectioning, are redefining the diagnostic potential of histopathology, pushing the field beyond static snapshots toward a more holistic, multidimensional understanding of disease.

Unlike single sections, which offer only a 2D view, serial sections provide a more comprehensive and three-dimensional representation of the tissue, significantly enhancing the understanding and analysis of lesion areas \cite{almagro2021tissue, kench2021generating}. The greatest advantage of serial sections is its ability to reconstruct the three-dimensional structure of the tissue. By superimposing multiple serial sections, researchers can accurately reproduce the layout and morphological features of the tissue in three-dimensional space \cite{larsen2021cellular}. This three-dimensional view significantly compensates for the limitations of a single section, allowing for a more comprehensive and realistic picture of the lesion as a whole. Additionally, serial sections can greatly reduce the risk of misdiagnosis or missed diagnosis. Tumors and other diseased tissues often show obvious heterogeneity. Serial sections allow for detailed observation of diseased tissues layer by layer, revealing complex changes within tumors \cite{seferbekova2023spatial}. Different layers may show varying patterns of invasiveness, which can be accurately captured only through this method, enabling more precise diagnostic and treatment decisions. While each section is static, analyzing multiple layers facilitates speculation on lesion progression across different stages, which is particularly relevant in cancer, inflammation, and regeneration studies. This approach also enhances diagnostic consistency and reliability, as it provides a comprehensive view of the tissue, allowing pathologists to analyze lesions from multiple perspectives \cite{jahn2020digital}. Consequently, interpretations of the same lesion by different pathologists are more consistent, reducing diagnostic bias linked to single-section analysis \cite{pallua2020future}.

% 连续切片如何生成
The process of preparing serial sections begins with fixing tissues to preserve structures, followed by dehydration to remove moisture. The tissues are then embedded in molten paraffin for cutting, with microtome thickness controlled at the micrometer level to maintain z-axis sequence information \cite{sanchez2022tissue}. After preparation, biological markers undergo processing for observation and digital analysis, typically involving H\&E staining and cytological examination. More complex markers require techniques such as immunohistochemistry and PCR. Genetic techniques like spatial transcriptomics, FISH, and comparative genomic hybridization can also be applied. The quality of sections is influenced by scanning and data organization methods. .

The large volume of image data from digital processing of serial tissue sections offers significant opportunities for pathological research, but also presents numerous challenges. The immense scale and high resolution of the data demand substantial computational resources \cite{hosseini2024computational}. Variability in tissue preparation and uneven staining lead to inconsistencies in image quality, complicating analysis. Additionally, diverse cell morphologies, overlap, and occlusion hinder cell segmentation and tracking \cite{bai2023deep}. Tasks such as three-dimensional reconstruction and alignment face technical difficulties in registering and stitching multiple slices. Extracting biologically meaningful features and conducting quantitative analyses are also challenging, as traditional image processing methods often yield limited results with complex biological images. Although deep learning shows promise in feature extraction, issues of model generalizability and interpretability remain. Furthermore, the integration of multimodal data, large-scale data annotation, and data privacy protection require further exploration \cite{pasolli2016machine}.

% 利用连续切片的场景和方法简单罗列一下
%Pathological serial sections serve as a pivotal tool in histopathology, facilitating a nuanced exploration of tissue samples across various scenarios. In tumor characterization, these sections enable the detailed examination of architectural features and invasive behavior, thereby contributing to a refined understanding of tumor biology and treatment responsiveness \cite{jahn2020digital}. Moreover, they play a crucial role in studying disease progression by allowing for the temporal analysis of structural and cellular changes, which can illuminate underlying mechanisms and potential biomarkers. Advanced techniques such as digital imaging and quantitative morphometric analysis further enhance the precision of assessing morphological variations, while immunohistochemistry and in situ hybridization enable the visualization of specific protein expressions and gene activity patterns within the tissue. The integration of data from serial sections into three-dimensional reconstructions enriches the spatial understanding of complex tissue structures, facilitating the identification of disease-specific features. Additionally, molecular techniques like PCR and FISH provide genetic insights that complement traditional histopathological assessments, ultimately fostering a comprehensive approach to biomarker discovery and histopathological grading across different disease states.

Pathological serial sections serve as a pivotal tool in histopathology, facilitating a nuanced exploration of tissue samples across various scenarios. In tumor characterization, these sections allow for the detailed examination of tissue architecture and invasive behavior, providing valuable insights into tumor biology and its response to treatment \cite{jahn2020digital}. They are essential for studying disease progression by enabling the temporal analysis of structural and cellular changes, which can help uncover underlying mechanisms and identify potential biomarkers. Advanced techniques such as digital imaging and quantitative morphometric analysis improve the accuracy of morphological assessments, while immunohistochemistry and in situ hybridization facilitate the visualization of specific protein expressions and gene activity within the tissue. The integration of serial section data into three-dimensional reconstructions enhances our spatial understanding of complex tissue structures, aiding in the identification of disease-specific features. Molecular techniques like PCR and FISH provide genetic insights that complement traditional histopathological methods, promoting a more comprehensive approach to biomarker discovery and histopathological grading across various disease states.

% \section{Registration for Serial Slices}\label{section2}
\section{Serial Sections Construction}\label{section2}
% \begin{figure}
%     \centering
%     \includegraphics[width=1\textwidth]{images/part2.pdf}
%     \caption{Caption part2}
%     \label{fig2}
% \end{figure}
Serial sectioning is widely employed in pathological analysis due to its capacity to provide multi-perspective structural information, which significantly enhances diagnostic efficacy \cite{liu2022novel}. Consequently, it is imperative to delineate the workflow for constructing serial sections (Fig \ref{fig1} a.). Specifically, tissue specimens must undergo standardized biochemical processing (e.g., fixation and staining) tailored to experimental objectives, followed by precision sectioning and subsequent imaging acquisition or multimodal data collection \cite{larsen2021cellular}.

\subsection{Preparation}

Traditional histopathology sample preparation involves embedding fixed, processed, and oriented tissue sections in a solidifying medium, followed by slicing into thin sections (3–4 µm) and staining to enhance contrast for cellular and tissue structure observation. Advanced methods, such as chemical staining, immunohistochemistry, in situ PCR, FISH, spatial transcriptomics, ultramicropathology, and flow cytometry, enable the analysis of ultrastructural and biomolecular components, providing insights into cell function. Tissue labeling introduces markers like stains, fluorescent tags, or nucleic acid probes to visualize specific structures, while sequencing DNA, RNA, or proteins facilitates gene expression analysis, mutation detection, and pathological studies, advancing disease understanding in molecular level.

\textbf{Histocytochemistry} staining techniques are indispensable for identifying and localizing specific biochemical components within tissues and cells, utilizing targeted chemical or biological reagents \cite{norris2013analysis}. Histochemical methods rely on characteristic chemical reactions of biomolecules, with protocol selection dictated by the target's biochemistry and tissue preparation requirements. Crucially, these techniques provide significant diagnostic and research insights: The periodic acid-Schiff (PAS) reaction is vital for detecting carbohydrates (e.g., mucins), aiding in the diagnosis of gastrointestinal and mucinous tumors. Lipid visualization requires frozen sections (avoiding solvent dissolution) and specialized stains (e.g., Oil Red O), key for studying metabolic disorders. Nucleic acid stains like the Feulgen reaction (DNA) and methyl green-pyronin (DNA/RNA) enable assessment of ploidy changes in tumors. Enzyme activity localization (e.g., acid phosphatase) reveals metabolic alterations in disease states. Connective tissue stains like Masson's trichrome are essential for differentiating collagen from muscle, critical in evaluating fibrosis and lesion morphology. Finally, silver staining of nucleolar organizer regions (AgNOR) serves as a valuable marker of cellular proliferation, assisting in distinguishing benign from malignant tumors. These targeted histochemical approaches are fundamental tools for elucidating tissue composition, cellular function, and pathological processes in both diagnostic and research contexts.

\textbf{Immunohistochemistry} (IHC) significantly enhances the sensitivity and specificity of pathological diagnosis by utilizing the precise binding affinity of antibodies, conjugated to detectable labels (fluorescent, chromogenic, isotopic, or colloidal gold), for specific tissue antigens \cite{atmar2023immunologic,hasic2022immunohistochemistry}. Its power lies in the strategic application of diverse labeling strategies and binding modes (direct antigen-antibody, affinity-based, polymer-enhanced) tailored to the target antigen and tissue context. This versatility makes IHC indispensable for elucidating complex antigen expression patterns, particularly evident in the detailed analysis of serial tissue sections (Fig \ref{fig1}.b), which reveals spatial heterogeneity and dynamic changes across tissue architecture.

Immunofluorescence (IF) achieves high-resolution spatial mapping of antigens by exciting fluorophore-conjugated antibodies, enabling precise co-localization studies and quantitative analysis of multiple targets within cells or tissues under fluorescence microscopy. The immunoenzyme labeling method (IELM), leveraging enzyme-antibody conjugates (e.g., horseradish peroxidase) and chromogenic substrates, provides robust, permanent, and high-contrast visualizations of antigen distribution under standard light microscopy. Amplification techniques (e.g., ABC, SP) further boost signal intensity and detection sensitivity, crucial for low-abundance targets, while also facilitating compatibility with electron microscopy. Immunocolloidal gold technology (ICGT), utilizing colloidal gold-labeled probes, offers exceptional versatility for multi-scale imaging. Its unique properties—minimal interference with biomolecule function and distinct electron density—make it the gold standard for ultra-structural localization in immunoelectron microscopy. Furthermore, its visible color under light microscopy, often enhanced by silver deposition, allows for correlative light and electron microscopic analysis of antigen distribution.

\textbf{Sequencing} understands cellular heterogeneity and dynamic processes within tissues requires capturing gene expression in situ while preserving spatial context. Spatial Transcriptomics (ST) addresses this fundamental challenge by enabling the precise mapping of transcriptional activity directly within intact tissue architecture, revealing how gene expression patterns define cellular identity, communication, differentiation states, and temporal dynamics across serial sections \cite{rao2021exploring, paik2020single}. This capability transforms the ability to decipher the functional organization of tissues and identify spatially regulated biomarkers critical for development, disease mechanisms, and potential therapeutic targets. Diverse methodological strategies achieve this goal, each offering distinct advantages for specific biological applications.

Techniques can be broadly categorized based on how they preserve spatial context. Microdissection-based approaches (e.g., LCM) isolate specific tissue regions or single cells under microscopic guidance for downstream sequencing, offering high cellular resolution and suitability for rare cell types. FISH-based methods directly visualize RNA molecules within fixed tissues using fluorescent probes, enabling subcellular localization and expression variability analysis. In Situ Sequencing (ISS) performs transcript sequencing directly on tissue sections, retaining spatial information without physical isolation. Finally, spatially barcoded capture methods (e.g., Visium, VisiumHD, Slide-seq) use spatially indexed arrays to map gene expression across whole tissue sections, offering unbiased, high-throughput spatial profiling ideal for tissue-wide discovery.

\textbf{Mass Cytometry} fundamentally transforms cellular heterogeneity analysis by utilizing stable metal isotope-tagged antibodies coupled with time-of-flight mass spectrometry. This paradigm overcomes the spectral limitations of fluorescence-based cytometry, enabling simultaneous quantification of 40+ protein markers per cell with near-zero signal overlap. Its core innovation extends beyond suspension analysis (Classic CyTOF)—which deciphers immune cell hierarchies and rare populations in dissociated tissues—to spatially resolved technologies: Imaging Mass Cytometry (IMC) and Multiplexed Ion Beam Imaging (MIBI). These techniques map protein expression directly on tissue sections via laser ablation or ion beams, preserving spatial context to reveal cellular neighborhoods, signaling gradients, and tumor-immune ecosystem dynamics at subcellular resolution. Crucially, cell barcoding with metal tags allows multiplexed sample processing, slashing batch effects while enabling high-throughput clinical cohort studies. By integrating with transcriptomics, mass cytometry unifies protein, spatial, and molecular data into systems-level models of disease mechanisms, driving breakthroughs in precision oncology (e.g., therapy resistance mapping), autoimmune disorder profiling, and stem cell niche deconvolution.

%The ISS padlock method is based on padlock detection, rolling circle amplification (RCA), and ligation chemistry sequencing. In intact tissue sections, mRNA is reverse transcribed into cDNA, and the mRNA is subsequently degraded by RNase H. The ISS padlock method is based on padlock probing, rolling-circle amplification (RCA), and sequencing by ligation chemistry. Within intact tissue sections, mRNA is reversely transcribed to cDNA, which is followed by mRNA degradation by RNase H. Fluorescence in situ sequencing (FISSEQ), like ISS padlock, is a sequencing method using reverse transcription, rolling circle amplification, and ligation techniques.Barcode in situ targeted sequencing (Barista-seq) is an improvement on the gap padlock probe methodology boasting a fivefold increase in efficiency, an increased read length of fifteen bases and is compatible with illumina sequencing platforms. Spatially-resolved transcript amplicon readout mapping (STARmap) utilizes a padlock probe with an additional primer which allows for direct amplification of mRNA, forgoing the need for reverse transcription. Similar to other padlock probe-based methods amplification occurs via rolling circle amplification. The DNA amplicons are chemically modified and embedded into a polymerized hydrogel within the cell. Captured RNA can then be sequenced in situ providing three-dimensional locations of the mRNA within each cell.\\

\subsection{Serial Sections Imaging Techniques}
% 从高分辨率数字扫描仪扫描的整个幻灯片图像提供了有关生物系统形态和功能特征的丰富信息。 因此，这种图像模式提供了对组织生物进展的潜在机制的见解。
Optical automated microscopy converts stained sections into a digital format, while electron microscopy is used to examine ultrastructural details. Multiplexed tissue imaging techniques can detect biomarkers in situ and track the start position of mRNA transcripts to the original tissue sections, thereby precisely positioning spatial transcriptomics data \cite{yang2024virtual}. High-resolution digital scanned images like the above provide rich information about the morphological and functional characteristics of biological systems, helping to gain a deeper understanding of the underlying mechanisms of tissue biology progress.

\begin{figure}[h!]
    \centering
    \includegraphics[width=\textwidth]{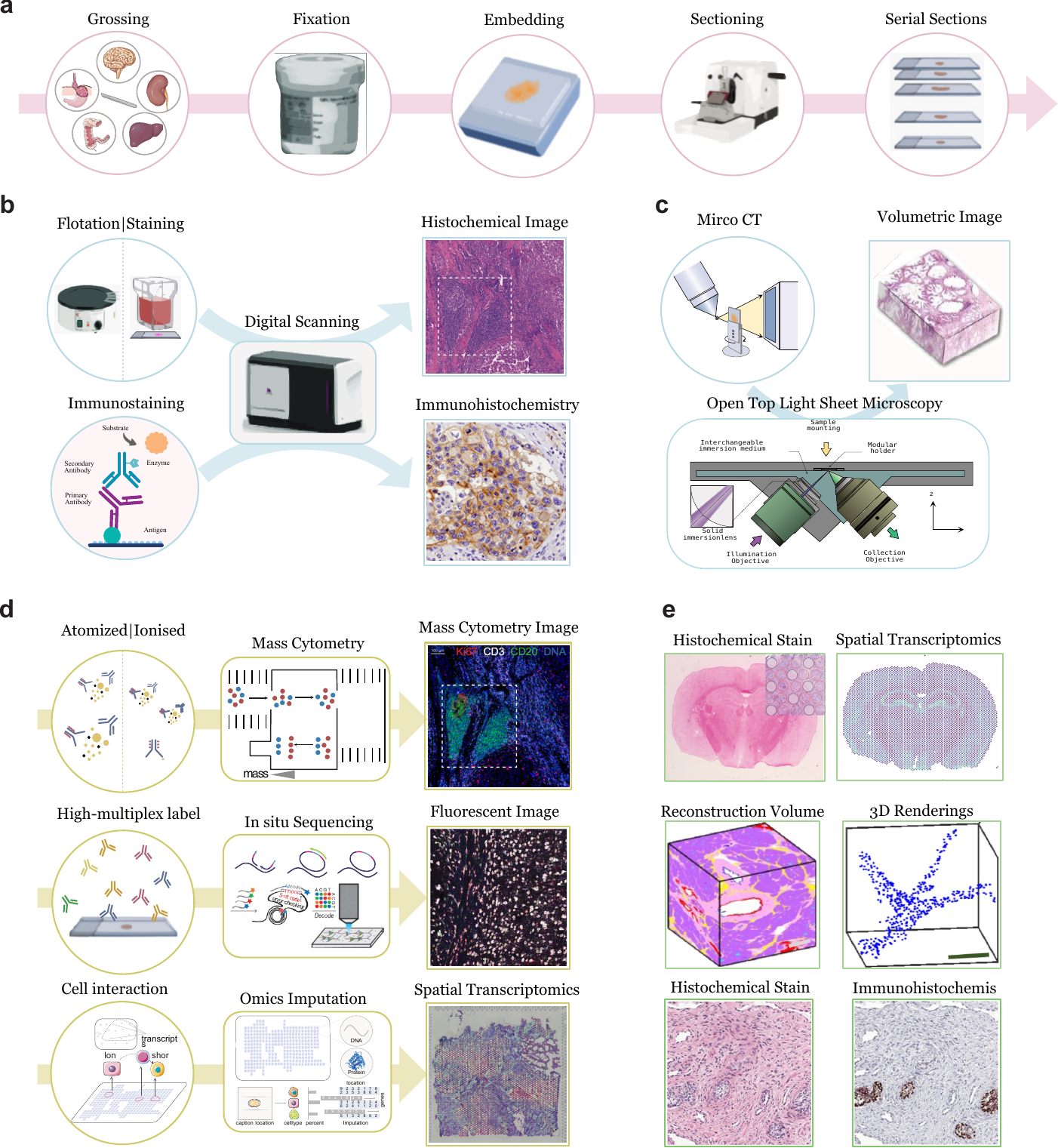}
    \caption{\textbf{Overview of Serial Sections Workflow.} \textbf{a.} Section Preparation. The sampled tissue undergoes gross processing, fixation, and embedding to preserve structural integrity. It is then embedded in a supportive medium, sectioned into serial thin slices with a microtome. \textbf{b.} Stain and Micro-Scanning. Tissue staining is primarily divided into chemical staining to highlight morphological structure and immunohistochemical staining to detect the expression and localization of specific proteins, with whole slide images captured through digital scanners. \textbf{c.} OTLS-based methods use thin light sheets to selectively excite fluorescence in samples, which is suitable for 3D imaging of large-sized cleared tissue samples. \textbf{d.} High-multiplex labeling enables the simultaneous detection of multiple biomarkers. Mass cytometry integrates flow cytometry and mass spectrometry to detect numerous markers. In situ sequencing uses fluorescent nucleotides and imaging to reveal spatial gene expression. Spatial omics uses barcode mapping to track spatial molecular information in tissues. \textbf{e}. Multimodal serial section pairs examples.}
    \label{fig1}
\end{figure}

% \subsubsection{Optical Digital Microscope Scan}
% 切片的软硬度和切片刀的锋利度都决定了切片质量，同时切片扫描潜经过硬脂酸处理后，对于减少切片上肉眼看不见的细小皱折有很大的作用，也可以大大提高切片及扫描速度
\textbf{Optical Digital Microscope} depends on specimen hardness and blade sharpness, with stearic acid treatment on the slicing stage reducing microscopic wrinkles and enhancing slicing quality and scanning speed (Fig \ref{fig1}.b). Serial section scanning electron microscopy (SEM) enables high-resolution 3D imaging with nanometer precision across large fields of view. Using an SEM equipped with an in situ ultramicrotome, this method sequentially acquires images as each sample layer is cut and exposed for imaging \cite{antao2024sample}.

Destructive imaging methods like knife-edge scanning microscopy and micro-optical sectioning tomography capture 2D images from serial sections to reconstruct 3D samples. In contrast, confocal and multiphoton laser scanning microscopy non-destructively generate 3D images by illuminating single points within thick samples and scanning spatially \cite{liu2021harnessing}. Light-sheet microscopy creates 3D images by illuminating 2D optical sections in transparent samples, with fluorescence imaged onto a high-speed camera. OTLS microscopy provides lateral imaging by moving the sample on a transparent holder, while a camera collects angled light-sheet images for rapid 3D visualization. This setup can also translate laterally and vertically to accommodate large tissue volumes (Fig \ref{fig1}.c) \cite{glaser2022hybrid}.
% Selected imaging methods for stacked pathology include knife-edge scanning microscopy and micro-optical sectioning tomography, which are destructive techniques that acquire two-dimensional images from serially sectioned specimens. Stacks of adjacent 2D images are used to reconstruct the 3D image of the sample. Confocal microscopy and multiphoton laser scanning microscopy typically illuminate single points within thick samples and perform spatial scanning in three directions, generating 3D images non-destructively over time. Light-sheet microscopy illuminates 2D optical sections within transparent thick samples, and the fluorescence generated within the light sheet is imaged orthogonally onto a sensitive high-speed camera. By scanning the 2D light sheet through the sample (or vice versa), 3D images can be rapidly generated plane by plane. The architecture of OTLS microscopy allows for unconstrained lateral imaging of one or multiple samples placed on a transparent sample holder, akin to a flatbed scanner for tissues. As the sample moves, the camera (via a collection objective) quickly gathers angled light-sheet images extending into the sample. For imaging large tissue volumes, the sample holder can translate laterally and vertically.

% \subsubsection{Ultra-micro Imaging}
\textbf{Ultra-micro Imaging}, including transmission electron microscopes (TEM) and scanning electron microscopes (SEM), enables ultrastructural observation of cell membranes, nuclei, organelles, and more. TEM examines internal tissue structures using fresh tissue samples embedded in resin and stained for imaging, while SEM focuses on surface structures, requiring dehydration and drying instead of sectioning \cite{mukhamadiyarov2021embedding}.

\textbf{The Laser Scanning Confocal Microscope (LSCM)} plays a crucial role in advancing the understanding of cellular structure and function. Widely used in the life sciences, LSCM enables high-resolution, three-dimensional visualization of biological samples and supports dynamic, real-time analysis of living cells. By capturing fluorescence signals with high precision, it allows researchers to explore the spatial organization and interactions of key biomolecules within cells and tissues. This capability is particularly valuable for studying gene expression, protein localization, and intracellular processes such as pH regulation and ion exchange. Beyond static observation, LSCM also provides insights into the dynamics of cellular behavior, including membrane fluidity and molecular diffusion. As a result, it serves not only as a visualization tool but also as a quantitative platform for probing the molecular underpinnings of physiological and pathological states, contributing significantly to fields such as cell biology, developmental biology, and pathology.

\textbf{Multiplexed and Flow Cytometer Imaging} represents a growing frontier in biomedical imaging, integrating established diagnostic techniques with emerging spatial and high-throughput technologies. Traditional imaging modalities—such as X-ray mammography, CT, MRI, PET, SPECT, and ultrasonography—are widely used for cancer diagnosis and staging. These methods typically address either single diagnostic endpoints or, in some cases, multiple parameters through combined imaging. Recently, modalities like optical imaging (OI) and photoacoustic imaging (PAI) have introduced intrinsic multiplexing capabilities, broadening the potential for comprehensive clinical insights \cite{aziz2020medical}.

High-multiplex tissue imaging (HMTI) extends beyond conventional immunohistology, enabling spatially resolved, single-cell analysis of protein expression on FFPE or fresh-frozen tissues \cite{einhaus2023high}. Among these, Imaging Mass Cytometry (IMC) has emerged as a key spatial biology platform. Based on cytometry by time-of-flight (CyTOF), IMC utilizes metal-tagged antibodies to detect over 40 markers simultaneously at subcellular resolution, circumventing the limitations of traditional immunofluorescence, such as autofluorescence and spectral overlap \cite{tan2020overview}. Combined with spatial transcriptomics, metabolomics, and machine learning, HMTI techniques like IMC are reshaping tissue diagnostics and biomarker discovery, offering unprecedented depth and precision in spatial molecular profiling.

\textbf{Time Dimensions}. Longitudinal pathology records contain rich multidimensional data, making them ideal for time series analysis. To support such applications, slide scanning technologies aim to balance image quality and acquisition speed. Static high-fidelity scanning ensures maximum image clarity by pausing at each point but is time-consuming. Dynamic scanning uses brief light pulses during continuous motion to reduce blur, though often at the cost of SNR. Line scanning or TDI further improves efficiency by using one-dimensional sensors to enhance SNR during motion. For reliable benchmarks, low-speed, fixed-exposure scans are used to obtain diffraction-limited ground truth, supplemented by offset scans to verify spatial fidelity. Modern workflows also incorporate multi-plane scanning and Z-stacking, which improve focus selection and enable 3D reconstruction. Efficient compression (e.g., HEVC) reduces storage needs without sacrificing detail. The Z-plane information, measuring depth in microns, is crucial for spatial accuracy and 3D modeling. Together, these innovations support a time-series-based pathology analysis framework that integrates spatial precision, efficient storage, and longitudinal insight.  %Longitudinal pathology records of patients contain multidimensional data, which are naturally suitable for time series analysis for prioritization. In the field of microscope slide scanning, several strategies are commonly used to balance time efficiency and image quality. The most basic is "static high-fidelity scanning" (Stop-and-Start), which pauses at each position to obtain the highest quality image, but is time-consuming. Another dynamic method is to use brief light pulses to illuminate the slide during movement to reduce motion blur, but this often comes at the expense of signal-to-noise ratio (SNR). A third line scanning or time delay integration (TDI) technique uses a one-dimensional sensor to continuously acquire signals, which can improve SNR during movement.  To ensure the reliability of the analytical benchmark, the same area was scanned at a low speed and fixed exposure to obtain images of diffraction-limited quality as ground truth, supplemented by additional static high-fidelity scans with lateral displacement to verify the spatial fidelity of these reference images.

\section{Techniques for Serial Sections}\label{section3}
\subsection{Registration}
\begin{figure}[h!]
    \centering
    \includegraphics[width=\textwidth]{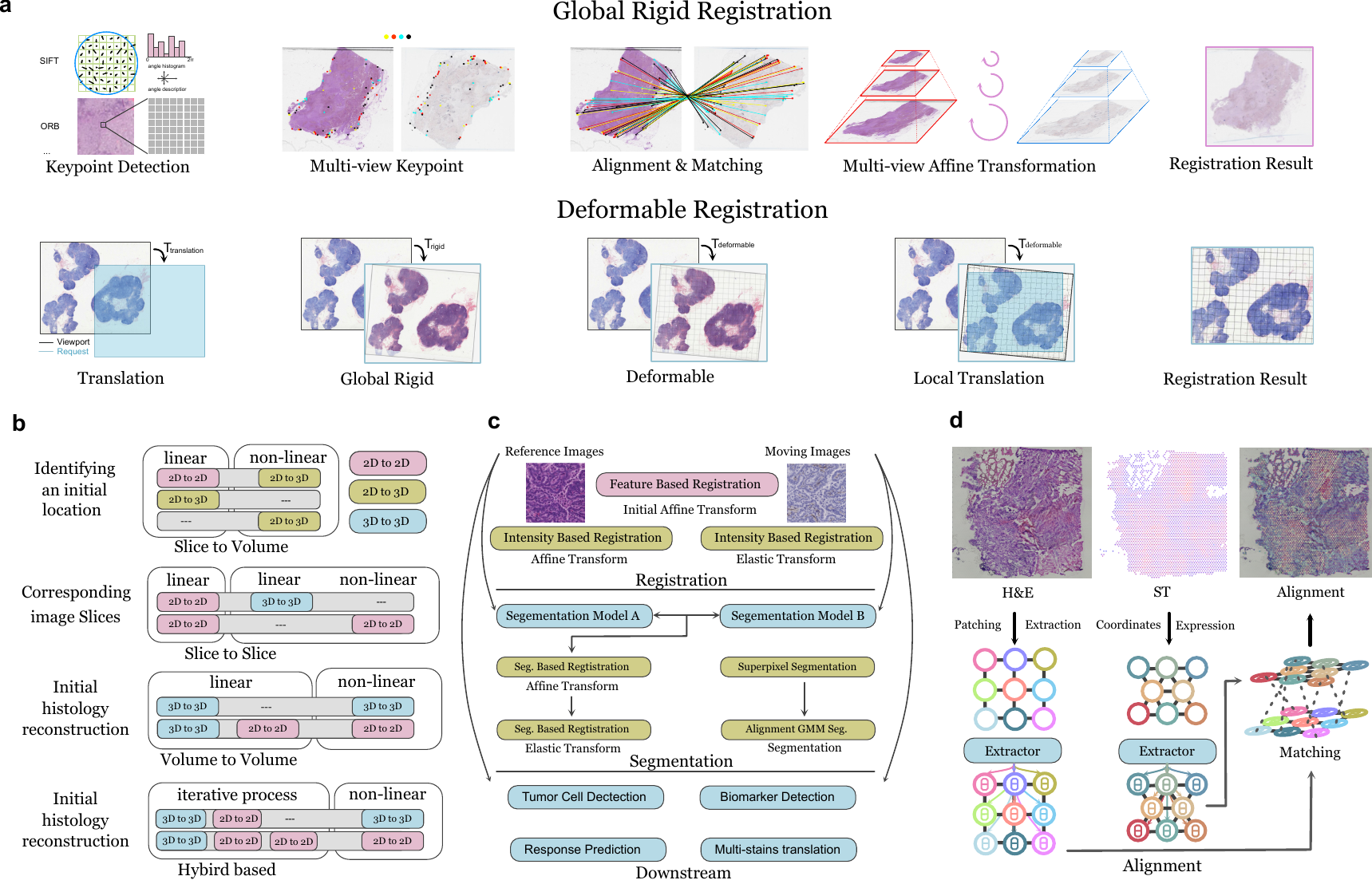}
    \caption{\textbf{Illustrative of Multi-view tasks for Serial Sections}. \textbf{a.} Slices Registration: Key point detection via ORB algorithm enables global rigid registration for initial alignment. Deformable and local translation registration then refines alignment, adapting to local deformation and structural changes. \textbf{b.} 3D Reconstruction Workflow: Converts 2D slices to 3D models using linear/nonlinear methods. Involves 2D-2D, 2D-3D, and 3D-3D correspondence, with initial slice position identification critical. \textbf{c.} Multi-stain Analysis. Begins with feature-based registration \& affine transform, incorporates intensity-based registration. Uses superpixel \& GMM segmentation for alignment. Includes elastic transformation \& segmentation-based registration, culminating in tumor cell detection and biomarker identification. \textbf{d.} Cross Modality. Cross-modality alignment workflow for histopathology imaging, illustrating the transition from histochemical-stained images to spatially registered gene expression via patch-based segmentation and multimodal matching.}
    \label{fig2}
\end{figure}

The registration of images involves transforming data into a common coordinate system to achieve alignment (Fig \ref{fig2}.).  Image registration is the process of aligning two or more images. Given a reference image $\mathcal{R}$ and a translation image $\mathcal{T}$, the goal is to find a suitable geometric transformation $y$ that makes $\mathcal{R}$ and $\mathcal{T}$ as similar as possible with respect to parameters of an image transformation model \cite{he2023deformable}:
\begin{equation}
    \mathcal{J}(w) = \mathcal{D}[\mathcal{T}(y(w,x)),\mathcal{R}] + S(w)
\end{equation}
The similarity metric $\mathcal{D}$ indicates the registration accuracy and $S(w)$ is a regularization term. This problem can be formulated as the minimization of the joint objective function $\mathcal{J}$ concerning the parameters of the image transformation model. However, when registering serial histopathological tissue sections, several challenging issues arise, including the introduction of tissue loss, noise, morphological deformation, and potential problems caused by data preparation in individual steps. Current work in this field integrates the advantages of methods based on spatial and global-to-local feature information. For heterogeneous serial sections, identifying registration key points and deformation fields is more challenging due to local and global similarity concerns. Additionally, differences in pixel intensity between the target and source images may occur due to variations in staining. 

Registration is crucial in various medical image processing tasks across different modalities. Rigid pre-alignment aligns serial slices to match their original positioning in the tissue sample block, addressing the challenge of combining rigid and deformable methods. Deformable registration follows established techniques, and careful microtome operation minimizes cross-section variation. The specimen and knife are rigidly mounted, compensating for intersection variation across the series. Each section is imaged as overlapping tiles, with section index 
$s$ assigned by the operator, though these coordinates are prone to measurement errors $\theta_t =(u,v)^T$. A section's pose relative to its neighbors involves an unknown rigid transformation $\mathbf{R}_s$. Additionally, sectioning and preparation, as well as the imaging process, introduce noticeable deformations both at the section level ($\mathbf{D}_s$) and tile level ($\mathbf{D}_t$). Assuming planar sections of constant thickness, scene and section spaces are connected by a permutation $\mathbf{P}_{zs}$ between the scene dimension 
$z$ and section index $s$, leading to the following mapping \cite{pichat2018survey}:
\begin{equation}
    \binom{u}{v}=\mathbf{R}_{s} \mathbf{D}_{t} \mathbf{D}_{s} \mathbf{P}_{z s}\left(\begin{array}{l}
x \\
y \\
z
\end{array}\right)+\mathbf{t}_{t}+\theta_{t}
\end{equation}
with $\mathbf{R}_s\mathbf{D}_t\mathbf{D}_s$ and $(x,y,z)^T$ being unknown. Without precise knowledge about the deforming process $\mathbf{D}_t\mathbf{D}_s$, the appropriate solution is that with minimal no-rigid deformation for all sections, a model being as-rigid-as-possible. This coincides with the researcher's request not to introduce unmotivated artificial deformation to the image data that would adulterate this observation. Declaring a tile being the unit of rigidity, we approximate $\mathbf{R}_s \mathbf{D}_t \mathbf{D}_s + \mathbf{t}_t$ by a rigid-per-tile transformation $\mathbf{R}_t$ such that 
\begin{equation}
    \binom{u}{v}=\mathbf{R}_{t} \mathbf{P}_{z s}\left(\begin{array}{l}
x \\
y \\
z
\end{array}\right)+\omega_{t}+\theta_{t}
\end{equation}
Let $T$ be the set of all tiles $t\in T$ and $R$ be the set of all rigid transformations $R_t \in R$, then the best configuration $R$ is that minimizing the sum of all relative transfer errors $\epsilon$ of pairwise overlapping tiles:
\begin{equation}
    arg \min_{R}\sum_{t\in T} (\sum_{o \in T \backslash \{t\}} \epsilon_{to})
\end{equation}
where \cite{zhang2020point} proposes an accurate non-rigid registration method optimized via extracted points. Most similarity measures rely on features from the fixed and moving images. Control points are defined on a regular grid overlaid on the fixed image, with a cubic multi-dimensional B-spline polynomial applied. The B-spline coefficient, control point spacing, and the set of all control points within the compact support are considered. The objective function is optimized with respect to these control points, with spacing set at several pixels. A 3rd-order B-spline interpolation is used during deformable registration. Marzahl et al. \cite{marzahl2021robust} introduce a multi-scale approach based on a quad-tree (QT), with several termination criteria that makes the algorithm particularly insensitive to tissue artefacts and that further allows to estimate a piece-wise affine transformation. Various optimization approaches and heuristics have been proposed to discover the optimal alignment. Schwarz et al. \cite{schwarz2007non} applied symmetric compositive unimodal Demons, symmetric compositive MIND Demons, and TPS interpolation using the best matches.
% \subsubsection{Intensity and keypoint-based Alignment}
% \subsubsection{Multi-cue Alignment}
% 两大类配准方法是基于强度的方法和基于关键点的方法。 基于强度的方法试图最小化移动源和目标的图像之间的强度梯度，相比之下，基于关键点的策略使用点特征提取器，如定向FAST和悬转BRIEF, 较为传统的是缩放不变特征变换 SIFT，然后匹配和对齐多个图像中的关键点，这两种情况下，提取的信息都用于计算刚性变换和变形场，以补偿连续切片之间的失真。 而有或没有不同染色剂的连续载玻片的配准需要非刚性配准，除了经典图像 处理的方法外，基于深度学习的方法也可以应用于整体或部分配准任务。
Image registration methods broadly fall into intensity-based and keypoint-based categories. Intensity-based approaches optimize alignment by minimizing pixel-wise intensity differences between source and target images, while keypoint-based methods extract distinctive features—such as ORB (Oriented FAST and Rotated BRIEF) \cite{rublee2011orb} or SIFT (Scale-Invariant Feature Transform) \cite{lowe2004distinctive}—and align corresponding keypoints. Both estimate rigid transformations or deformation fields to correct spatial discrepancies, commonly encountered in serial tissue sections. 

Beyond conventional techniques, deep learning and hybrid strategies enhance global and local alignment: Cifor et al. \cite{cifor2013hybrid} proposed block-matching using intensity-based similarity, while Schwier et al. \cite{schwier2013registration} addressed staining variability by registering distance-weighted vessel masks to reduce sensitivity to intensity artifacts. Their method optimizes squared intensity differences and applies computed displacement fields to original images, though misregistration persists in regions with homogeneous intensities or inconsistent staining. For molecular-level alignment of adjacent tissue sections, PASTE \cite{zeira2022alignment} integrates spatial transcriptomics (ST) data by computing pairwise alignments from transcriptional similarity and spatial proximity, fusing these into a coherent 3D tissue representation. In histopathological imaging, dyes exhibit complex overlapping absorption spectra despite appearing as distinct colors. Improved data normalization—such as channel-wise optimization based on image histogram distributions—reduces variability and enhances tissue patterns (e.g., nuclear morphology \cite{clarke2018d}), thereby strengthening both global feature matching and localized alignment robustness.

%提出的配准完全在距离加权的vessel掩码图像上执行, 因此由于相邻图像切片不受局部强度分布和染色特定变化的影响，可以使用基于强度的平方差和来优化配准过程，最后计算出位移应用于原始图像，但这种方法可能会对组织图像应用产生误导，不仅因为会出现目标中像素的强度值和转换源中错误配准像素的强度值相似，因为局部组织图像特征的相似性，也因为目标图像中像素强度和转换源中的准确的像素强度可能会因染色变化而显得不同。

% 对于生物图像，尽管使用的染料被可视化为具有不同的颜色，但所得染色剂实际上具有复杂的重叠吸收光谱。而改进的数据归一化过程能够减少图像变化并增强组织模式，这极大地有利于全局特征匹配和基于局部区域的定向匹配过程。它根据对图像直方图分布的分析自动优化各通道，以有效试别主簇的动态范围并处理类似于长尾问题，能够较好地增强组织图像特征如核模式

\subsection{Consecutive Slices for 3D Analysis}\label{section3.2}
3D reconstruction originated in embryology, enabling the visualization of the spatial relationships within developing systems and organs (Fig \ref{fig4}.b). This technique has since expanded to biomedicine, where the examination of individual stained sections offers only a partial understanding of normal and abnormal tissues. However, despite providing visual insight, achieving realistic reconstruction without prior knowledge of tissue morphology remains elusive. The development of 3D medical imaging has made this structural ground truth accessible. Moreover, combining non-invasive imaging with histology creates valuable opportunities to link macroscopic information with the underlying microscopic characteristics of tissues by establishing spatial correspondences. Image registration is a technique that allows the automation of this process, and we describe reconstruction methods and applications that rely on it.
% Multimodal分块写作流程总结：
% 写介绍什么是3D重建，主要做个什么事
% 介绍了具体任务分类 ：分类分割检测：任务
% 主要做个什么事的每一步：具体方法有哪些(构图+图卷积)
% 第二段写了另一种场景
% \subsubsection{Applications of 3D Reconstruction} 
Analyzing three-dimensional (3D) pathological models offers a new dimension of insight, revealing the intricate morphology of histological specimens that can often be missed in traditional 2D images. Critical information, such as the spatial structure of blood vessels, is frequently lost in 2D analysis. Consequently, there is a growing imperative to reassemble consecutive slices into a cohesive 3D volume, enhancing the understanding and accuracy in histopathological diagnostics \cite{liang2015liver, tolkach2018three, duanmu2021spatial, almagro2025ai}. 

The application of 3D reconstruction using consecutive slices is highly versatile. For example, Arslan et al. \cite{arslan2023efficient} builds 3D vascular models to help predict and understand the metabolic states of cancer cells within the tumor. Kiemen et al. \cite{kiemen2024high} use the 3D reconstruction of consecutive tissue slices to investigate the true structures of normal ducts, Pancreatic ductal adenocarcinoma (PDAC), and pancreatic intraepithelial neoplasia (PanIN), revealing significant changes in the human pancreatic architecture during tumorigenesis. Deng ei al. \cite{deng2021map3d} automatically identifying and associating large-scale cross-sections of 3D objects like glomerular from routine consecutive sectioning and WSI.

The challenges of 3D reconstruction primarily stem from the histological sectioning process, which often introduces artifacts and distortions such as holes, foldings, tears, and complex deformations. Additionally, unbalanced staining and variations in data appearance further complicate the task. Consequently, many researchers have dedicated significant efforts to developing algorithms for 3D reconstruction from consecutive slices.

%3D重建方法主要分为两个步骤，paired registration and 3D 渲染重建
% \subsubsection{Methodology}

% Most 3D reconstruction methods consist of two main steps: paired-slide registration and the subsequent final 3D reconstruction with the whole slide series.

\textbf{Paired-Slide Registration}.
% In \cite{deng2021map3d}, authors used scale-invariant feature transform \cite{lowe2004distinctive}, speeded-up robust features \cite{bay2006surf} and recent Graph Neural Network based SuperGlue \cite{sarlin2020superglue} for affine registration, and advanced normalization tools \cite{avants2008symmetric} for non-rigid registration. Gatenbee et al. \cite{gatenbee2023virtual} processed slides into single-channel images normalized them, and used BRISK \cite{simonyan2014learning} and VGG \cite{li2003foreground} for feature detection and matching. They estimated transformation matrices from matched coordinates \cite{umeyama1991least} and used Tukey’s box plot for secondary filtering. Non-rigid and micro-registration were performed using Deep Flow, SimpleElastix, or Groupwise SimpleElastix \cite{klein2009elastix,shamonin2014fast,marstal2016simpleelastix,weinzaepfel2013deepflow} for high and higher resolutions, respectively. In addition, many approaches enhance this process by segmenting or detecting specific regions of interest relevant to the task such as vessels, or color devolution at hand. Wu et al. \cite{wu2023general} trained a network utilizing a hybrid architecture consisting of two branches: a gated axial transformer and a CNN-transformer combination, serving as the backbone to extract both global and local information. Then they generate layered point clouds based on down-sampled semantic segmentation results. Finally, they integrated the ICP (Iterative Closest Point) strategy to register the segmented point clouds effectively.
Paired-slide registration, particularly between consecutive tissue sections, has evolved from classical feature-based methods to hybrid deep learning and geometry-aware strategies. Early works such as Deng et al. \cite{deng2021map3d} employed classical local feature detectors like SIFT \cite{lowe2004distinctive}, SURF \cite{bay2006surf}, and GNN-based SuperGlue \cite{sarlin2020superglue} to establish correspondences, followed by affine and non-rigid transformations using ANTs \cite{avants2008symmetric}. These methods are interpretable and modular, yet may struggle under severe tissue deformation or staining variation.

Gatenbee et al. \cite{gatenbee2023virtual} further emphasized normalization and simplification, converting slides to single-channel images to reduce inter-slide variability. They combined handcrafted features (BRISK \cite{simonyan2014learning}, VGG \cite{li2003foreground}) with geometric filtering (e.g., Tukey's box plot), and leveraged hierarchical registration—using DeepFlow and (Groupwise) SimpleElastix \cite{klein2009elastix,shamonin2014fast,marstal2016simpleelastix,weinzaepfel2013deepflow}—to improve robustness across resolutions. This pipeline reflects a pragmatic, resolution-aware design, but still relies heavily on accurate initial correspondences.

Recent trends point toward integrating semantic priors and geometric reasoning. Wu et al. \cite{wu2023general} proposed a hybrid architecture that combines a gated axial transformer with a CNN-transformer backbone to capture global-local tissue structures. By generating layered point clouds from down-sampled segmentation maps and applying ICP-based refinement, their method bridges pixel- and geometry-level representations. This approach is especially promising for scenarios with limited appearance consistency but preserved anatomical topology. The field is shifting from purely intensity- or feature-based matching toward semantically enriched, multi-scale, and geometry-aware registration frameworks. Incorporating domain-specific cues—such as vessel structures or stain decomposition—offers further potential for tailored, pathology-aware alignment.

\textbf{3D Volume Reconstruction}
Compared to standard registration methods, a crucial step in 3D reconstruction using consecutive slicess is determining how to register the whole slide series to achieve the final 3D registration result or render the 3D outcome. Deng et al. \cite{deng2021map3d} proposed quality-aware whole series registration method with cycle-consistent registration failure detection, to perform automatic QA to classify the quality of the registration on the entire series as “good”, “acceptable”, or “bad” with paired slides. Gatenbee et al. \cite{gatenbee2023virtual} ordered slides based on their feature similarity, such that the most similar images are adjacent to one another in the z-stack and then performing pairwise registration. Kiemen et al. \cite{kiemen2022coda} created multilabeled images by the cell detected by CODA pipeline were consolidated into a 3D matrix using the H$\&$E image registration results. Wu ei al. \cite{wu2023general} Develop a hybrid approach combining rendering and interpolation: rendering stacks existing data to form an initial vascular volume, while interpolation fills in gaps between the H$\&$E-rendered tissue sections.

The results of 3D reconstruction can be further analyzed to explore various aspects, such as the microscopic anatomical structures \cite{kiemen2022coda}, prognostic differences among different 3D structures \cite{wentzensen2007combined}, and their impact on grading and heterogeneity \cite{tolkach2018three}. In addition, Deng et al. \cite{deng2021map3d} uses image registration for affinity estimation in the multi-object tracking of glomeruli detected by CircleNet. Jeon et al. \cite{jeong2010interactive} enables interactive histology of ultra-large microscopy image stacks using a locally adaptive data structure and demand-driven processing. It efficiently handles massive images (e.g., 30GB confocal datasets) without global pyramids by dynamically fetching required data. A GPU-based 3D texture compression exploits slice similarity for real-time decompression with adjustable quality and bitrate.

\subsection{Modalities Translation}
% \subsubsection{Stain Independent Segmentation}
\label{section3.2}
% 介绍使用连续切片多种染色数据做染色无关分割
\textbf{Stain Independent Segmentation}. Stains are essential for the analysis of histological images~\cite{multi-stains-bib1}. The combined information derived from applying different stains to the same tissue structure is highly valuable in medical research~\cite{multi-stains-bib2, multi-stains-bib3} but practically unfeasible. Typically, different stains are applied on serial slicess of the same specimen and the valuable structures are then segmented for further analysis. An automatic and precise segmentation pipeline for any stain would ensure more consistent and effective analysis across various experiments and laboratories. However, here comes the multi-stain segmentation challenge. It is difficult to develop a segmentation method that performs well across different stains of serial slicess due to the various color and texture characteristics of similar tissue structures when stained differently. Consequently, most segmentation algorithms are stain-specific~\cite{multi-stains-bib4, multi-stains-bib5, multi-stains-bib6} rather than stain-independent. Currently, there are three methodologic concepts to achieve stain-independent segmentation: image registration, stain translation, and stain augmentation.

\textbf{Registration-based}. Previous works have already focused on the combination of registration and segmentation~\cite{multi-stains-bib7, multi-stains-bib8}. The image registration methodologic concept of stain-independent segmentation was first proposed by Gupta et al.~\cite{multi-stains-bib9}. The authors proposed a pipeline by registering an arbitrarily stained WSI with a differently stained WSI that has an existing trained model, allowing the direct transfer of the segmentation mask. While this approach enables the segmentation of arbitrarily stained WSIs, it relies on serial slices with similar image content, which are generally not available.

\textbf{Translation-based}. Recently, image-to-image translation approaches~\cite{multi-stains-bib10, multi-stains-bib11}, especially generative adversarial networks(GAN)~\cite{multi-stains-bib10}, provide a new perspective for image analysis. These approaches achieve the stain translation of images from one stain to another~\cite{multi-stains-bib12, multi-stains-bib13}, or from unstained to stained~\cite{multi-stains-bib14, multi-stains-bib15, multi-stains-bib16}. Making use of an unpaired image-to-image translation approach, Gadermayr et al.~\cite{multi-stains-bib17} proposed the stain translation-based method and investigated two different pipelines for performing stain-independent segmentation. Concretely, one pipeline directly translates the target stain into a simulated version of the analyzable stain with a pre-trained segmentation model. The other one applies a reverse stain translation to generate fake target stain data, thereby retraining a new segmentation model for the target stain analysis. Both pipelines, especially the former, exhibit good performance. Further integrating unsupervised segmentation, Gadermayr et al.~\cite{multi-stains-bib18} comprehensively investigated four pipelines of the stain translation-based method.
% The results exhibit good performance especially for the first approach and provide evidence that the direction of translation plays a crucial role considering the final segmentation accuracy.

\textbf{Augmentation-based}. Different with image registration-based and stain translation-based methods, stain augmentation-based method translate the analyzable stain with a pre-trained segmentation model into the target stains for data augmentation, thereby retraining a new segmentation model that suits all stains. Bouteldja et al.~\cite{multi-stains-bib19} were the first to propose this method, presenting a comprehensive comparison with image registration-based and stain translation-based methods. The results demonstrate that stain augmentation significantly outperforms the other two methods.

%联合分割 特指多染色场景

% 不同类别有不同信息

\subsection{Cross-Modality Tasks}
%多模态
Multimodal data overcome the limitations of traditional staining, with cross-modal tasks playing a crucial role in linking molecular, spatial, and histological features.

In order to understand the interplay between distinct cell types in complex organ tissues, Ke et al. \cite{ke2013situ} developed an in situ sequencing method for parallel targeted analysis of short RNA fragments in morphologically preserved cells and tissue. It uses a pyramid approach to subpixel registration based on the intensity between a merged image of all signals and the general stain from the first hybridization step \cite{thevenaz1998pyramid}.  Besides, Wang et al. \cite{wang2018three} proposed and applied a technology for 3D intact-tissue RNA sequencing, termed STARmap, which integrates hydrogel-tissue chemistry, targeted signal amplification, and in situ sequencing.  For image data processing, STARmap first implements a three-dimensional fast Fourier transform to accomplish  image registration, where all images are registered to the first round of sequencing. After that, cells are identified by a Laplacian of Gaussian filter, quantifying the expression of each gene and providing other cellular analysis. Furthermore, spatial transcriptomics analysis based on serial sections has advanced the understanding of complex disease mechanisms. Chen et al. \cite{chen2020spatial} exploited in situ sequencing on mouse and human brain sections, which contains two outer immunostained sections and one middle section for spatial transciptomics, to confirm the majority of the observed alterations at the cellular level. It provides an unprecedented approach to untangle the dysregulated cellular network in the vicinity of pathogenic hallmarks of AD and other brain diseases. 

Moreover, there also exists many researches that focus on how to more effectively align the serial spatial transcriptomics data, pursuing more comprehensive identification of cell types and gene expression, which also helps with the data sparsity problem in individual slides. For example, Zeira et al. \cite{zeira2022alignment} introduced PASTE, a method that computes pairwise alignments of slices using the fused Gromov-Wasserstein optimal transport algorithm that models both transcriptional similarity and physical distances between spots, to align and integrate spatial transciptomics data from multiple adjacent tissue slices. Subsequently, Liu et al. \cite{liu2023partial} optimized PASTE, named as PASTE2, which performs partial pairwise alignment, only selecting and aligning a subset of spots. This problem is solved by a conditional gradient algorithm. PASTE2 also includes a model selection procedure to estimate the fraction of overlap between slides, and optionally uses information from histological images that accompany some spatial transcriptomics experiments. In particular, Xia et al. \cite{xia2023spatial} proposed SLAT, a graph adversarial matching strategy. SLAT is the first algorithm capable of aligning heterogeneous spatial data across distinct technologies and modalities. Utilizing multimodal data from serial sections can significantly enhance the richness of biological datasets and improve model generalization, leading to better performance in various biological analyses and applications.

\section{Summary}\label{section4}
\subsection{Literatures}
% 这篇工作调研了超过150篇病理连续切片相关的工作，并总结了8大子分类领域的兴衰随时间的变化情况，其中配准领域正处于转折时期，3D和虚拟染色正逐渐进入高速发展时期，这一趋势也可以从关键词趋势发展中看出。从词云中也可以看出贯穿连续切片的始终是配准相关的方法，这也是目前的2D成像方式决定的。
This study systematically surveyed more than 150 representative research works related to pathological continuous sections , covering the major progress in this field over the past decade. By classifying, counting and analyzing the timeline of the literature, the authors summarized the development and evolution trends of continuous section research in different directions and summarized eight major sub-research fields, including Registration, 3D Reconstruction, pathology survey, Segmentation, Virtual strain, Tools and Diagnosis, etc. shown in Fig\ref{fig3}.a.

As a basic link in continuous section research, section registration has long been a research hotspot, but in recent years, it is in a critical transition period from traditional rigid/affine registration to more intelligent, robust and automated registration strategies. More and more works have begun to explore registration methods based on deep learning, graph neural networks and cross-modal alignment. At the same time, 3D reconstruction and virtual staining, as emerging and forward-looking directions in recent years, are gradually entering a rapid development stage, especially in the context of high-throughput imaging and multimodal fusion, showing broad application potential. The time evolution trend of literature keywords also confirms this judgment. The relevant high-frequency keywords have increased significantly in recent years, showing that the research community pays high attention to these directions.

In addition, through the analysis of the keyword cloud, it can be observed that regardless of the research focus, almost all research work related to serial sections involves the core link of "registration". This phenomenon is closely related to the current mainstream 2D imaging method - due to the inevitable differences in position, angle, deformation, etc. between tissue sections, reliable image registration becomes the prerequisite for any subsequent cross-section analysis (such as three-dimensional modeling, molecular tracking, spatial map construction, etc.). Therefore, it can be considered that "registration" is not only one of the most core technical issues in serial section research, but also a key link throughout the development of this field.
\subsection{Series Slides Datasets}

In pathology, AI has potential to streamline workflows, reduce reading time, and enhance diagnostic accuracy, yet its clinical adoption is limited. A key barrier is the lack of suitable training data, particularly datasets of serial sections from the same tissue. These datasets are crucial for analyzing biological tissue information within the same spatial context but are challenging to obtain at scale. 
We counted the staining conditions of the survey datasets in Fig \ref{fig3}. b,among which H\&E staining is the most common because of its low cost. Sample size distribution: The number of samples of serial sections is mostly less than 100, and the proportion of data declared public is also relatively small, limiting AI’s effectiveness in the field \cite{panayides2020ai}.

% \subsubsection{MultiStaining}
%  连续切片的染色数据集通常图片尺寸大尺度多，组织器官类型单一，层数多, 染色模式复杂多样。 ANHIR 提出了对用不同染料染色的组织病理学组织样本的二维显微镜图像进行自动非线性图像配准，每种染色的外观不同，纹理重复。
\textbf{MultiStaining}. Staining-based datasets of serial histological sections often consist of large-scale images with many layers, focused on a single tissue or organ type, complex and diverse staining patterns. The ANHIR dataset addresses the automatic nonlinear image registration of 2D microscopic images of histopathological samples stained with different dyes, each of which presents a distinct appearance and repetitive textures. The ACROBAT dataset \cite{weitz2024acrobat} aims to register whole slide images (WSIs) of tissue sections stained with one of four conventional diagnostic IHC stains—ER, PGR, HER2, and Ki67—with corresponding WSIs stained with H\&E. WSI registration is particularly challenging due to the gigapixel scale of these images. Additionally, the ultra-thin tissue sections, only a few microns thick, are prone to deformation during sample preparation, and certain tissue regions present in one WSI may be absent in others from the same sample, especially when the sections are not consecutive. Moreover, the appearance of tissues varies significantly between different staining techniques.
These datasets primarily focused on high-resolution imaging in histopathology, incorporating various staining techniques.  Like WHS and MOST, employ Nissl staining to capture detailed brain structures in mice, with resolutions ranging from 0.35$\mu$m to 1$\mu$m. Additionally, the BPS dataset combines colocalized fluorescent labeling with counterstaining to enhance structural visibility. \cite{Liu_2022_CVPR} proposes a breast cancer immunohistochemical (BCI) benchmark attempting to synthesize IHC data directly with the paired hematoxylin and eosin (H\&E) stained images. This is an image-to-image translation task that builds a mapping between two domains (H\&E and IHC) under registration. Given a H\&E image, the algorithm should predict the corresponding IHC image. In addition, \cite{paknezhad2020regional} dataset double-stained and visualized the complete immune tissue collected from three patients with renal clear cell carcinoma. Although it consists of only three acquired image stacks, 20 blood vessels with different shapes and directions from different areas of the image were selected for registration, and tissue tearing, folding, and missing were frequently observed in all four image stacks. This brings a series of challenges to subsequent tasks.
% 此外M Paknezhad et al.数据集对收集的三名肾透明细胞癌患者的完整免疫组织进行双染色和可视化，虽然仅由三个获取的图像堆栈组成，但选择了20条具有不同形状和方向且来自图像不同区域的血管组合进行配准，且在所有四个图像堆栈中都经常观察到组织撕裂、折叠和缺失。
% \subsubsection

\textbf{Stacks in sequencing}. Spatial transcriptomics (ST) enables the measurement of mRNA expression across thousands of spots within a tissue section, while simultaneously capturing the 2D spatial coordinates of each spot. State-of-the-art technologies, such as 10$\times$ Genomics Visium, can generate datasets with up to 5,000 spots, each encompassing 1–30 cells \cite{zheng2017massively}. Numerous computational approaches have been proposed to analyze such data, including methods for identifying spatial gene expression patterns, detecting spatially variable genes, and inferring intercellular communication. However, meeting key technical requirements—such as efficiency, signal strength, accuracy, gene scalability, and broad applicability—remains a significant challenge.

In addition to ST, other technologies such as single-molecule fluorescence in situ hybridization (smFISH) can also capture both gene expression and spatial localization. Yet, a major limitation lies in the anatomical mismatch between tissue sections and array layouts, complicating the direct comparison of spatial coordinates across sections and hindering integration with other spatial or transcriptomic data. Furthermore, current ST platforms lack true single-cell resolution. To address this, hybrid approaches have been proposed that combine unbiased ST techniques with high-resolution in situ hybridization methods, such as RNAscope and multi-molecule ISS.

We curated and summarized spatial transcriptomics datasets across multiple species and tissues. Human datasets include the dorsolateral prefrontal cortex and embryonic heart from the ST platform, as well as a HER2-positive breast tumor dataset. Mouse datasets span multiple brain regions (e.g., anterolateral motor cortex, visual cortex, and primary cortex in 3D from STARmap \cite{wang2018three}; hypothalamic preoptic area from MERFISH \cite{wang2018multiplexed}; whole brain and olfactory bulb from Visium \cite{ortiz2020molecular}), as well as lymph node and skin slices. Drosophila embryo and Stereo-seq-derived scRNA-seq datasets \cite{wang2022high} are also included.

Recent advances in spatial omics technologies have expanded the resolution and scope of spatial analysis. Techniques based on fluorescent in situ hybridization—such as seqFISH, MERFISH \cite{moffitt2018molecular}, seqFISH+ \cite{chee2019nih}, and Xenium—as well as spatial barcoding approaches—such as 10$\times$ Visium \cite{zheng2017massively}, HDST, Slide-seqV2, Stereo-seq, and spatial-ATAC-seq—have enabled the investigation of cell positioning, cell–cell communication, and tissue function across serial sections. These methods have revealed intricate heterogeneity in tissue architecture, providing valuable insights into both physiological and pathological mechanisms.

Specifically, 10$\times$ Visium was applied to serial sections of the human dorsolateral prefrontal cortex, each with around 3,000 spots at 50$\mu$m resolution and >20,000 detected genes \cite{lewis2021spatial}. MERFISH was used to profile the mouse hypothalamic preoptic area, achieving subcellular resolution and detecting 151 genes \cite{moffitt2018molecular}. Stereo-seq profiled serial sections of E15.5 mouse embryos, capturing $\ge$100,000 single cells per section, covering 25 cell types and detecting $\ge$20,000 genes. In our analysis, all available sections were utilized (9 for Visium, 12 for MERFISH, and 4 for Stereo-seq), with statistical summaries computed across sections. 

\textbf{3D and Atlases datasets}. The quality of histological data from 3D and Atases is often inconsistent, with some datasets being extremely sparse, limited to thin sections from only a few serial slices. Additionally, the direction of serial slices can vary across datasets, further affecting the overall data quality. Poor contrast in histological sections also contributes to these challenges. 

Atlases are predominantly constructed using 3D brain datasets, as mapping brain structures in three dimensions is essential. High-resolution microscopic imaging of these datasets has significantly advanced the field, revealing pronounced heterogeneity and non-uniformity. For instance, the Allen CCFv3 dataset \cite{wang2020allen} provides a three-dimensional digital brain atlas with a resolution of 100 $\mu$m. Six data subsets were constructed from 1,675 STP brains, encompassing four resolution levels. Other datasets include the Waxholm Space (WHS) of the Nissl-stained C57BL/6J mouse brain, the Nissl-stained C57BL/6 adult mouse brain acquired through MOST imaging \cite{staeger2020three}, the Thy1-GFP M-line transgenic mouse brain imaged with BPS, a problematic dataset with imaging artifacts, and a dual-color labeled dataset featuring brain-wide fluorescent-labeled neurons and counterstained cell bodies with 2 $\mu$m axial and 0.32 $\mu$m horizontal resolution, as well as an autofluorescent C57BL/6 mouse brain imaged via STP microscopy \cite{johnson2010waxholm}. The preprocessing of 3D reconstruction data from liver needle biopsies includes sectioning liver tissue into ten 0.5 × 0.5 cm slices, embedding in paraffin, and staining 4 $\mu$m slides with Masson's trichrome \cite{li2022virtual}. After mordanting in Bouin's solution, slides are stained with hematoxylin, Biebrich scarlet acid fuchsin, and phosphotungstic acid, then scanned at 40x magnification. The WSIs (0.4437 $\mu$m/pixel resolution) highlight hepatocytes, collagen fibers, and blood vessels in pink, blue, and white, respectively.
% [Virtual Liver Needle Biopsy] 肝穿刺活检数据较少，将肝组织切成 10 个 0.5 × 0.5 cm2 的连续切片，包埋在石蜡块中，并用 Masson 三色染色进行组织学评估。将块切成 4 $\mu$m 厚的载玻片。将组织在 Bouin 溶液中媒染、苏木精用 Biebrich 猩红色酸品红染色。将玻片用磷钨酸/磷钨酸溶液染色， 40 倍扫描载玻片图像。所得 WSI 的物理分辨率为 4.437 × 10−1 $\mu$m/像素。每张图像都是从载玻片中的全尺寸肝组织获取的，其中正常的肝细胞、胶原纤维和血管以粉红色、蓝色和白色突出显示。

CODA is a methodological dataset that reconstructs large-scale (up to several cubic centimeters) tissues with subcellular resolution using serial sections stained with hematoxylin and eosin \cite{kiemen2022coda}. The dataset includes 3D reconstructions of pancreatic, skin, lung, and liver tissues stored in a matrix format (M × N × Z). These matrix data are then converted into a triangular mesh format in XML, composed of vertex triplets and triangular faces, as well as cuboidal mesh files in STL format. After importing the large STL files into zBrush, mesh integrity is first checked by filling all holes, removing intersecting faces, and eliminating duplicate vertices that were accidentally created during export. This workflow demonstrates the reconstruction of 3D data from pancreatic, skin, lung, and liver tissues. These data are relevant to the research and its objectives, particularly when considering important parameters for registration. The number and spacing of histological sections relative to the resolution of medical images should be taken into account, as large morphological changes between sections can violate the smoothness assumption and may necessitate individual consideration.
% Atlases通常以脑部3D数据集居多，在三个维度上绘制大脑结构图谱是必要的，但这种数据的高分辨率显微成像促进了快速发展，并具有明显的异质性和非均匀性。For instance, the Allen CCFv3 dataset, provides a three-dimensional digital brain atlas with a resolution of 100$\mu$m. 构建了6个数据子集， 1,675 STP brains. It includes four resolution levels、Waxholm Space (WHS) of the C57BL/6J Nissl-stained Mouse Brain、 Nissl-Stained C57BL/6 Adult Mouse Brain (MOST Imaging)、 Thy1-GFP M-Line Transgenic Mouse Brain (BPS Imaging)、 Problematic Dataset with Imaging Artifacts, the dual-color labeled dataset features brain-wide fluorescent-labeled neurons and counterstained cell bodies with a 2 $\mu$m axial and 0.32 $\mu$m horizontal resolution、 Autofluorescent C57BL/6 Mouse Brain (STP Microscopy)
%CODA是一种使用连续切片的苏木精和伊红染色的组织切片以亚细胞分辨率重建超大(高达几厘米立方）组织的方法学数据集， 创建了以矩阵(M × N × Z）格式存储的 3D 胰腺、皮肤、肺和肝脏组织数据集， 这些矩阵数据被转换为由顶点三元组和三角形面组成的三角网格格式的xml，以及长方体网格的stl 文件，将大型 stl 文件导入 zBrush 后，首先检查网格完整性，以填充所有孔洞、删除相交面，并移除在导出过程中偶然创建的重复顶点，以此展示了在胰腺、皮肤、肺和肝脏组织中重建三维数据
% 这些数据将与研究及目标相关，在对配准很重要的参数中，应考虑相对于医学图像分辨率的组织学切片的数量和间距，in the sense that its sections exhibit large morphological changes and therefore break the smoothness assumption, may also require to consider them individually. 
% \subsubsection{Heterogenous slices}
% 

\subsection{Quantitative Metrics}

In the analysis of serial sections in histopathology, quantitative metrics are essential for evaluating registration accuracy, alignment effectiveness, and image quality. These metrics have been proposed over a long period of time and target a wide range of tasks with varying complexity. (Fig \ref{fig3}.c) Overlap between binary segmentations is assessed using the Dice Similarity Coefficient (DSC) and Intersection over Union (IoU), with DSC offering sensitivity for small objects and IoU robustness for larger structures; the Jaccard coefficient provides scale-invariant shape similarity. Boundary alignment is quantified via Mean Surface Distance (MSD), Average Symmetric Surface Distance (ASSD), and Hausdorff Distance (HD), which captures maximum boundary mismatch. Statistical and structural similarity leverages Mutual Information (MI), Normalized Mutual Information (NMI) for grayscale-invariant alignment, Structural Similarity Index (SSIM), Contrast-Structure Similarity (CSS), and Normalized Cross-Correlation (NCC) for rotation/scale robustness. Physically plausible deformations are ensured by monitoring Negative Jacobian Determinants (NJD). Classification performance employs accuracy (ACC), precision, recall, specificity, F1-score, and the balanced Matthews Correlation Coefficient (MCC). Specialized tasks utilize Target Registration Error (TRE) for landmark-based accuracy and the Modality Independent Neighborhood Descriptor (MIND) for cross-modal registration. Virtual staining quality is evaluated using SSIM and Peak Signal-to-Noise Ratio (PSNR) for reconstruction fidelity, while Fréchet Inception Distance (FID) measures realism in re-stained images. Statistical significance of biological variability is tested via Mann-Whitney U, Wilcoxon signed-rank, and Kruskal-Wallis tests.

\section{Challenges}\label{section5}
\subsection{Heterogeneity in serial tissue sections}
In the preprocessing stage, variations in section quality, staining methods, and microscope settings due to differing pathologist techniques introduce human variability, complicating image standardization, registration, and analysis stability \cite{brixtel2022whole}. Artifacts like tears and folds present significant challenges, often resulting in the exclusion of affected sections. Tear correction requires verifying no tissue loss. If automatically determinable, the task involves restoring in-plane continuity of separated structures. Fold detection currently relies on basic assumptions about color and brightness but could benefit from geometric considerations. Determining fold layer count is particularly challenging. Even if known, fold correction involves separating intra-layer structures and rearranging the tissue to its pre-fold configuration, potentially approximated from adjacent sections \cite{salvi2021impact,liu2021harnessing}. 
\begin{itemize}
    \item Pathological sections often employ a variety of staining techniques, and even when the same method is used, differences in staining concentration, duration, temperature, and reagent batches lead to significant variations in color, contrast, and brightness, complicating tissue visualization and analysis. Inconsistent staining impedes algorithmic recognition of tissue types in automated segmentation or classification tasks \cite{priego2020automatic}.
    \item Ideally, serial sections should maintain uniform thickness, but variations caused by microtome adjustments or operator inexperience distort tissue morphology, complicating image registration and 3D reconstruction. Furthermore, differences in microscope settings, such as light intensity, focus, and magnification, across pathologists affect image quality, challenging the consistency required for image processing algorithms \cite{jahn2020digital}.
    \item Different tissue-handling techniques during sectioning may induce damage or deformation, further complicating image registration. Manual annotations also vary among pathologists, introducing subjectivity that impacts model training and generalization. Mechanical errors during section preparation may cause spatial misalignments, such as translation or rotation between slices, further complicating registration and 3D reconstruction tasks \cite{tian2021tissue}.
\end{itemize}

\subsection{Real-world Opportunity}
The real-world application of serial tissue sections presents numerous opportunities, particularly in pathological diagnosis and analysis (Fig \ref{fig4}). By integrating multiplex immunohistochemistry (IHC), it enables the diagnostic evaluation of molecular biomarkers. Through multi-staining and cross-modal registration of serial tissue sections, it becomes possible to localize lesions in cases where single-data interpretations are inconclusive. This approach enhances diagnostic accuracy by leveraging complementary information from multiple staining modalities, thereby improving the identification and characterization of pathological regions.

Cancer research focuses on tumor microenvironment (TME) analysis. By applying multiplex staining and cross-modal registration to consecutive tissue sections, researchers can investigate the spatial interactions among different cell types, such as tumor cells, immune cells, and stromal cells, to better understand the complexity of the TME. For instance, in melanoma or lung cancer studies, the expression patterns of immune checkpoints (e.g., PD-L1, CTLA-4) in immune-infiltrated tumors can be compared.

In infectious disease research, spatial dissemination of SARS-CoV-2 in COVID-19 can be analyzed using cross-modal section registration (IHC-FISH-H\&E) to examine how the virus spreads within alveoli, blood vessels, and other tissues. Additionally, this approach enables the study of immune cell infiltration patterns and the spatial distribution of inflammatory cytokines.

The process of embryonic development can be reconstructed in both spatial and temporal dimensions, enabling the analysis of how cells differentiate into various tissues. For instance, in zebrafish or mouse embryos, the integration of immunofluorescence and spatial transcriptomic data allows for the reconstruction of cell migration trajectories and the investigation of dynamic changes in signaling pathways, thereby advancing the study of developmental biology. In tissue regeneration research, registration-based reconstruction techniques can be employed to analyze the spatiotemporal dynamics of different cell types during injury repair processes. This approach provides valuable insights into the mechanisms underlying tissue regeneration and repair.

The registration of serial tissue sections, combined with spatial transcriptomics, can be utilized to construct high-resolution brain atlases. For instance, the Allen Brain Atlas employs 3D reconstruction techniques to map gene expression patterns in the brains of mice, rats, and humans. In studies of neural networks in organisms such as fruit flies, rats, or primates, the integration of serial sectioning with fluorescent labeling enables the analysis of neural connectivity patterns across different brain regions. This approach aids in understanding processes such as learning and memory, as well as the mechanisms underlying neurodegenerative diseases like Alzheimer's. Additionally, it facilitates the investigation of pathological spread patterns, providing critical insights into disease progression.

\subsection{Hardware Challenges}
Storage requirements and performance optimization for datasets of serial tissue sections are key challenges, particularly in addressing the issues posed by whole slide imaging (WSI) and DICOM limitations. Scanning tissue at 40$\times$ magnification with a resolution of 0.2 $\mu$m per pixel for a 20 mm $\times$ 15 mm serial sections typically generates approximately 15-20 GB of RGB WSI data \cite{brixtel2022whole}. In addition, solutions exist for scanning samples at 100$\times$ magnification, which significantly increases data volume. Despite the DICOM object definition for visible light microscopy, WSI files are often too large (both in image pixel dimensions and byte size) to fit within certain built-in constraints of the DICOM standard: an unsigned 16-bit integer specifies image dimensions, and an unsigned 32-bit integer specifies the size of each data element, including pixel data. Comparing focus quality across different scanners using algorithms is another area requiring further work, as pathologists manually identify out-of-focus (OOF) regions in different image patches \cite{ba2022framework}. Further efforts are needed to develop focus quality metrics that are independent of the scanner or tissue type.

Visualizing 3D histology introduces challenges that extend beyond the traditional uses of WSI and other medical 3D renderings. Key requirements for an effective visualization system include: managing extreme data sizes, such as volumes of 100,000 $\times$ 100,000 $\times$ 100 voxels; enabling frequent zooming interactions between low and high magnification; handling highly anisotropic voxels; preserving the natural colors of stained tissue, which are diagnostically meaningful to pathologists; and addressing the issue of background rendering \cite{hanna2020whole}. Backgrounds in WSI are naturally opaque white, leading to errors in representing blank areas in 3D. Moreover, the system must provide contextual orientation and position information, even in close-up views where dense data may obscure directional features. It should also allow arbitrary sectioning without introducing orientation issues in detailed views, and offer all necessary operations without overwhelming users with excessive interactions. Furthermore, automating common data extraction processes and improving parallel computing performance, such as GPU utilization, are crucial \cite{weiss2021tutorial,zhou2021review}.

Data privacy also remains a major barrier to accessing histopathology data. Federated learning offers a solution by enabling collaborative learning without centralizing data, thereby addressing privacy and governance issues. Tissue serial section datasets are often limited due to size storage and processing requirements: large image datasets (such as whole tissue section images) can be very large, involving hundreds of GB or TB of data, and when shared publicly, they bring technical challenges to storage and transmission, medical privacy and ethical issues, risks of data abuse, and other limitations \cite{roy2022demystifying,lu2022federated}. In addition, tissue imaging techniques are numerous and extensive, and imaging equipment is varied, which brings difficulties to unification \cite{bai2023deep}.
\subsection{Evaluation Complexity}
Evaluation systems for serial section tasks vary. In the registration process, mutual information-based methods assess alignment by establishing pixel correspondences between images. However, this approach may not capture meaningful relationships between multiplex immunohistochemistry (mIHC) images. Intensity-based methods also face limitations, leading us to favor feature-based approaches in related work. When focusing on landmarks as the sole measure of accuracy, it is important to note that landmarks are placed at locations recognizable by human observers, which often have higher contrast \cite{hill2000across}. This can disproportionately influence distance measurements, potentially biasing evaluations by underestimating registration errors in larger, lower-contrast regions. Particularly in low-contrast areas, regularizers help maintain the smoothness of deformations, a crucial factor in assessing image registration quality. Regularity or smoothness of deformation is another standard for evaluating the quality of image registration \cite{varadhan2013framework}.

Since the cellular structures and compositions of serial sections are not identical, a scalable framework is needed to collect samples for training deep neural networks \cite{levy2020topological}. Such a framework would allow the network to learn deep texture patterns that facilitate image alignment by matching large sets of images with reliable alignment scores. In spatial omics alignment, methods combining spatial graph convolution and adversarial matching have shown promise \cite{bressan2023dawn}. However, we note that other topology-based metrics have been proposed for performance evaluation, such as edge scores that quantify spatial neighbor retention. We argue that these metrics may inherently suffer from bias, as they only measure alignment continuity, which can be particularly problematic in cases involving structural changes.

More precise 3D tumor size measurements have the potential to improve prognosis estimation and enable more accurate risk-based treatment prescriptions \cite{farrokh2019accuracy}. Larger-scale studies are needed to test the relationship between alternative tumor size measurements and clinical outcomes, including lymph node involvement and local or distant recurrence.

\section{Outlook}\label{section6}
The analytical and technological evolution of serial section histopathology is reshaping how we model spatial biology in both research and clinical contexts. At the computational level, non-rigid registration methods are transitioning from fixed-weight schemes to adaptive parameterizations—such as B-spline free-form deformation (FFD) with embedded spatial regularization—which effectively reduce tissue warping while preserving structural fidelity \cite{akossi2021image}. When tissue loss occurs due to re-staining, intensity-based methods offer complementary robustness for local transformations. These advancements lay the groundwork for more accurate alignment in large-scale 3D tissue reconstructions.

At the same time, spatial omics platforms Stereo-seq \cite{moffitt2018molecular} are beginning to capture subcellular compartmentalization within tissue sections, yet they remain inherently 2D. To address this, the integration of spatial omics across serial sections enables molecular cartography in three dimensions, bridging the gap between static molecular snapshots and dynamic biological processes. Recent work in multimodal image harmonization and cross-modal feature space alignment \cite{jaume2024multistain} provides a path forward for extracting consistent features across diverse staining protocols and imaging modalities.

These biological and computational advances are converging to support scalable 3D histopathology, which requires both dense sectioning and high-throughput analysis. GPU-accelerated pipelines now make it feasible to register and analyze terabyte-scale whole-slide image (WSI) datasets \cite{pichat2018survey}, while machine learning-driven feature selection enables biologically meaningful interpretations, such as genotype-phenotype associations and 3D biomarker discovery \cite{bermingham2015application}. However, practical challenges persist—most notably, memory limitations in full-resolution 3D rendering, temporal constraints in capturing molecular dynamics, and the difficulty of integrating spatial omics with traditional histology.

Historically, manual alignment for diagnostic slide reading was labor-intensive and time-consuming. Subsequent development of semi-automated affine registration reduced this burden but still required iterative refinement. Future advancements aim to overcome these bottlenecks through full automation and innovations in visualization. AI-assisted sectioning and fully automated stainers hold promise for streamlining tissue preparation and minimizing human-induced variability \cite{schouten2024full}. Concurrently, high-resolution visualization systems will enable interactive navigation within gigapixel-scale tissue volumes, facilitating targeted sub-volume examination and intuitive multi-scale exploration \cite{beyer2022survey}. These systems are evolving towards a "3D microscope" paradigm, thereby enhancing both research and diagnostic applications. Regarding deployment, cloud platforms offer accessible interfaces for non-expert users to perform predictions and basic analyses. However, seamless integration into existing image processing workflows and advanced model customization necessitates the continued importance of local deployment. Emerging open-source pipelines \cite{kashyap2021survey} are addressing this need by providing modular tools for segmentation, quantification, and spatial analysis of multiplexed image data.

~\\
\noindent\textbf{Acknowledgments}\\
This work was supported in part by the Ministry of Science and Technology of the People's Republic of China under Grant STI2030-Major Projects2021ZD0201900, in part by the National Natural Science Foundation of China under Grant 62371409, and National Natural Science Foundation of China (Grant No. 62371409) and Fujian Provincial Natural Science Foundation of China (Grant No. 2023J01005).

\noindent\textbf{Author contributions}\\
Authors' contributions are listed in accordance with the CRediT framework contributor roles taxonomy (see details in \href{https://credit.niso.org/contributor-roles-defined/}{https://credit.niso.org/contributor-roles-defined/}). ZZ and LW conceptualized the study. Data curation was done by ZZ, MC and JL, equally. Formal analysis was done by all authors. Funding acquisition was done by LW. Investigation design the search strategy was done by ZZ MC and JL equally. MC, PQ and YH focused on methodology. Project administration was done by LW and ZZ. Supervision was done by LW and ZZ. Visualization was done by ZZ. Writing of original draft was done by ZZ, MC, JL, PQ, YH. Writing of review \& editing was done by HZ, MC and ZZ. All authors contributed to discussions, provided critical feedback, and approved the final version of the paper.

\bibliography{sn-bibliography}% common bib file


\begin{thebibliography}{148}
\ifx \bisbn   \undefined \def \bisbn  #1{ISBN #1}\fi
\ifx \binits  \undefined \def \binits#1{#1}\fi
\ifx \bauthor  \undefined \def \bauthor#1{#1}\fi
\ifx \batitle  \undefined \def \batitle#1{#1}\fi
\ifx \bjtitle  \undefined \def \bjtitle#1{#1}\fi
\ifx \bvolume  \undefined \def \bvolume#1{\textbf{#1}}\fi
\ifx \byear  \undefined \def \byear#1{#1}\fi
\ifx \bissue  \undefined \def \bissue#1{#1}\fi
\ifx \bfpage  \undefined \def \bfpage#1{#1}\fi
\ifx \blpage  \undefined \def \blpage #1{#1}\fi
\ifx \burl  \undefined \def \burl#1{\textsf{#1}}\fi
\ifx \doiurl  \undefined \def \doiurl#1{\url{https://doi.org/#1}}\fi
\ifx \betal  \undefined \def \betal{\textit{et al.}}\fi
\ifx \binstitute  \undefined \def \binstitute#1{#1}\fi
\ifx \binstitutionaled  \undefined \def \binstitutionaled#1{#1}\fi
\ifx \bctitle  \undefined \def \bctitle#1{#1}\fi
\ifx \beditor  \undefined \def \beditor#1{#1}\fi
\ifx \bpublisher  \undefined \def \bpublisher#1{#1}\fi
\ifx \bbtitle  \undefined \def \bbtitle#1{#1}\fi
\ifx \bedition  \undefined \def \bedition#1{#1}\fi
\ifx \bseriesno  \undefined \def \bseriesno#1{#1}\fi
\ifx \blocation  \undefined \def \blocation#1{#1}\fi
\ifx \bsertitle  \undefined \def \bsertitle#1{#1}\fi
\ifx \bsnm \undefined \def \bsnm#1{#1}\fi
\ifx \bsuffix \undefined \def \bsuffix#1{#1}\fi
\ifx \bparticle \undefined \def \bparticle#1{#1}\fi
\ifx \barticle \undefined \def \barticle#1{#1}\fi
\bibcommenthead
\ifx \bconfdate \undefined \def \bconfdate #1{#1}\fi
\ifx \botherref \undefined \def \botherref #1{#1}\fi
\ifx \url \undefined \def \url#1{\textsf{#1}}\fi
\ifx \bchapter \undefined \def \bchapter#1{#1}\fi
\ifx \bbook \undefined \def \bbook#1{#1}\fi
\ifx \bcomment \undefined \def \bcomment#1{#1}\fi
\ifx \oauthor \undefined \def \oauthor#1{#1}\fi
\ifx \citeauthoryear \undefined \def \citeauthoryear#1{#1}\fi
\ifx \endbibitem  \undefined \def \endbibitem {}\fi
\ifx \bconflocation  \undefined \def \bconflocation#1{#1}\fi
\ifx \arxivurl  \undefined \def \arxivurl#1{\textsf{#1}}\fi
\csname PreBibitemsHook\endcsname

%%% 1
\bibitem[\protect\citeauthoryear{Herrmann et~al.}{2014}]{herrmann2014three}
\begin{barticle}
\bauthor{\bsnm{Herrmann}, \binits{D.}},
\bauthor{\bsnm{Conway}, \binits{J.R.}},
\bauthor{\bsnm{Vennin}, \binits{C.}},
\bauthor{\bsnm{Magenau}, \binits{A.}},
\bauthor{\bsnm{Hughes}, \binits{W.E.}},
\bauthor{\bsnm{Morton}, \binits{J.P.}},
\bauthor{\bsnm{Timpson}, \binits{P.}}:
\batitle{Three-dimensional cancer models mimic cell--matrix interactions in the tumour microenvironment}.
\bjtitle{Carcinogenesis}
\bvolume{35}(\bissue{8}),
\bfpage{1671}--\blpage{1679}
(\byear{2014})
\end{barticle}
\endbibitem

%%% 2
\bibitem[\protect\citeauthoryear{Cakir et~al.}{2024}]{cakir2024determination}
\begin{botherref}
\oauthor{\bsnm{Cakir}, \binits{C.}},
\oauthor{\bsnm{Inci}, \binits{E.}},
\oauthor{\bsnm{Kilinc}, \binits{F.}},
\oauthor{\bsnm{Yildiz}, \binits{O.}}:
Determination of the diagnostic accuracy of peritoneal biopsy with an 18g cutting needle under ultrasonography guidance and the contribution of ct findings to diagnosis before biopsy (our 8-year clinical experience).
Clinical Radiology
(2024)
\end{botherref}
\endbibitem

%%% 3
\bibitem[\protect\citeauthoryear{Cuenca et~al.}{2020}]{cuenca2020interpretation}
\begin{barticle}
\bauthor{\bsnm{Cuenca}, \binits{N.}},
\bauthor{\bsnm{Ortuno-Lizaran}, \binits{I.}},
\bauthor{\bsnm{Sanchez-Saez}, \binits{X.}},
\bauthor{\bsnm{Kutsyr}, \binits{O.}},
\bauthor{\bsnm{Albertos-Arranz}, \binits{H.}},
\bauthor{\bsnm{Fernandez-Sanchez}, \binits{L.}},
\bauthor{\bsnm{Martinez-Gil}, \binits{N.}},
\bauthor{\bsnm{Noailles}, \binits{A.}},
\bauthor{\bsnm{L{\'o}pez-Garrido}, \binits{J.A.}},
\bauthor{\bsnm{Lopez-Galvez}, \binits{M.}}, \betal:
\batitle{Interpretation of oct and octa images from a histological approach: clinical and experimental implications}.
\bjtitle{Progress in retinal and eye research}
\bvolume{77},
\bfpage{100828}
(\byear{2020})
\end{barticle}
\endbibitem

%%% 4
\bibitem[\protect\citeauthoryear{Veluponnar et~al.}{2024}]{veluponnar2024resection}
\begin{barticle}
\bauthor{\bsnm{Veluponnar}, \binits{D.}},
\bauthor{\bsnm{Dashtbozorg}, \binits{B.}},
\bauthor{\bsnm{Guimaraes}, \binits{M.D.S.}},
\bauthor{\bsnm{Peeters}, \binits{M.-J.T.V.}},
\bauthor{\bsnm{Boer}, \binits{L.L.d.}},
\bauthor{\bsnm{Ruers}, \binits{T.J.}}:
\batitle{Resection ratios and tumor eccentricity in breast-conserving surgery specimens for surgical accuracy assessment}.
\bjtitle{Cancers}
\bvolume{16}(\bissue{10}),
\bfpage{1813}
(\byear{2024})
\end{barticle}
\endbibitem

%%% 5
\bibitem[\protect\citeauthoryear{Garrison et~al.}{2022}]{garrison2022trends}
\begin{bchapter}
\bauthor{\bsnm{Garrison}, \binits{L.A.}},
\bauthor{\bsnm{Kolesar}, \binits{I.}},
\bauthor{\bsnm{Viola}, \binits{I.}},
\bauthor{\bsnm{Hauser}, \binits{H.}},
\bauthor{\bsnm{Bruckner}, \binits{S.}}:
\bctitle{Trends \& opportunities in visualization for physiology: A multiscale overview}.
In: \bbtitle{Computer Graphics Forum},
vol. \bseriesno{41},
pp. \bfpage{609}--\blpage{643}
(\byear{2022}).
\bcomment{Wiley Online Library}
\end{bchapter}
\endbibitem

%%% 6
\bibitem[\protect\citeauthoryear{Bai et~al.}{2023}]{bai2023deep}
\begin{barticle}
\bauthor{\bsnm{Bai}, \binits{B.}},
\bauthor{\bsnm{Yang}, \binits{X.}},
\bauthor{\bsnm{Li}, \binits{Y.}},
\bauthor{\bsnm{Zhang}, \binits{Y.}},
\bauthor{\bsnm{Pillar}, \binits{N.}},
\bauthor{\bsnm{Ozcan}, \binits{A.}}:
\batitle{Deep learning-enabled virtual histological staining of biological samples}.
\bjtitle{Light: Science \& Applications}
\bvolume{12}(\bissue{1}),
\bfpage{57}
(\byear{2023})
\end{barticle}
\endbibitem

%%% 7
\bibitem[\protect\citeauthoryear{Weissleder and Lee}{2020}]{weissleder2020automated}
\begin{barticle}
\bauthor{\bsnm{Weissleder}, \binits{R.}},
\bauthor{\bsnm{Lee}, \binits{H.}}:
\batitle{Automated molecular-image cytometry and analysis in modern oncology}.
\bjtitle{Nature Reviews Materials}
\bvolume{5}(\bissue{6}),
\bfpage{409}--\blpage{422}
(\byear{2020})
\end{barticle}
\endbibitem

%%% 8
\bibitem[\protect\citeauthoryear{Gurcan et~al.}{2009}]{gurcan2009histopathological}
\begin{barticle}
\bauthor{\bsnm{Gurcan}, \binits{M.N.}},
\bauthor{\bsnm{Boucheron}, \binits{L.E.}},
\bauthor{\bsnm{Can}, \binits{A.}},
\bauthor{\bsnm{Madabhushi}, \binits{A.}},
\bauthor{\bsnm{Rajpoot}, \binits{N.M.}},
\bauthor{\bsnm{Yener}, \binits{B.}}:
\batitle{Histopathological image analysis: A review}.
\bjtitle{IEEE reviews in biomedical engineering}
\bvolume{2},
\bfpage{147}--\blpage{171}
(\byear{2009})
\end{barticle}
\endbibitem

%%% 9
\bibitem[\protect\citeauthoryear{Bui et~al.}{2019}]{bui2019digital}
\begin{botherref}
\oauthor{\bsnm{Bui}, \binits{M.M.}},
\oauthor{\bsnm{Asa}, \binits{S.L.}},
\oauthor{\bsnm{Pantanowitz}, \binits{L.}},
\oauthor{\bsnm{Parwani}, \binits{A.}},
\oauthor{\bsnm{Laak}, \binits{J.}},
\oauthor{\bsnm{Ung}, \binits{C.}},
\oauthor{\bsnm{Balis}, \binits{U.}},
\oauthor{\bsnm{Isaacs}, \binits{M.}},
\oauthor{\bsnm{Glassy}, \binits{E.}},
\oauthor{\bsnm{Manning}, \binits{L.}}:
Digital and computational pathology: bring the future into focus.
Journal of Pathology Informatics
\textbf{10}
(2019)
\end{botherref}
\endbibitem

%%% 10
\bibitem[\protect\citeauthoryear{Al-Thelaya et~al.}{2023}]{al2023applications}
\begin{barticle}
\bauthor{\bsnm{Al-Thelaya}, \binits{K.}},
\bauthor{\bsnm{Gilal}, \binits{N.U.}},
\bauthor{\bsnm{Alzubaidi}, \binits{M.}},
\bauthor{\bsnm{Majeed}, \binits{F.}},
\bauthor{\bsnm{Agus}, \binits{M.}},
\bauthor{\bsnm{Schneider}, \binits{J.}},
\bauthor{\bsnm{Househ}, \binits{M.}}:
\batitle{Applications of discriminative and deep learning feature extraction methods for whole slide image analysis: A survey}.
\bjtitle{Journal of Pathology Informatics}
\bvolume{14},
\bfpage{100335}
(\byear{2023})
\end{barticle}
\endbibitem

%%% 11
\bibitem[\protect\citeauthoryear{Almagro et~al.}{2021}]{almagro2021tissue}
\begin{barticle}
\bauthor{\bsnm{Almagro}, \binits{J.}},
\bauthor{\bsnm{Messal}, \binits{H.A.}},
\bauthor{\bsnm{Zaw~Thin}, \binits{M.}},
\bauthor{\bsnm{Rheenen}, \binits{J.}},
\bauthor{\bsnm{Behrens}, \binits{A.}}:
\batitle{Tissue clearing to examine tumour complexity in three dimensions}.
\bjtitle{Nature Reviews Cancer}
\bvolume{21}(\bissue{11}),
\bfpage{718}--\blpage{730}
(\byear{2021})
\end{barticle}
\endbibitem

%%% 12
\bibitem[\protect\citeauthoryear{Kench and Cooper}{2021}]{kench2021generating}
\begin{barticle}
\bauthor{\bsnm{Kench}, \binits{S.}},
\bauthor{\bsnm{Cooper}, \binits{S.J.}}:
\batitle{Generating three-dimensional structures from a two-dimensional slice with generative adversarial network-based dimensionality expansion}.
\bjtitle{Nature Machine Intelligence}
\bvolume{3}(\bissue{4}),
\bfpage{299}--\blpage{305}
(\byear{2021})
\end{barticle}
\endbibitem

%%% 13
\bibitem[\protect\citeauthoryear{Larsen et~al.}{2021}]{larsen2021cellular}
\begin{barticle}
\bauthor{\bsnm{Larsen}, \binits{N.Y.}},
\bauthor{\bsnm{Li}, \binits{X.}},
\bauthor{\bsnm{Tan}, \binits{X.}},
\bauthor{\bsnm{Ji}, \binits{G.}},
\bauthor{\bsnm{Lin}, \binits{J.}},
\bauthor{\bsnm{Rajkowska}, \binits{G.}},
\bauthor{\bsnm{M{\o}ller}, \binits{J.}},
\bauthor{\bsnm{Vihrs}, \binits{N.}},
\bauthor{\bsnm{Sporring}, \binits{J.}},
\bauthor{\bsnm{Sun}, \binits{F.}}, \betal:
\batitle{Cellular 3d-reconstruction and analysis in the human cerebral cortex using automatic serial sections}.
\bjtitle{Communications Biology}
\bvolume{4}(\bissue{1}),
\bfpage{1030}
(\byear{2021})
\end{barticle}
\endbibitem

%%% 14
\bibitem[\protect\citeauthoryear{Seferbekova et~al.}{2023}]{seferbekova2023spatial}
\begin{barticle}
\bauthor{\bsnm{Seferbekova}, \binits{Z.}},
\bauthor{\bsnm{Lomakin}, \binits{A.}},
\bauthor{\bsnm{Yates}, \binits{L.R.}},
\bauthor{\bsnm{Gerstung}, \binits{M.}}:
\batitle{Spatial biology of cancer evolution}.
\bjtitle{Nature Reviews Genetics}
\bvolume{24}(\bissue{5}),
\bfpage{295}--\blpage{313}
(\byear{2023})
\end{barticle}
\endbibitem

%%% 15
\bibitem[\protect\citeauthoryear{Jahn et~al.}{2020}]{jahn2020digital}
\begin{barticle}
\bauthor{\bsnm{Jahn}, \binits{S.W.}},
\bauthor{\bsnm{Plass}, \binits{M.}},
\bauthor{\bsnm{Moinfar}, \binits{F.}}:
\batitle{Digital pathology: advantages, limitations and emerging perspectives}.
\bjtitle{Journal of clinical medicine}
\bvolume{9}(\bissue{11}),
\bfpage{3697}
(\byear{2020})
\end{barticle}
\endbibitem

%%% 16
\bibitem[\protect\citeauthoryear{Pallua et~al.}{2020}]{pallua2020future}
\begin{barticle}
\bauthor{\bsnm{Pallua}, \binits{J.}},
\bauthor{\bsnm{Brunner}, \binits{A.}},
\bauthor{\bsnm{Zelger}, \binits{B.}},
\bauthor{\bsnm{Schirmer}, \binits{M.}},
\bauthor{\bsnm{Haybaeck}, \binits{J.}}:
\batitle{The future of pathology is digital}.
\bjtitle{Pathology-Research and Practice}
\bvolume{216}(\bissue{9}),
\bfpage{153040}
(\byear{2020})
\end{barticle}
\endbibitem

%%% 17
\bibitem[\protect\citeauthoryear{S{\'a}nchez-Porras et~al.}{2022}]{sanchez2022tissue}
\begin{bchapter}
\bauthor{\bsnm{S{\'a}nchez-Porras}, \binits{D.}},
\bauthor{\bsnm{Bermejo-Casares}, \binits{F.}},
\bauthor{\bsnm{Carmona}, \binits{R.}},
\bauthor{\bsnm{Weiss}, \binits{T.}},
\bauthor{\bsnm{Campos}, \binits{F.}},
\bauthor{\bsnm{Carriel}, \binits{V.}}:
\bctitle{Tissue fixation and processing for the histological identification of lipids}.
In: \bbtitle{Histochemistry of Single Molecules: Methods and Protocols},
pp. \bfpage{175}--\blpage{186}.
\bpublisher{Springer}, \blocation{???}
(\byear{2022})
\end{bchapter}
\endbibitem

%%% 18
\bibitem[\protect\citeauthoryear{Hosseini et~al.}{2024}]{hosseini2024computational}
\begin{botherref}
\oauthor{\bsnm{Hosseini}, \binits{M.S.}},
\oauthor{\bsnm{Bejnordi}, \binits{B.E.}},
\oauthor{\bsnm{Trinh}, \binits{V.Q.-H.}},
\oauthor{\bsnm{Chan}, \binits{L.}},
\oauthor{\bsnm{Hasan}, \binits{D.}},
\oauthor{\bsnm{Li}, \binits{X.}},
\oauthor{\bsnm{Yang}, \binits{S.}},
\oauthor{\bsnm{Kim}, \binits{T.}},
\oauthor{\bsnm{Zhang}, \binits{H.}},
\oauthor{\bsnm{Wu}, \binits{T.}}, et al.:
Computational pathology: a survey review and the way forward.
Journal of Pathology Informatics,
100357
(2024)
\end{botherref}
\endbibitem

%%% 19
\bibitem[\protect\citeauthoryear{Pasolli et~al.}{2016}]{pasolli2016machine}
\begin{barticle}
\bauthor{\bsnm{Pasolli}, \binits{E.}},
\bauthor{\bsnm{Truong}, \binits{D.T.}},
\bauthor{\bsnm{Malik}, \binits{F.}},
\bauthor{\bsnm{Waldron}, \binits{L.}},
\bauthor{\bsnm{Segata}, \binits{N.}}:
\batitle{Machine learning meta-analysis of large metagenomic datasets: tools and biological insights}.
\bjtitle{PLoS computational biology}
\bvolume{12}(\bissue{7}),
\bfpage{1004977}
(\byear{2016})
\end{barticle}
\endbibitem

%%% 20
\bibitem[\protect\citeauthoryear{Liu et~al.}{2022}]{liu2022novel}
\begin{barticle}
\bauthor{\bsnm{Liu}, \binits{J.}},
\bauthor{\bsnm{Wu}, \binits{X.}},
\bauthor{\bsnm{Xu}, \binits{C.}},
\bauthor{\bsnm{Ma}, \binits{M.}},
\bauthor{\bsnm{Zhao}, \binits{J.}},
\bauthor{\bsnm{Li}, \binits{M.}},
\bauthor{\bsnm{Yu}, \binits{Q.}},
\bauthor{\bsnm{Hao}, \binits{X.}},
\bauthor{\bsnm{Wang}, \binits{G.}},
\bauthor{\bsnm{Wei}, \binits{B.}}, \betal:
\batitle{A novel method for observing tumor margin in hepatoblastoma based on microstructure 3d reconstruction}.
\bjtitle{Fetal and Pediatric Pathology}
\bvolume{41}(\bissue{3}),
\bfpage{371}--\blpage{380}
(\byear{2022})
\end{barticle}
\endbibitem

%%% 21
\bibitem[\protect\citeauthoryear{Norris and Caprioli}{2013}]{norris2013analysis}
\begin{barticle}
\bauthor{\bsnm{Norris}, \binits{J.L.}},
\bauthor{\bsnm{Caprioli}, \binits{R.M.}}:
\batitle{Analysis of tissue specimens by matrix-assisted laser desorption/ionization imaging mass spectrometry in biological and clinical research}.
\bjtitle{Chemical reviews}
\bvolume{113}(\bissue{4}),
\bfpage{2309}--\blpage{2342}
(\byear{2013})
\end{barticle}
\endbibitem

%%% 22
\bibitem[\protect\citeauthoryear{Atmar and Ramani}{2023}]{atmar2023immunologic}
\begin{bchapter}
\bauthor{\bsnm{Atmar}, \binits{R.L.}},
\bauthor{\bsnm{Ramani}, \binits{S.}}:
\bctitle{Immunologic detection and characterization}.
In: \bbtitle{Viral Infections of Humans: Epidemiology and Control},
pp. \bfpage{1}--\blpage{30}.
\bpublisher{Springer}, \blocation{???}
(\byear{2023})
\end{bchapter}
\endbibitem

%%% 23
\bibitem[\protect\citeauthoryear{Hasic}{2022}]{hasic2022immunohistochemistry}
\begin{botherref}
\oauthor{\bsnm{Hasic}, \binits{E.}}:
Immunohistochemistry fundamentals.
Immunohistochemistry: A Technical Guide to Current Practices,
1
(2022)
\end{botherref}
\endbibitem

%%% 24
\bibitem[\protect\citeauthoryear{Rao et~al.}{2021}]{rao2021exploring}
\begin{barticle}
\bauthor{\bsnm{Rao}, \binits{A.}},
\bauthor{\bsnm{Barkley}, \binits{D.}},
\bauthor{\bsnm{Fran{\c{c}}a}, \binits{G.S.}},
\bauthor{\bsnm{Yanai}, \binits{I.}}:
\batitle{Exploring tissue architecture using spatial transcriptomics}.
\bjtitle{Nature}
\bvolume{596}(\bissue{7871}),
\bfpage{211}--\blpage{220}
(\byear{2021})
\end{barticle}
\endbibitem

%%% 25
\bibitem[\protect\citeauthoryear{Paik et~al.}{2020}]{paik2020single}
\begin{barticle}
\bauthor{\bsnm{Paik}, \binits{D.T.}},
\bauthor{\bsnm{Cho}, \binits{S.}},
\bauthor{\bsnm{Tian}, \binits{L.}},
\bauthor{\bsnm{Chang}, \binits{H.Y.}},
\bauthor{\bsnm{Wu}, \binits{J.C.}}:
\batitle{Single-cell rna sequencing in cardiovascular development, disease and medicine}.
\bjtitle{Nature Reviews Cardiology}
\bvolume{17}(\bissue{8}),
\bfpage{457}--\blpage{473}
(\byear{2020})
\end{barticle}
\endbibitem

%%% 26
\bibitem[\protect\citeauthoryear{Yang et~al.}{2024}]{yang2024virtual}
\begin{barticle}
\bauthor{\bsnm{Yang}, \binits{X.}},
\bauthor{\bsnm{Bai}, \binits{B.}},
\bauthor{\bsnm{Zhang}, \binits{Y.}},
\bauthor{\bsnm{Aydin}, \binits{M.}},
\bauthor{\bsnm{Li}, \binits{Y.}},
\bauthor{\bsnm{Selcuk}, \binits{S.Y.}},
\bauthor{\bsnm{Casteleiro~Costa}, \binits{P.}},
\bauthor{\bsnm{Guo}, \binits{Z.}},
\bauthor{\bsnm{Fishbein}, \binits{G.A.}},
\bauthor{\bsnm{Atlan}, \binits{K.}}, \betal:
\batitle{Virtual birefringence imaging and histological staining of amyloid deposits in label-free tissue using autofluorescence microscopy and deep learning}.
\bjtitle{Nature Communications}
\bvolume{15}(\bissue{1}),
\bfpage{7978}
(\byear{2024})
\end{barticle}
\endbibitem

%%% 27
\bibitem[\protect\citeauthoryear{Antao et~al.}{2024}]{antao2024sample}
\begin{barticle}
\bauthor{\bsnm{Antao}, \binits{N.V.}},
\bauthor{\bsnm{Sall}, \binits{J.}},
\bauthor{\bsnm{Petzold}, \binits{C.}},
\bauthor{\bsnm{Ekiert}, \binits{D.C.}},
\bauthor{\bsnm{Bhabha}, \binits{G.}},
\bauthor{\bsnm{Liang}, \binits{F.-X.}}:
\batitle{Sample preparation and data collection for serial block face scanning electron microscopy of mammalian cell monolayers}.
\bjtitle{Plos one}
\bvolume{19}(\bissue{8}),
\bfpage{0301284}
(\byear{2024})
\end{barticle}
\endbibitem

%%% 28
\bibitem[\protect\citeauthoryear{Liu et~al.}{2021}]{liu2021harnessing}
\begin{barticle}
\bauthor{\bsnm{Liu}, \binits{J.T.}},
\bauthor{\bsnm{Glaser}, \binits{A.K.}},
\bauthor{\bsnm{Bera}, \binits{K.}},
\bauthor{\bsnm{True}, \binits{L.D.}},
\bauthor{\bsnm{Reder}, \binits{N.P.}},
\bauthor{\bsnm{Eliceiri}, \binits{K.W.}},
\bauthor{\bsnm{Madabhushi}, \binits{A.}}:
\batitle{Harnessing non-destructive 3d pathology}.
\bjtitle{Nature biomedical engineering}
\bvolume{5}(\bissue{3}),
\bfpage{203}--\blpage{218}
(\byear{2021})
\end{barticle}
\endbibitem

%%% 29
\bibitem[\protect\citeauthoryear{Glaser et~al.}{2022}]{glaser2022hybrid}
\begin{barticle}
\bauthor{\bsnm{Glaser}, \binits{A.K.}},
\bauthor{\bsnm{Bishop}, \binits{K.W.}},
\bauthor{\bsnm{Barner}, \binits{L.A.}},
\bauthor{\bsnm{Susaki}, \binits{E.A.}},
\bauthor{\bsnm{Kubota}, \binits{S.I.}},
\bauthor{\bsnm{Gao}, \binits{G.}},
\bauthor{\bsnm{Serafin}, \binits{R.B.}},
\bauthor{\bsnm{Balaram}, \binits{P.}},
\bauthor{\bsnm{Turschak}, \binits{E.}},
\bauthor{\bsnm{Nicovich}, \binits{P.R.}}, \betal:
\batitle{A hybrid open-top light-sheet microscope for versatile multi-scale imaging of cleared tissues}.
\bjtitle{Nature methods}
\bvolume{19}(\bissue{5}),
\bfpage{613}--\blpage{619}
(\byear{2022})
\end{barticle}
\endbibitem

%%% 30
\bibitem[\protect\citeauthoryear{Mukhamadiyarov et~al.}{2021}]{mukhamadiyarov2021embedding}
\begin{barticle}
\bauthor{\bsnm{Mukhamadiyarov}, \binits{R.A.}},
\bauthor{\bsnm{Bogdanov}, \binits{L.A.}},
\bauthor{\bsnm{Glushkova}, \binits{T.V.}},
\bauthor{\bsnm{Shishkova}, \binits{D.K.}},
\bauthor{\bsnm{Kostyunin}, \binits{A.E.}},
\bauthor{\bsnm{Koshelev}, \binits{V.A.}},
\bauthor{\bsnm{Shabaev}, \binits{A.R.}},
\bauthor{\bsnm{Frolov}, \binits{A.V.}},
\bauthor{\bsnm{Stasev}, \binits{A.N.}},
\bauthor{\bsnm{Lyapin}, \binits{A.A.}}, \betal:
\batitle{Embedding and backscattered scanning electron microscopy: a detailed protocol for the whole-specimen, high-resolution analysis of cardiovascular tissues}.
\bjtitle{Frontiers in Cardiovascular Medicine}
\bvolume{8},
\bfpage{739549}
(\byear{2021})
\end{barticle}
\endbibitem

%%% 31
\bibitem[\protect\citeauthoryear{Aziz et~al.}{2020}]{aziz2020medical}
\begin{barticle}
\bauthor{\bsnm{Aziz}, \binits{A.}},
\bauthor{\bsnm{Pane}, \binits{S.}},
\bauthor{\bsnm{Iacovacci}, \binits{V.}},
\bauthor{\bsnm{Koukourakis}, \binits{N.}},
\bauthor{\bsnm{Czarske}, \binits{J.}},
\bauthor{\bsnm{Menciassi}, \binits{A.}},
\bauthor{\bsnm{Medina-S{\'a}nchez}, \binits{M.}},
\bauthor{\bsnm{Schmidt}, \binits{O.G.}}:
\batitle{Medical imaging of microrobots: Toward in vivo applications}.
\bjtitle{ACS nano}
\bvolume{14}(\bissue{9}),
\bfpage{10865}--\blpage{10893}
(\byear{2020})
\end{barticle}
\endbibitem

%%% 32
\bibitem[\protect\citeauthoryear{Einhaus et~al.}{2023}]{einhaus2023high}
\begin{barticle}
\bauthor{\bsnm{Einhaus}, \binits{J.}},
\bauthor{\bsnm{Rochwarger}, \binits{A.}},
\bauthor{\bsnm{Mattern}, \binits{S.}},
\bauthor{\bsnm{Gaudilli{\`e}re}, \binits{B.}},
\bauthor{\bsnm{Sch{\"u}rch}, \binits{C.M.}}:
\batitle{High-multiplex tissue imaging in routine pathology—are we there yet?}
\bjtitle{Virchows Archiv}
\bvolume{482}(\bissue{5}),
\bfpage{801}--\blpage{812}
(\byear{2023})
\end{barticle}
\endbibitem

%%% 33
\bibitem[\protect\citeauthoryear{Tan et~al.}{2020}]{tan2020overview}
\begin{barticle}
\bauthor{\bsnm{Tan}, \binits{W.C.C.}},
\bauthor{\bsnm{Nerurkar}, \binits{S.N.}},
\bauthor{\bsnm{Cai}, \binits{H.Y.}},
\bauthor{\bsnm{Ng}, \binits{H.H.M.}},
\bauthor{\bsnm{Wu}, \binits{D.}},
\bauthor{\bsnm{Wee}, \binits{Y.T.F.}},
\bauthor{\bsnm{Lim}, \binits{J.C.T.}},
\bauthor{\bsnm{Yeong}, \binits{J.}},
\bauthor{\bsnm{Lim}, \binits{T.K.H.}}:
\batitle{Overview of multiplex immunohistochemistry/immunofluorescence techniques in the era of cancer immunotherapy}.
\bjtitle{Cancer Communications}
\bvolume{40}(\bissue{4}),
\bfpage{135}--\blpage{153}
(\byear{2020})
\end{barticle}
\endbibitem

%%% 34
\bibitem[\protect\citeauthoryear{He et~al.}{2023}]{he2023deformable}
\begin{barticle}
\bauthor{\bsnm{He}, \binits{Z.}},
\bauthor{\bsnm{He}, \binits{Y.}},
\bauthor{\bsnm{Cao}, \binits{W.}}:
\batitle{Deformable image registration with attention-guided fusion of multi-scale deformation fields}.
\bjtitle{Applied Intelligence}
\bvolume{53}(\bissue{3}),
\bfpage{2936}--\blpage{2950}
(\byear{2023})
\end{barticle}
\endbibitem

%%% 35
\bibitem[\protect\citeauthoryear{Pichat et~al.}{2018}]{pichat2018survey}
\begin{barticle}
\bauthor{\bsnm{Pichat}, \binits{J.}},
\bauthor{\bsnm{Iglesias}, \binits{J.E.}},
\bauthor{\bsnm{Yousry}, \binits{T.}},
\bauthor{\bsnm{Ourselin}, \binits{S.}},
\bauthor{\bsnm{Modat}, \binits{M.}}:
\batitle{A survey of methods for 3d histology reconstruction}.
\bjtitle{Medical image analysis}
\bvolume{46},
\bfpage{73}--\blpage{105}
(\byear{2018})
\end{barticle}
\endbibitem

%%% 36
\bibitem[\protect\citeauthoryear{Zhang et~al.}{2020}]{zhang2020point}
\begin{bchapter}
\bauthor{\bsnm{Zhang}, \binits{J.}},
\bauthor{\bsnm{Li}, \binits{Z.}},
\bauthor{\bsnm{Yu}, \binits{Q.}}:
\bctitle{Point-based registration for multi-stained histology images}.
In: \bbtitle{2020 IEEE 5th International Conference on Image, Vision and Computing (ICIVC)},
pp. \bfpage{92}--\blpage{96}
(\byear{2020}).
\bcomment{IEEE}
\end{bchapter}
\endbibitem

%%% 37
\bibitem[\protect\citeauthoryear{Marzahl et~al.}{2021}]{marzahl2021robust}
\begin{bchapter}
\bauthor{\bsnm{Marzahl}, \binits{C.}},
\bauthor{\bsnm{Wilm}, \binits{F.}},
\bauthor{\bsnm{Tharun}, \binits{L.}},
\bauthor{\bsnm{Perner}, \binits{S.}},
\bauthor{\bsnm{Bertram}, \binits{C.A.}},
\bauthor{\bsnm{Kr{\"o}ger}, \binits{C.}},
\bauthor{\bsnm{Voigt}, \binits{J.}},
\bauthor{\bsnm{Klopfleisch}, \binits{R.}},
\bauthor{\bsnm{Maier}, \binits{A.}},
\bauthor{\bsnm{Aubreville}, \binits{M.}}, \betal:
\bctitle{Robust quad-tree based registration on whole slide images}.
In: \bbtitle{MICCAI Workshop on Computational Pathology},
pp. \bfpage{181}--\blpage{190}
(\byear{2021}).
\bcomment{PMLR}
\end{bchapter}
\endbibitem

%%% 38
\bibitem[\protect\citeauthoryear{Schwarz}{2007}]{schwarz2007non}
\begin{botherref}
\oauthor{\bsnm{Schwarz}, \binits{L.A.}}:
Non-rigid registration using free-form deformations.
Technische Universit{\"a}t M{\"u}nchen
\textbf{6}(4)
(2007)
\end{botherref}
\endbibitem

%%% 39
\bibitem[\protect\citeauthoryear{Rublee et~al.}{2011}]{rublee2011orb}
\begin{bchapter}
\bauthor{\bsnm{Rublee}, \binits{E.}},
\bauthor{\bsnm{Rabaud}, \binits{V.}},
\bauthor{\bsnm{Konolige}, \binits{K.}},
\bauthor{\bsnm{Bradski}, \binits{G.}}:
\bctitle{Orb: An efficient alternative to sift or surf}.
In: \bbtitle{2011 International Conference on Computer Vision},
pp. \bfpage{2564}--\blpage{2571}
(\byear{2011}).
\bcomment{Ieee}
\end{bchapter}
\endbibitem

%%% 40
\bibitem[\protect\citeauthoryear{Lowe}{2004}]{lowe2004distinctive}
\begin{barticle}
\bauthor{\bsnm{Lowe}, \binits{D.G.}}:
\batitle{Distinctive image features from scale-invariant keypoints}.
\bjtitle{International journal of computer vision}
\bvolume{60},
\bfpage{91}--\blpage{110}
(\byear{2004})
\end{barticle}
\endbibitem

%%% 41
\bibitem[\protect\citeauthoryear{Cifor et~al.}{2013}]{cifor2013hybrid}
\begin{barticle}
\bauthor{\bsnm{Cifor}, \binits{A.}},
\bauthor{\bsnm{Risser}, \binits{L.}},
\bauthor{\bsnm{Chung}, \binits{D.}},
\bauthor{\bsnm{Anderson}, \binits{E.M.}},
\bauthor{\bsnm{Schnabel}, \binits{J.A.}}:
\batitle{Hybrid feature-based diffeomorphic registration for tumor tracking in 2-d liver ultrasound images}.
\bjtitle{IEEE transactions on medical imaging}
\bvolume{32}(\bissue{9}),
\bfpage{1647}--\blpage{1656}
(\byear{2013})
\end{barticle}
\endbibitem

%%% 42
\bibitem[\protect\citeauthoryear{Schwier et~al.}{2013}]{schwier2013registration}
\begin{barticle}
\bauthor{\bsnm{Schwier}, \binits{M.}},
\bauthor{\bsnm{B{\"o}hler}, \binits{T.}},
\bauthor{\bsnm{Hahn}, \binits{H.K.}},
\bauthor{\bsnm{Dahmen}, \binits{U.}},
\bauthor{\bsnm{Dirsch}, \binits{O.}}:
\batitle{Registration of histological whole slide images guided by vessel structures}.
\bjtitle{Journal of pathology informatics}
\bvolume{4}(\bissue{2}),
\bfpage{10}
(\byear{2013})
\end{barticle}
\endbibitem

%%% 43
\bibitem[\protect\citeauthoryear{Zeira et~al.}{2022}]{zeira2022alignment}
\begin{barticle}
\bauthor{\bsnm{Zeira}, \binits{R.}},
\bauthor{\bsnm{Land}, \binits{M.}},
\bauthor{\bsnm{Strzalkowski}, \binits{A.}},
\bauthor{\bsnm{Raphael}, \binits{B.J.}}:
\batitle{Alignment and integration of spatial transcriptomics data}.
\bjtitle{Nature Methods}
\bvolume{19}(\bissue{5}),
\bfpage{567}--\blpage{575}
(\byear{2022})
\end{barticle}
\endbibitem

%%% 44
\bibitem[\protect\citeauthoryear{Clarke et~al.}{2018}]{clarke2018d}
\begin{barticle}
\bauthor{\bsnm{Clarke}, \binits{E.L.}},
\bauthor{\bsnm{Revie}, \binits{C.}},
\bauthor{\bsnm{Brettle}, \binits{D.}},
\bauthor{\bsnm{Shires}, \binits{M.}},
\bauthor{\bsnm{Jackson}, \binits{P.}},
\bauthor{\bsnm{Cochrane}, \binits{R.}},
\bauthor{\bsnm{Wilson}, \binits{R.}},
\bauthor{\bsnm{Mello-Thoms}, \binits{C.}},
\bauthor{\bsnm{Treanor}, \binits{D.}}:
\batitle{D evelopment of a novel tissue-mimicking color calibration slide for digital microscopy}.
\bjtitle{Color Research \& Application}
\bvolume{43}(\bissue{2}),
\bfpage{184}--\blpage{197}
(\byear{2018})
\end{barticle}
\endbibitem

%%% 45
\bibitem[\protect\citeauthoryear{Liang et~al.}{2015}]{liang2015liver}
\begin{bchapter}
\bauthor{\bsnm{Liang}, \binits{Y.}},
\bauthor{\bsnm{Wang}, \binits{F.}},
\bauthor{\bsnm{Treanor}, \binits{D.}},
\bauthor{\bsnm{Magee}, \binits{D.}},
\bauthor{\bsnm{Teodoro}, \binits{G.}},
\bauthor{\bsnm{Zhu}, \binits{Y.}},
\bauthor{\bsnm{Kong}, \binits{J.}}:
\bctitle{Liver whole slide image analysis for 3d vessel reconstruction}.
In: \bbtitle{2015 IEEE 12th International Symposium on Biomedical Imaging (ISBI)},
pp. \bfpage{182}--\blpage{185}
(\byear{2015}).
\bcomment{IEEE}
\end{bchapter}
\endbibitem

%%% 46
\bibitem[\protect\citeauthoryear{Tolkach et~al.}{2018}]{tolkach2018three}
\begin{barticle}
\bauthor{\bsnm{Tolkach}, \binits{Y.}},
\bauthor{\bsnm{Thomann}, \binits{S.}},
\bauthor{\bsnm{Kristiansen}, \binits{G.}}:
\batitle{Three-dimensional reconstruction of prostate cancer architecture with serial immunohistochemical sections: hallmarks of tumour growth, tumour compartmentalisation, and implications for grading and heterogeneity}.
\bjtitle{Histopathology}
\bvolume{72}(\bissue{6}),
\bfpage{1051}--\blpage{1059}
(\byear{2018})
\end{barticle}
\endbibitem

%%% 47
\bibitem[\protect\citeauthoryear{Duanmu et~al.}{2021}]{duanmu2021spatial}
\begin{bchapter}
\bauthor{\bsnm{Duanmu}, \binits{H.}},
\bauthor{\bsnm{Bhattarai}, \binits{S.}},
\bauthor{\bsnm{Li}, \binits{H.}},
\bauthor{\bsnm{Cheng}, \binits{C.C.}},
\bauthor{\bsnm{Wang}, \binits{F.}},
\bauthor{\bsnm{Teodoro}, \binits{G.}},
\bauthor{\bsnm{Janssen}, \binits{E.A.}},
\bauthor{\bsnm{Gogineni}, \binits{K.}},
\bauthor{\bsnm{Subhedar}, \binits{P.}},
\bauthor{\bsnm{Aneja}, \binits{R.}}, \betal:
\bctitle{Spatial attention-based deep learning system for breast cancer pathological complete response prediction with serial histopathology images in multiple stains}.
In: \bbtitle{Medical Image Computing and Computer Assisted Intervention--MICCAI 2021: 24th International Conference, Strasbourg, France, September 27--October 1, 2021, Proceedings, Part VIII 24},
pp. \bfpage{550}--\blpage{560}
(\byear{2021}).
\bcomment{Springer}
\end{bchapter}
\endbibitem

%%% 48
\bibitem[\protect\citeauthoryear{Almagro-P{\'e}rez et~al.}{2025}]{almagro2025ai}
\begin{botherref}
\oauthor{\bsnm{Almagro-P{\'e}rez}, \binits{C.}},
\oauthor{\bsnm{Song}, \binits{A.H.}},
\oauthor{\bsnm{Weishaupt}, \binits{L.}},
\oauthor{\bsnm{Kim}, \binits{A.}},
\oauthor{\bsnm{Jaume}, \binits{G.}},
\oauthor{\bsnm{Williamson}, \binits{D.F.}},
\oauthor{\bsnm{Hemker}, \binits{K.}},
\oauthor{\bsnm{Lu}, \binits{M.Y.}},
\oauthor{\bsnm{Singh}, \binits{K.}},
\oauthor{\bsnm{Chen}, \binits{B.}}, et al.:
Ai-driven 3d spatial transcriptomics.
arXiv preprint arXiv:2502.17761
(2025)
\end{botherref}
\endbibitem

%%% 49
\bibitem[\protect\citeauthoryear{Arslan et~al.}{2023}]{arslan2023efficient}
\begin{bchapter}
\bauthor{\bsnm{Arslan}, \binits{J.}},
\bauthor{\bsnm{Ounissi}, \binits{M.}},
\bauthor{\bsnm{Luo}, \binits{H.}},
\bauthor{\bsnm{Lacroix}, \binits{M.}},
\bauthor{\bsnm{Dupr{\'e}}, \binits{P.}},
\bauthor{\bsnm{Kumar}, \binits{P.}},
\bauthor{\bsnm{Hodgkinson}, \binits{A.}},
\bauthor{\bsnm{Dandou}, \binits{S.}},
\bauthor{\bsnm{Larive}, \binits{R.}},
\bauthor{\bsnm{Pignodel}, \binits{C.}}, \betal:
\bctitle{Efficient 3d reconstruction of whole slide images in melanoma}.
In: \bbtitle{Medical Imaging 2023: Digital and Computational Pathology},
vol. \bseriesno{12471},
pp. \bfpage{463}--\blpage{475}
(\byear{2023}).
\bcomment{SPIE}
\end{bchapter}
\endbibitem

%%% 50
\bibitem[\protect\citeauthoryear{Kiemen et~al.}{2024}]{kiemen2024high}
\begin{barticle}
\bauthor{\bsnm{Kiemen}, \binits{A.L.}},
\bauthor{\bsnm{Forjaz}, \binits{A.}},
\bauthor{\bsnm{Sousa}, \binits{R.}},
\bauthor{\bsnm{Han}, \binits{K.S.}},
\bauthor{\bsnm{Hruban}, \binits{R.H.}},
\bauthor{\bsnm{Wood}, \binits{L.D.}},
\bauthor{\bsnm{Wu}, \binits{P.}},
\bauthor{\bsnm{Wirtz}, \binits{D.}}:
\batitle{High-resolution 3d printing of pancreatic ductal microanatomy enabled by serial histology}.
\bjtitle{Advanced Materials Technologies}
\bvolume{9}(\bissue{6}),
\bfpage{2301837}
(\byear{2024})
\end{barticle}
\endbibitem

%%% 51
\bibitem[\protect\citeauthoryear{Deng et~al.}{2021}]{deng2021map3d}
\begin{barticle}
\bauthor{\bsnm{Deng}, \binits{R.}},
\bauthor{\bsnm{Yang}, \binits{H.}},
\bauthor{\bsnm{Jha}, \binits{A.}},
\bauthor{\bsnm{Lu}, \binits{Y.}},
\bauthor{\bsnm{Chu}, \binits{P.}},
\bauthor{\bsnm{Fogo}, \binits{A.B.}},
\bauthor{\bsnm{Huo}, \binits{Y.}}:
\batitle{Map3d: registration-based multi-object tracking on 3d serial whole slide images}.
\bjtitle{IEEE transactions on medical imaging}
\bvolume{40}(\bissue{7}),
\bfpage{1924}--\blpage{1933}
(\byear{2021})
\end{barticle}
\endbibitem

%%% 52
\bibitem[\protect\citeauthoryear{Bay et~al.}{2006}]{bay2006surf}
\begin{bchapter}
\bauthor{\bsnm{Bay}, \binits{H.}},
\bauthor{\bsnm{Tuytelaars}, \binits{T.}},
\bauthor{\bsnm{Van~Gool}, \binits{L.}}:
\bctitle{Surf: Speeded up robust features}.
In: \bbtitle{Computer Vision--ECCV 2006: 9th European Conference on Computer Vision, Graz, Austria, May 7-13, 2006. Proceedings, Part I 9},
pp. \bfpage{404}--\blpage{417}
(\byear{2006}).
\bcomment{Springer}
\end{bchapter}
\endbibitem

%%% 53
\bibitem[\protect\citeauthoryear{Sarlin et~al.}{2020}]{sarlin2020superglue}
\begin{bchapter}
\bauthor{\bsnm{Sarlin}, \binits{P.-E.}},
\bauthor{\bsnm{DeTone}, \binits{D.}},
\bauthor{\bsnm{Malisiewicz}, \binits{T.}},
\bauthor{\bsnm{Rabinovich}, \binits{A.}}:
\bctitle{Superglue: Learning feature matching with graph neural networks}.
In: \bbtitle{Proceedings of the IEEE/CVF Conference on Computer Vision and Pattern Recognition},
pp. \bfpage{4938}--\blpage{4947}
(\byear{2020})
\end{bchapter}
\endbibitem

%%% 54
\bibitem[\protect\citeauthoryear{Avants et~al.}{2008}]{avants2008symmetric}
\begin{barticle}
\bauthor{\bsnm{Avants}, \binits{B.B.}},
\bauthor{\bsnm{Epstein}, \binits{C.L.}},
\bauthor{\bsnm{Grossman}, \binits{M.}},
\bauthor{\bsnm{Gee}, \binits{J.C.}}:
\batitle{Symmetric diffeomorphic image registration with cross-correlation: evaluating automated labeling of elderly and neurodegenerative brain}.
\bjtitle{Medical image analysis}
\bvolume{12}(\bissue{1}),
\bfpage{26}--\blpage{41}
(\byear{2008})
\end{barticle}
\endbibitem

%%% 55
\bibitem[\protect\citeauthoryear{Gatenbee et~al.}{2023}]{gatenbee2023virtual}
\begin{barticle}
\bauthor{\bsnm{Gatenbee}, \binits{C.D.}},
\bauthor{\bsnm{Baker}, \binits{A.-M.}},
\bauthor{\bsnm{Prabhakaran}, \binits{S.}},
\bauthor{\bsnm{Swinyard}, \binits{O.}},
\bauthor{\bsnm{Slebos}, \binits{R.J.}},
\bauthor{\bsnm{Mandal}, \binits{G.}},
\bauthor{\bsnm{Mulholland}, \binits{E.}},
\bauthor{\bsnm{Andor}, \binits{N.}},
\bauthor{\bsnm{Marusyk}, \binits{A.}},
\bauthor{\bsnm{Leedham}, \binits{S.}}, \betal:
\batitle{Virtual alignment of pathology image series for multi-gigapixel whole slide images}.
\bjtitle{Nature communications}
\bvolume{14}(\bissue{1}),
\bfpage{4502}
(\byear{2023})
\end{barticle}
\endbibitem

%%% 56
\bibitem[\protect\citeauthoryear{Simonyan et~al.}{2014}]{simonyan2014learning}
\begin{barticle}
\bauthor{\bsnm{Simonyan}, \binits{K.}},
\bauthor{\bsnm{Vedaldi}, \binits{A.}},
\bauthor{\bsnm{Zisserman}, \binits{A.}}:
\batitle{Learning local feature descriptors using convex optimisation}.
\bjtitle{IEEE Transactions on Pattern Analysis and Machine Intelligence}
\bvolume{36}(\bissue{8}),
\bfpage{1573}--\blpage{1585}
(\byear{2014})
\end{barticle}
\endbibitem

%%% 57
\bibitem[\protect\citeauthoryear{Li et~al.}{2003}]{li2003foreground}
\begin{bchapter}
\bauthor{\bsnm{Li}, \binits{L.}},
\bauthor{\bsnm{Huang}, \binits{W.}},
\bauthor{\bsnm{Gu}, \binits{I.Y.}},
\bauthor{\bsnm{Tian}, \binits{Q.}}:
\bctitle{Foreground object detection from videos containing complex background}.
In: \bbtitle{Proceedings of the Eleventh ACM International Conference on Multimedia},
pp. \bfpage{2}--\blpage{10}
(\byear{2003})
\end{bchapter}
\endbibitem

%%% 58
\bibitem[\protect\citeauthoryear{Klein et~al.}{2009}]{klein2009elastix}
\begin{barticle}
\bauthor{\bsnm{Klein}, \binits{S.}},
\bauthor{\bsnm{Staring}, \binits{M.}},
\bauthor{\bsnm{Murphy}, \binits{K.}},
\bauthor{\bsnm{Viergever}, \binits{M.A.}},
\bauthor{\bsnm{Pluim}, \binits{J.P.}}:
\batitle{Elastix: a toolbox for intensity-based medical image registration}.
\bjtitle{IEEE transactions on medical imaging}
\bvolume{29}(\bissue{1}),
\bfpage{196}--\blpage{205}
(\byear{2009})
\end{barticle}
\endbibitem

%%% 59
\bibitem[\protect\citeauthoryear{Shamonin et~al.}{2014}]{shamonin2014fast}
\begin{barticle}
\bauthor{\bsnm{Shamonin}, \binits{D.P.}},
\bauthor{\bsnm{Bron}, \binits{E.E.}},
\bauthor{\bsnm{Lelieveldt}, \binits{B.P.}},
\bauthor{\bsnm{Smits}, \binits{M.}},
\bauthor{\bsnm{Klein}, \binits{S.}},
\bauthor{\bsnm{Staring}, \binits{M.}},
\bauthor{\bsnm{Initiative}, \binits{A.D.N.}}:
\batitle{Fast parallel image registration on cpu and gpu for diagnostic classification of alzheimer's disease}.
\bjtitle{Frontiers in neuroinformatics}
\bvolume{7},
\bfpage{50}
(\byear{2014})
\end{barticle}
\endbibitem

%%% 60
\bibitem[\protect\citeauthoryear{Marstal et~al.}{2016}]{marstal2016simpleelastix}
\begin{bchapter}
\bauthor{\bsnm{Marstal}, \binits{K.}},
\bauthor{\bsnm{Berendsen}, \binits{F.}},
\bauthor{\bsnm{Staring}, \binits{M.}},
\bauthor{\bsnm{Klein}, \binits{S.}}:
\bctitle{Simpleelastix: A user-friendly, multi-lingual library for medical image registration}.
In: \bbtitle{Proceedings of the IEEE Conference on Computer Vision and Pattern Recognition Workshops},
pp. \bfpage{134}--\blpage{142}
(\byear{2016})
\end{bchapter}
\endbibitem

%%% 61
\bibitem[\protect\citeauthoryear{Weinzaepfel et~al.}{2013}]{weinzaepfel2013deepflow}
\begin{bchapter}
\bauthor{\bsnm{Weinzaepfel}, \binits{P.}},
\bauthor{\bsnm{Revaud}, \binits{J.}},
\bauthor{\bsnm{Harchaoui}, \binits{Z.}},
\bauthor{\bsnm{Schmid}, \binits{C.}}:
\bctitle{Deepflow: Large displacement optical flow with deep matching}.
In: \bbtitle{Proceedings of the IEEE International Conference on Computer Vision},
pp. \bfpage{1385}--\blpage{1392}
(\byear{2013})
\end{bchapter}
\endbibitem

%%% 62
\bibitem[\protect\citeauthoryear{Wu et~al.}{2023}]{wu2023general}
\begin{bchapter}
\bauthor{\bsnm{Wu}, \binits{Q.}},
\bauthor{\bsnm{Shen}, \binits{Y.}},
\bauthor{\bsnm{Ke}, \binits{J.}}:
\bctitle{A general computationally-efficient 3d reconstruction pipeline for multiple images with point clouds}.
In: \bbtitle{International Conference on Medical Image Computing and Computer-Assisted Intervention},
pp. \bfpage{193}--\blpage{202}
(\byear{2023}).
\bcomment{Springer}
\end{bchapter}
\endbibitem

%%% 63
\bibitem[\protect\citeauthoryear{Kiemen et~al.}{2022}]{kiemen2022coda}
\begin{barticle}
\bauthor{\bsnm{Kiemen}, \binits{A.L.}},
\bauthor{\bsnm{Braxton}, \binits{A.M.}},
\bauthor{\bsnm{Grahn}, \binits{M.P.}},
\bauthor{\bsnm{Han}, \binits{K.S.}},
\bauthor{\bsnm{Babu}, \binits{J.M.}},
\bauthor{\bsnm{Reichel}, \binits{R.}},
\bauthor{\bsnm{Jiang}, \binits{A.C.}},
\bauthor{\bsnm{Kim}, \binits{B.}},
\bauthor{\bsnm{Hsu}, \binits{J.}},
\bauthor{\bsnm{Amoa}, \binits{F.}}, \betal:
\batitle{Coda: quantitative 3d reconstruction of large tissues at cellular resolution}.
\bjtitle{Nature Methods}
\bvolume{19}(\bissue{11}),
\bfpage{1490}--\blpage{1499}
(\byear{2022})
\end{barticle}
\endbibitem

%%% 64
\bibitem[\protect\citeauthoryear{Wentzensen et~al.}{2007}]{wentzensen2007combined}
\begin{barticle}
\bauthor{\bsnm{Wentzensen}, \binits{N.}},
\bauthor{\bsnm{Braumann}, \binits{U.-D.}},
\bauthor{\bsnm{Einenkel}, \binits{J.}},
\bauthor{\bsnm{Horn}, \binits{L.-C.}},
\bauthor{\bsnm{Doeberitz}, \binits{M.v.K.}},
\bauthor{\bsnm{L{\"o}ffler}, \binits{M.}},
\bauthor{\bsnm{Kuska}, \binits{J.-P.}}:
\batitle{Combined serial section-based 3d reconstruction of cervical carcinoma invasion using h\&e/p16ink4a/cd3 alternate staining}.
\bjtitle{Cytometry Part A: the journal of the International Society for Analytical Cytology}
\bvolume{71}(\bissue{5}),
\bfpage{327}--\blpage{333}
(\byear{2007})
\end{barticle}
\endbibitem

%%% 65
\bibitem[\protect\citeauthoryear{Jeong et~al.}{2010}]{jeong2010interactive}
\begin{barticle}
\bauthor{\bsnm{Jeong}, \binits{W.-K.}},
\bauthor{\bsnm{Beyer}, \binits{J.}},
\bauthor{\bsnm{Hadwiger}, \binits{M.}},
\bauthor{\bsnm{Vazquez}, \binits{A.}},
\bauthor{\bsnm{Pfister}, \binits{H.}},
\bauthor{\bsnm{Whitaker}, \binits{R.}}:
\batitle{Interactive histology of large-scale biomedical image stacks}.
\bjtitle{IEEE Transactions on Visualization and Computer Graphics}
\bvolume{16}(\bissue{6}),
\bfpage{1386}--\blpage{1395}
(\byear{2010})
\doiurl{10.1109/TVCG.2010.168}
\end{barticle}
\endbibitem

%%% 66
\bibitem[\protect\citeauthoryear{Novakovic et~al.}{2012}]{multi-stains-bib1}
\begin{barticle}
\bauthor{\bsnm{Novakovic}, \binits{Z.S.}},
\bauthor{\bsnm{Durdov}, \binits{M.G.}},
\bauthor{\bsnm{Puljak}, \binits{L.}},
\bauthor{\bsnm{Saraga}, \binits{M.}},
\bauthor{\bsnm{Ljutic}, \binits{D.}},
\bauthor{\bsnm{Filipovic}, \binits{T.}},
\bauthor{\bsnm{Pastar}, \binits{Z.}},
\bauthor{\bsnm{Bendic}, \binits{A.}},
\bauthor{\bsnm{Vukojevic}, \binits{K.}}:
\batitle{The interstitial expression of alpha-smooth muscle actin in glomerulonephritis is associated with renal function}.
\bjtitle{Medical science monitor: international medical journal of experimental and clinical research}
\bvolume{18}(\bissue{4}),
\bfpage{235}
(\byear{2012})
\end{barticle}
\endbibitem

%%% 67
\bibitem[\protect\citeauthoryear{Moles~Lopez et~al.}{2015}]{multi-stains-bib2}
\begin{barticle}
\bauthor{\bsnm{Moles~Lopez}, \binits{X.}},
\bauthor{\bsnm{Barbot}, \binits{P.}},
\bauthor{\bsnm{Van~Eycke}, \binits{Y.-R.}},
\bauthor{\bsnm{Verset}, \binits{L.}},
\bauthor{\bsnm{Tr{\'e}pant}, \binits{A.-L.}},
\bauthor{\bsnm{Larbanoix}, \binits{L.}},
\bauthor{\bsnm{Salmon}, \binits{I.}},
\bauthor{\bsnm{Decaestecker}, \binits{C.}}:
\batitle{Registration of whole immunohistochemical slide images: an efficient way to characterize biomarker colocalization}.
\bjtitle{Journal of the American Medical Informatics Association}
\bvolume{22}(\bissue{1}),
\bfpage{86}--\blpage{99}
(\byear{2015})
\end{barticle}
\endbibitem

%%% 68
\bibitem[\protect\citeauthoryear{Song et~al.}{2013}]{multi-stains-bib3}
\begin{barticle}
\bauthor{\bsnm{Song}, \binits{Y.}},
\bauthor{\bsnm{Treanor}, \binits{D.}},
\bauthor{\bsnm{Bulpitt}, \binits{A.J.}},
\bauthor{\bsnm{Magee}, \binits{D.R.}}:
\batitle{3d reconstruction of multiple stained histology images}.
\bjtitle{Journal of pathology informatics}
\bvolume{4}(\bissue{2}),
\bfpage{7}
(\byear{2013})
\end{barticle}
\endbibitem

%%% 69
\bibitem[\protect\citeauthoryear{Kato et~al.}{2015}]{multi-stains-bib4}
\begin{barticle}
\bauthor{\bsnm{Kato}, \binits{T.}},
\bauthor{\bsnm{Relator}, \binits{R.}},
\bauthor{\bsnm{Ngouv}, \binits{H.}},
\bauthor{\bsnm{Hirohashi}, \binits{Y.}},
\bauthor{\bsnm{Takaki}, \binits{O.}},
\bauthor{\bsnm{Kakimoto}, \binits{T.}},
\bauthor{\bsnm{Okada}, \binits{K.}}:
\batitle{Segmental hog: new descriptor for glomerulus detection in kidney microscopy image}.
\bjtitle{Bmc Bioinformatics}
\bvolume{16},
\bfpage{1}--\blpage{16}
(\byear{2015})
\end{barticle}
\endbibitem

%%% 70
\bibitem[\protect\citeauthoryear{Gadermayr et~al.}{2017}]{multi-stains-bib5}
\begin{botherref}
\oauthor{\bsnm{Gadermayr}, \binits{M.}},
\oauthor{\bsnm{Dombrowski}, \binits{A.-K.}},
\oauthor{\bsnm{Klinkhammer}, \binits{B.M.}},
\oauthor{\bsnm{Boor}, \binits{P.}},
\oauthor{\bsnm{Merhof}, \binits{D.}}:
Cnn cascades for segmenting whole slide images of the kidney.
arXiv preprint arXiv:1708.00251
(2017)
\end{botherref}
\endbibitem

%%% 71
\bibitem[\protect\citeauthoryear{Samsi et~al.}{2012}]{multi-stains-bib6}
\begin{bchapter}
\bauthor{\bsnm{Samsi}, \binits{S.}},
\bauthor{\bsnm{Jarjour}, \binits{W.N.}},
\bauthor{\bsnm{Krishnamurthy}, \binits{A.}}:
\bctitle{Glomeruli segmentation in h\&e stained tissue using perceptual organization}.
In: \bbtitle{2012 IEEE Signal Processing in Medicine and Biology Symposium (SPMB)},
pp. \bfpage{1}--\blpage{5}
(\byear{2012}).
\bcomment{IEEE}
\end{bchapter}
\endbibitem

%%% 72
\bibitem[\protect\citeauthoryear{Ens et~al.}{2009}]{multi-stains-bib7}
\begin{bchapter}
\bauthor{\bsnm{Ens}, \binits{K.}},
\bauthor{\bsnm{Berg}, \binits{J.}},
\bauthor{\bsnm{Fischer}, \binits{B.}}:
\bctitle{A communication term for the combined registration and segmentation}.
In: \bbtitle{4th European Conference of the International Federation for Medical and Biological Engineering: ECIFMBE 2008 23--27 November 2008 Antwerp, Belgium},
pp. \bfpage{673}--\blpage{675}
(\byear{2009}).
\bcomment{Springer}
\end{bchapter}
\endbibitem

%%% 73
\bibitem[\protect\citeauthoryear{Kybic and Borovec}{2014}]{multi-stains-bib8}
\begin{bchapter}
\bauthor{\bsnm{Kybic}, \binits{J.}},
\bauthor{\bsnm{Borovec}, \binits{J.}}:
\bctitle{Automatic simultaneous segmentation and fast registration of histological images}.
In: \bbtitle{2014 IEEE 11th International Symposium on Biomedical Imaging (ISBI)},
pp. \bfpage{774}--\blpage{777}
(\byear{2014}).
\bcomment{IEEE}
\end{bchapter}
\endbibitem

%%% 74
\bibitem[\protect\citeauthoryear{Gupta et~al.}{2018}]{multi-stains-bib9}
\begin{bchapter}
\bauthor{\bsnm{Gupta}, \binits{L.}},
\bauthor{\bsnm{Klinkhammer}, \binits{B.M.}},
\bauthor{\bsnm{Boor}, \binits{P.}},
\bauthor{\bsnm{Merhof}, \binits{D.}},
\bauthor{\bsnm{Gadermayr}, \binits{M.}}:
\bctitle{Stain independent segmentation of whole slide images: A case study in renal histology}.
In: \bbtitle{2018 IEEE 15th International Symposium on Biomedical Imaging (ISBI 2018)},
pp. \bfpage{1360}--\blpage{1364}
(\byear{2018}).
\bcomment{IEEE}
\end{bchapter}
\endbibitem

%%% 75
\bibitem[\protect\citeauthoryear{Goodfellow et~al.}{2014a}]{multi-stains-bib10}
\begin{botherref}
\oauthor{\bsnm{Goodfellow}, \binits{I.}},
\oauthor{\bsnm{Pouget-Abadie}, \binits{J.}},
\oauthor{\bsnm{Mirza}, \binits{M.}},
\oauthor{\bsnm{Xu}, \binits{B.}},
\oauthor{\bsnm{Warde-Farley}, \binits{D.}},
\oauthor{\bsnm{Ozair}, \binits{S.}},
\oauthor{\bsnm{Courville}, \binits{A.}},
\oauthor{\bsnm{Bengio}, \binits{Y.}}:
Generative adversarial nets.
Advances in neural information processing systems
\textbf{27}
(2014)
\end{botherref}
\endbibitem

%%% 76
\bibitem[\protect\citeauthoryear{Goodfellow et~al.}{2014b}]{multi-stains-bib11}
\begin{botherref}
\oauthor{\bsnm{Goodfellow}, \binits{I.}},
\oauthor{\bsnm{Pouget-Abadie}, \binits{J.}},
\oauthor{\bsnm{Mirza}, \binits{M.}},
\oauthor{\bsnm{Xu}, \binits{B.}},
\oauthor{\bsnm{Warde-Farley}, \binits{D.}},
\oauthor{\bsnm{Ozair}, \binits{S.}},
\oauthor{\bsnm{Courville}, \binits{A.}},
\oauthor{\bsnm{Bengio}, \binits{Y.}}:
Generative adversarial nets.
Advances in neural information processing systems
\textbf{27}
(2014)
\end{botherref}
\endbibitem

%%% 77
\bibitem[\protect\citeauthoryear{Liu et~al.}{2022}]{multi-stains-bib12}
\begin{bchapter}
\bauthor{\bsnm{Liu}, \binits{S.}},
\bauthor{\bsnm{Zhu}, \binits{C.}},
\bauthor{\bsnm{Xu}, \binits{F.}},
\bauthor{\bsnm{Jia}, \binits{X.}},
\bauthor{\bsnm{Shi}, \binits{Z.}},
\bauthor{\bsnm{Jin}, \binits{M.}}:
\bctitle{Bci: Breast cancer immunohistochemical image generation through pyramid pix2pix}.
In: \bbtitle{Proceedings of the IEEE/CVF Conference on Computer Vision and Pattern Recognition},
pp. \bfpage{1815}--\blpage{1824}
(\byear{2022})
\end{bchapter}
\endbibitem

%%% 78
\bibitem[\protect\citeauthoryear{Li et~al.}{2023}]{multi-stains-bib13}
\begin{bchapter}
\bauthor{\bsnm{Li}, \binits{F.}},
\bauthor{\bsnm{Hu}, \binits{Z.}},
\bauthor{\bsnm{Chen}, \binits{W.}},
\bauthor{\bsnm{Kak}, \binits{A.}}:
\bctitle{Adaptive supervised patchnce loss for learning h\&e-to-ihc stain translation with inconsistent groundtruth image pairs}.
In: \bbtitle{International Conference on Medical Image Computing and Computer-Assisted Intervention},
pp. \bfpage{632}--\blpage{641}
(\byear{2023}).
\bcomment{Springer}
\end{bchapter}
\endbibitem

%%% 79
\bibitem[\protect\citeauthoryear{Asaf et~al.}{2024}]{multi-stains-bib14}
\begin{barticle}
\bauthor{\bsnm{Asaf}, \binits{M.Z.}},
\bauthor{\bsnm{Rao}, \binits{B.}},
\bauthor{\bsnm{Akram}, \binits{M.U.}},
\bauthor{\bsnm{Khawaja}, \binits{S.G.}},
\bauthor{\bsnm{Khan}, \binits{S.}},
\bauthor{\bsnm{Truong}, \binits{T.M.}},
\bauthor{\bsnm{Sekhon}, \binits{P.}},
\bauthor{\bsnm{Khan}, \binits{I.J.}},
\bauthor{\bsnm{Abbasi}, \binits{M.S.}}:
\batitle{Dual contrastive learning based image-to-image translation of unstained skin tissue into virtually stained h\&e images}.
\bjtitle{Scientific Reports}
\bvolume{14}(\bissue{1}),
\bfpage{2335}
(\byear{2024})
\end{barticle}
\endbibitem

%%% 80
\bibitem[\protect\citeauthoryear{Khan et~al.}{2023}]{multi-stains-bib15}
\begin{botherref}
\oauthor{\bsnm{Khan}, \binits{U.}},
\oauthor{\bsnm{Koivukoski}, \binits{S.}},
\oauthor{\bsnm{Valkonen}, \binits{M.}},
\oauthor{\bsnm{Latonen}, \binits{L.}},
\oauthor{\bsnm{Ruusuvuori}, \binits{P.}}:
The effect of neural network architecture on virtual h\&e staining: Systematic assessment of histological feasibility.
Patterns
\textbf{4}(5)
(2023)
\end{botherref}
\endbibitem

%%% 81
\bibitem[\protect\citeauthoryear{Koivukoski et~al.}{2023}]{multi-stains-bib16}
\begin{barticle}
\bauthor{\bsnm{Koivukoski}, \binits{S.}},
\bauthor{\bsnm{Khan}, \binits{U.}},
\bauthor{\bsnm{Ruusuvuori}, \binits{P.}},
\bauthor{\bsnm{Latonen}, \binits{L.}}:
\batitle{Unstained tissue imaging and virtual hematoxylin and eosin staining of histologic whole slide images}.
\bjtitle{Laboratory Investigation}
\bvolume{103}(\bissue{5}),
\bfpage{100070}
(\byear{2023})
\end{barticle}
\endbibitem

%%% 82
\bibitem[\protect\citeauthoryear{Gadermayr et~al.}{2018}]{multi-stains-bib17}
\begin{bchapter}
\bauthor{\bsnm{Gadermayr}, \binits{M.}},
\bauthor{\bsnm{Appel}, \binits{V.}},
\bauthor{\bsnm{Klinkhammer}, \binits{B.M.}},
\bauthor{\bsnm{Boor}, \binits{P.}},
\bauthor{\bsnm{Merhof}, \binits{D.}}:
\bctitle{Which way round? a study on the performance of stain-translation for segmenting arbitrarily dyed histological images}.
In: \bbtitle{Medical Image Computing and Computer Assisted Intervention--MICCAI 2018: 21st International Conference, Granada, Spain, September 16-20, 2018, Proceedings, Part II 11},
pp. \bfpage{165}--\blpage{173}
(\byear{2018}).
\bcomment{Springer}
\end{bchapter}
\endbibitem

%%% 83
\bibitem[\protect\citeauthoryear{Gadermayr et~al.}{2019}]{multi-stains-bib18}
\begin{barticle}
\bauthor{\bsnm{Gadermayr}, \binits{M.}},
\bauthor{\bsnm{Gupta}, \binits{L.}},
\bauthor{\bsnm{Appel}, \binits{V.}},
\bauthor{\bsnm{Boor}, \binits{P.}},
\bauthor{\bsnm{Klinkhammer}, \binits{B.M.}},
\bauthor{\bsnm{Merhof}, \binits{D.}}:
\batitle{Generative adversarial networks for facilitating stain-independent supervised and unsupervised segmentation: a study on kidney histology}.
\bjtitle{IEEE transactions on medical imaging}
\bvolume{38}(\bissue{10}),
\bfpage{2293}--\blpage{2302}
(\byear{2019})
\end{barticle}
\endbibitem

%%% 84
\bibitem[\protect\citeauthoryear{Bouteldja et~al.}{2023}]{multi-stains-bib19}
\begin{barticle}
\bauthor{\bsnm{Bouteldja}, \binits{N.}},
\bauthor{\bsnm{H{\"o}lscher}, \binits{D.L.}},
\bauthor{\bsnm{Klinkhammer}, \binits{B.M.}},
\bauthor{\bsnm{Buelow}, \binits{R.D.}},
\bauthor{\bsnm{Lotz}, \binits{J.}},
\bauthor{\bsnm{Weiss}, \binits{N.}},
\bauthor{\bsnm{Daniel}, \binits{C.}},
\bauthor{\bsnm{Amann}, \binits{K.}},
\bauthor{\bsnm{Boor}, \binits{P.}}:
\batitle{Stain-independent deep learning--based analysis of digital kidney histopathology}.
\bjtitle{The American Journal of Pathology}
\bvolume{193}(\bissue{1}),
\bfpage{73}--\blpage{83}
(\byear{2023})
\end{barticle}
\endbibitem

%%% 85
\bibitem[\protect\citeauthoryear{Ke et~al.}{2013}]{ke2013situ}
\begin{barticle}
\bauthor{\bsnm{Ke}, \binits{R.}},
\bauthor{\bsnm{Mignardi}, \binits{M.}},
\bauthor{\bsnm{Pacureanu}, \binits{A.}},
\bauthor{\bsnm{Svedlund}, \binits{J.}},
\bauthor{\bsnm{Botling}, \binits{J.}},
\bauthor{\bsnm{W{\"a}hlby}, \binits{C.}},
\bauthor{\bsnm{Nilsson}, \binits{M.}}:
\batitle{In situ sequencing for rna analysis in preserved tissue and cells}.
\bjtitle{Nature methods}
\bvolume{10}(\bissue{9}),
\bfpage{857}--\blpage{860}
(\byear{2013})
\end{barticle}
\endbibitem

%%% 86
\bibitem[\protect\citeauthoryear{Thevenaz et~al.}{1998}]{thevenaz1998pyramid}
\begin{barticle}
\bauthor{\bsnm{Thevenaz}, \binits{P.}},
\bauthor{\bsnm{Ruttimann}, \binits{U.E.}},
\bauthor{\bsnm{Unser}, \binits{M.}}:
\batitle{A pyramid approach to subpixel registration based on intensity}.
\bjtitle{IEEE transactions on image processing}
\bvolume{7}(\bissue{1}),
\bfpage{27}--\blpage{41}
(\byear{1998})
\end{barticle}
\endbibitem

%%% 87
\bibitem[\protect\citeauthoryear{Wang et~al.}{2018}]{wang2018three}
\begin{barticle}
\bauthor{\bsnm{Wang}, \binits{X.}},
\bauthor{\bsnm{Allen}, \binits{W.E.}},
\bauthor{\bsnm{Wright}, \binits{M.A.}},
\bauthor{\bsnm{Sylwestrak}, \binits{E.L.}},
\bauthor{\bsnm{Samusik}, \binits{N.}},
\bauthor{\bsnm{Vesuna}, \binits{S.}},
\bauthor{\bsnm{Evans}, \binits{K.}},
\bauthor{\bsnm{Liu}, \binits{C.}},
\bauthor{\bsnm{Ramakrishnan}, \binits{C.}},
\bauthor{\bsnm{Liu}, \binits{J.}}, \betal:
\batitle{Three-dimensional intact-tissue sequencing of single-cell transcriptional states}.
\bjtitle{Science}
\bvolume{361}(\bissue{6400}),
\bfpage{5691}
(\byear{2018})
\end{barticle}
\endbibitem

%%% 88
\bibitem[\protect\citeauthoryear{Chen et~al.}{2020}]{chen2020spatial}
\begin{barticle}
\bauthor{\bsnm{Chen}, \binits{W.-T.}},
\bauthor{\bsnm{Lu}, \binits{A.}},
\bauthor{\bsnm{Craessaerts}, \binits{K.}},
\bauthor{\bsnm{Pavie}, \binits{B.}},
\bauthor{\bsnm{Frigerio}, \binits{C.S.}},
\bauthor{\bsnm{Corthout}, \binits{N.}},
\bauthor{\bsnm{Qian}, \binits{X.}},
\bauthor{\bsnm{Lal{\'a}kov{\'a}}, \binits{J.}},
\bauthor{\bsnm{K{\"u}hnemund}, \binits{M.}},
\bauthor{\bsnm{Voytyuk}, \binits{I.}}, \betal:
\batitle{Spatial transcriptomics and in situ sequencing to study alzheimer’s disease}.
\bjtitle{Cell}
\bvolume{182}(\bissue{4}),
\bfpage{976}--\blpage{991}
(\byear{2020})
\end{barticle}
\endbibitem

%%% 89
\bibitem[\protect\citeauthoryear{Liu et~al.}{2023}]{liu2023partial}
\begin{barticle}
\bauthor{\bsnm{Liu}, \binits{X.}},
\bauthor{\bsnm{Zeira}, \binits{R.}},
\bauthor{\bsnm{Raphael}, \binits{B.J.}}:
\batitle{Partial alignment of multislice spatially resolved transcriptomics data}.
\bjtitle{Genome Research}
\bvolume{33}(\bissue{7}),
\bfpage{1124}--\blpage{1132}
(\byear{2023})
\end{barticle}
\endbibitem

%%% 90
\bibitem[\protect\citeauthoryear{Xia et~al.}{2023}]{xia2023spatial}
\begin{barticle}
\bauthor{\bsnm{Xia}, \binits{C.-R.}},
\bauthor{\bsnm{Cao}, \binits{Z.-J.}},
\bauthor{\bsnm{Tu}, \binits{X.-M.}},
\bauthor{\bsnm{Gao}, \binits{G.}}:
\batitle{Spatial-linked alignment tool (slat) for aligning heterogenous slices}.
\bjtitle{Nature Communications}
\bvolume{14}(\bissue{1}),
\bfpage{7236}
(\byear{2023})
\end{barticle}
\endbibitem

%%% 91
\bibitem[\protect\citeauthoryear{Panayides et~al.}{2020}]{panayides2020ai}
\begin{barticle}
\bauthor{\bsnm{Panayides}, \binits{A.S.}},
\bauthor{\bsnm{Amini}, \binits{A.}},
\bauthor{\bsnm{Filipovic}, \binits{N.D.}},
\bauthor{\bsnm{Sharma}, \binits{A.}},
\bauthor{\bsnm{Tsaftaris}, \binits{S.A.}},
\bauthor{\bsnm{Young}, \binits{A.}},
\bauthor{\bsnm{Foran}, \binits{D.}},
\bauthor{\bsnm{Do}, \binits{N.}},
\bauthor{\bsnm{Golemati}, \binits{S.}},
\bauthor{\bsnm{Kurc}, \binits{T.}}, \betal:
\batitle{Ai in medical imaging informatics: current challenges and future directions}.
\bjtitle{IEEE journal of biomedical and health informatics}
\bvolume{24}(\bissue{7}),
\bfpage{1837}--\blpage{1857}
(\byear{2020})
\end{barticle}
\endbibitem

%%% 92
\bibitem[\protect\citeauthoryear{Weitz et~al.}{2024}]{weitz2024acrobat}
\begin{barticle}
\bauthor{\bsnm{Weitz}, \binits{P.}},
\bauthor{\bsnm{Valkonen}, \binits{M.}},
\bauthor{\bsnm{Solorzano}, \binits{L.}},
\bauthor{\bsnm{Carr}, \binits{C.}},
\bauthor{\bsnm{Kartasalo}, \binits{K.}},
\bauthor{\bsnm{Boissin}, \binits{C.}},
\bauthor{\bsnm{Koivukoski}, \binits{S.}},
\bauthor{\bsnm{Kuusela}, \binits{A.}},
\bauthor{\bsnm{Rasic}, \binits{D.}},
\bauthor{\bsnm{Feng}, \binits{Y.}}, \betal:
\batitle{The acrobat 2022 challenge: automatic registration of breast cancer tissue}.
\bjtitle{Medical image analysis}
\bvolume{97},
\bfpage{103257}
(\byear{2024})
\end{barticle}
\endbibitem

%%% 93
\bibitem[\protect\citeauthoryear{Liu et~al.}{2022}]{Liu_2022_CVPR}
\begin{bchapter}
\bauthor{\bsnm{Liu}, \binits{S.}},
\bauthor{\bsnm{Zhu}, \binits{C.}},
\bauthor{\bsnm{Xu}, \binits{F.}},
\bauthor{\bsnm{Jia}, \binits{X.}},
\bauthor{\bsnm{Shi}, \binits{Z.}},
\bauthor{\bsnm{Jin}, \binits{M.}}:
\bctitle{Bci: Breast cancer immunohistochemical image generation through pyramid pix2pix}.
In: \bbtitle{Proceedings of the IEEE/CVF Conference on Computer Vision and Pattern Recognition (CVPR) Workshops},
pp. \bfpage{1815}--\blpage{1824}
(\byear{2022})
\end{bchapter}
\endbibitem

%%% 94
\bibitem[\protect\citeauthoryear{Paknezhad et~al.}{2020}]{paknezhad2020regional}
\begin{barticle}
\bauthor{\bsnm{Paknezhad}, \binits{M.}},
\bauthor{\bsnm{Loh}, \binits{S.Y.M.}},
\bauthor{\bsnm{Choudhury}, \binits{Y.}},
\bauthor{\bsnm{Koh}, \binits{V.K.C.}},
\bauthor{\bsnm{Yong}, \binits{T.T.K.}},
\bauthor{\bsnm{Tan}, \binits{H.S.}},
\bauthor{\bsnm{Kanesvaran}, \binits{R.}},
\bauthor{\bsnm{Tan}, \binits{P.H.}},
\bauthor{\bsnm{Peng}, \binits{J.Y.S.}},
\bauthor{\bsnm{Yu}, \binits{W.}}, \betal:
\batitle{Regional registration of whole slide image stacks containing major histological artifacts}.
\bjtitle{BMC bioinformatics}
\bvolume{21}(\bissue{1}),
\bfpage{558}
(\byear{2020})
\end{barticle}
\endbibitem

%%% 95
\bibitem[\protect\citeauthoryear{Zheng et~al.}{2017}]{zheng2017massively}
\begin{barticle}
\bauthor{\bsnm{Zheng}, \binits{G.X.}},
\bauthor{\bsnm{Terry}, \binits{J.M.}},
\bauthor{\bsnm{Belgrader}, \binits{P.}},
\bauthor{\bsnm{Ryvkin}, \binits{P.}},
\bauthor{\bsnm{Bent}, \binits{Z.W.}},
\bauthor{\bsnm{Wilson}, \binits{R.}},
\bauthor{\bsnm{Ziraldo}, \binits{S.B.}},
\bauthor{\bsnm{Wheeler}, \binits{T.D.}},
\bauthor{\bsnm{McDermott}, \binits{G.P.}},
\bauthor{\bsnm{Zhu}, \binits{J.}}, \betal:
\batitle{Massively parallel digital transcriptional profiling of single cells}.
\bjtitle{Nature communications}
\bvolume{8}(\bissue{1}),
\bfpage{14049}
(\byear{2017})
\end{barticle}
\endbibitem

%%% 96
\bibitem[\protect\citeauthoryear{Wang et~al.}{2018}]{wang2018multiplexed}
\begin{barticle}
\bauthor{\bsnm{Wang}, \binits{G.}},
\bauthor{\bsnm{Moffitt}, \binits{J.R.}},
\bauthor{\bsnm{Zhuang}, \binits{X.}}:
\batitle{Multiplexed imaging of high-density libraries of rnas with merfish and expansion microscopy}.
\bjtitle{Scientific reports}
\bvolume{8}(\bissue{1}),
\bfpage{4847}
(\byear{2018})
\end{barticle}
\endbibitem

%%% 97
\bibitem[\protect\citeauthoryear{Ortiz et~al.}{2020}]{ortiz2020molecular}
\begin{barticle}
\bauthor{\bsnm{Ortiz}, \binits{C.}},
\bauthor{\bsnm{Navarro}, \binits{J.F.}},
\bauthor{\bsnm{Jurek}, \binits{A.}},
\bauthor{\bsnm{M{\"a}rtin}, \binits{A.}},
\bauthor{\bsnm{Lundeberg}, \binits{J.}},
\bauthor{\bsnm{Meletis}, \binits{K.}}:
\batitle{Molecular atlas of the adult mouse brain}.
\bjtitle{Science advances}
\bvolume{6}(\bissue{26}),
\bfpage{3446}
(\byear{2020})
\end{barticle}
\endbibitem

%%% 98
\bibitem[\protect\citeauthoryear{Wang et~al.}{2022}]{wang2022high}
\begin{barticle}
\bauthor{\bsnm{Wang}, \binits{M.}},
\bauthor{\bsnm{Hu}, \binits{Q.}},
\bauthor{\bsnm{Lv}, \binits{T.}},
\bauthor{\bsnm{Wang}, \binits{Y.}},
\bauthor{\bsnm{Lan}, \binits{Q.}},
\bauthor{\bsnm{Xiang}, \binits{R.}},
\bauthor{\bsnm{Tu}, \binits{Z.}},
\bauthor{\bsnm{Wei}, \binits{Y.}},
\bauthor{\bsnm{Han}, \binits{K.}},
\bauthor{\bsnm{Shi}, \binits{C.}}, \betal:
\batitle{High-resolution 3d spatiotemporal transcriptomic maps of developing drosophila embryos and larvae}.
\bjtitle{Developmental Cell}
\bvolume{57}(\bissue{10}),
\bfpage{1271}--\blpage{1283}
(\byear{2022})
\end{barticle}
\endbibitem

%%% 99
\bibitem[\protect\citeauthoryear{Moffitt et~al.}{2018}]{moffitt2018molecular}
\begin{barticle}
\bauthor{\bsnm{Moffitt}, \binits{J.R.}},
\bauthor{\bsnm{Bambah-Mukku}, \binits{D.}},
\bauthor{\bsnm{Eichhorn}, \binits{S.W.}},
\bauthor{\bsnm{Vaughn}, \binits{E.}},
\bauthor{\bsnm{Shekhar}, \binits{K.}},
\bauthor{\bsnm{Perez}, \binits{J.D.}},
\bauthor{\bsnm{Rubinstein}, \binits{N.D.}},
\bauthor{\bsnm{Hao}, \binits{J.}},
\bauthor{\bsnm{Regev}, \binits{A.}},
\bauthor{\bsnm{Dulac}, \binits{C.}}, \betal:
\batitle{Molecular, spatial, and functional single-cell profiling of the hypothalamic preoptic region}.
\bjtitle{Science}
\bvolume{362}(\bissue{6416}),
\bfpage{5324}
(\byear{2018})
\end{barticle}
\endbibitem

%%% 100
\bibitem[\protect\citeauthoryear{Chee~Huat and Long}{2019}]{chee2019nih}
\begin{botherref}
\oauthor{\bsnm{Chee~Huat}, \binits{E.}},
\oauthor{\bsnm{Long}, \binits{C.}}:
Nih3t3 point locations for rna seqfish+ experiments.
Nature Zenodo
(2019)
\end{botherref}
\endbibitem

%%% 101
\bibitem[\protect\citeauthoryear{Lewis et~al.}{2021}]{lewis2021spatial}
\begin{barticle}
\bauthor{\bsnm{Lewis}, \binits{S.M.}},
\bauthor{\bsnm{Asselin-Labat}, \binits{M.-L.}},
\bauthor{\bsnm{Nguyen}, \binits{Q.}},
\bauthor{\bsnm{Berthelet}, \binits{J.}},
\bauthor{\bsnm{Tan}, \binits{X.}},
\bauthor{\bsnm{Wimmer}, \binits{V.C.}},
\bauthor{\bsnm{Merino}, \binits{D.}},
\bauthor{\bsnm{Rogers}, \binits{K.L.}},
\bauthor{\bsnm{Naik}, \binits{S.H.}}:
\batitle{Spatial omics and multiplexed imaging to explore cancer biology}.
\bjtitle{Nature methods}
\bvolume{18}(\bissue{9}),
\bfpage{997}--\blpage{1012}
(\byear{2021})
\end{barticle}
\endbibitem

%%% 102
\bibitem[\protect\citeauthoryear{Wang et~al.}{2020}]{wang2020allen}
\begin{barticle}
\bauthor{\bsnm{Wang}, \binits{Q.}},
\bauthor{\bsnm{Ding}, \binits{S.-L.}},
\bauthor{\bsnm{Li}, \binits{Y.}},
\bauthor{\bsnm{Royall}, \binits{J.}},
\bauthor{\bsnm{Feng}, \binits{D.}},
\bauthor{\bsnm{Lesnar}, \binits{P.}},
\bauthor{\bsnm{Graddis}, \binits{N.}},
\bauthor{\bsnm{Naeemi}, \binits{M.}},
\bauthor{\bsnm{Facer}, \binits{B.}},
\bauthor{\bsnm{Ho}, \binits{A.}}, \betal:
\batitle{The allen mouse brain common coordinate framework: a 3d reference atlas}.
\bjtitle{Cell}
\bvolume{181}(\bissue{4}),
\bfpage{936}--\blpage{953}
(\byear{2020})
\end{barticle}
\endbibitem

%%% 103
\bibitem[\protect\citeauthoryear{St{\ae}ger et~al.}{2020}]{staeger2020three}
\begin{barticle}
\bauthor{\bsnm{St{\ae}ger}, \binits{F.F.}},
\bauthor{\bsnm{Mortensen}, \binits{K.N.}},
\bauthor{\bsnm{Nielsen}, \binits{M.S.N.}},
\bauthor{\bsnm{Sigurdsson}, \binits{B.}},
\bauthor{\bsnm{Kaufmann}, \binits{L.K.}},
\bauthor{\bsnm{Hirase}, \binits{H.}},
\bauthor{\bsnm{Nedergaard}, \binits{M.}}:
\batitle{A three-dimensional, population-based average of the c57bl/6 mouse brain from dapi-stained coronal slices}.
\bjtitle{Scientific Data}
\bvolume{7}(\bissue{1}),
\bfpage{235}
(\byear{2020})
\end{barticle}
\endbibitem

%%% 104
\bibitem[\protect\citeauthoryear{Johnson et~al.}{2010}]{johnson2010waxholm}
\begin{barticle}
\bauthor{\bsnm{Johnson}, \binits{G.A.}},
\bauthor{\bsnm{Badea}, \binits{A.}},
\bauthor{\bsnm{Brandenburg}, \binits{J.}},
\bauthor{\bsnm{Cofer}, \binits{G.}},
\bauthor{\bsnm{Fubara}, \binits{B.}},
\bauthor{\bsnm{Liu}, \binits{S.}},
\bauthor{\bsnm{Nissanov}, \binits{J.}}:
\batitle{Waxholm space: an image-based reference for coordinating mouse brain research}.
\bjtitle{Neuroimage}
\bvolume{53}(\bissue{2}),
\bfpage{365}--\blpage{372}
(\byear{2010})
\end{barticle}
\endbibitem

%%% 105
\bibitem[\protect\citeauthoryear{Li et~al.}{2022}]{li2022virtual}
\begin{barticle}
\bauthor{\bsnm{Li}, \binits{Q.}},
\bauthor{\bsnm{Wang}, \binits{F.}},
\bauthor{\bsnm{Chen}, \binits{Y.}},
\bauthor{\bsnm{Chen}, \binits{H.}},
\bauthor{\bsnm{Wu}, \binits{S.}},
\bauthor{\bsnm{Farris}, \binits{A.B.}},
\bauthor{\bsnm{Jiang}, \binits{Y.}},
\bauthor{\bsnm{Kong}, \binits{J.}}:
\batitle{Virtual liver needle biopsy from reconstructed three-dimensional histopathological images: Quantification of sampling error}.
\bjtitle{Computers in Biology and Medicine}
\bvolume{147},
\bfpage{105764}
(\byear{2022})
\end{barticle}
\endbibitem

%%% 106
\bibitem[\protect\citeauthoryear{Brixtel et~al.}{2022}]{brixtel2022whole}
\begin{barticle}
\bauthor{\bsnm{Brixtel}, \binits{R.}},
\bauthor{\bsnm{Bougleux}, \binits{S.}},
\bauthor{\bsnm{L{\'e}zoray}, \binits{O.}},
\bauthor{\bsnm{Caillot}, \binits{Y.}},
\bauthor{\bsnm{Lemoine}, \binits{B.}},
\bauthor{\bsnm{Fontaine}, \binits{M.}},
\bauthor{\bsnm{Nebati}, \binits{D.}},
\bauthor{\bsnm{Renouf}, \binits{A.}}:
\batitle{Whole slide image quality in digital pathology: review and perspectives}.
\bjtitle{IEEE Access}
\bvolume{10},
\bfpage{131005}--\blpage{131035}
(\byear{2022})
\end{barticle}
\endbibitem

%%% 107
\bibitem[\protect\citeauthoryear{Salvi et~al.}{2021}]{salvi2021impact}
\begin{barticle}
\bauthor{\bsnm{Salvi}, \binits{M.}},
\bauthor{\bsnm{Acharya}, \binits{U.R.}},
\bauthor{\bsnm{Molinari}, \binits{F.}},
\bauthor{\bsnm{Meiburger}, \binits{K.M.}}:
\batitle{The impact of pre-and post-image processing techniques on deep learning frameworks: A comprehensive review for digital pathology image analysis}.
\bjtitle{Computers in Biology and Medicine}
\bvolume{128},
\bfpage{104129}
(\byear{2021})
\end{barticle}
\endbibitem

%%% 108
\bibitem[\protect\citeauthoryear{Priego-Torres et~al.}{2020}]{priego2020automatic}
\begin{barticle}
\bauthor{\bsnm{Priego-Torres}, \binits{B.M.}},
\bauthor{\bsnm{Sanchez-Morillo}, \binits{D.}},
\bauthor{\bsnm{Fernandez-Granero}, \binits{M.A.}},
\bauthor{\bsnm{Garcia-Rojo}, \binits{M.}}:
\batitle{Automatic segmentation of whole-slide h\&e stained breast histopathology images using a deep convolutional neural network architecture}.
\bjtitle{Expert Systems With Applications}
\bvolume{151},
\bfpage{113387}
(\byear{2020})
\end{barticle}
\endbibitem

%%% 109
\bibitem[\protect\citeauthoryear{Tian et~al.}{2021}]{tian2021tissue}
\begin{barticle}
\bauthor{\bsnm{Tian}, \binits{T.}},
\bauthor{\bsnm{Yang}, \binits{Z.}},
\bauthor{\bsnm{Li}, \binits{X.}}:
\batitle{Tissue clearing technique: Recent progress and biomedical applications}.
\bjtitle{Journal of anatomy}
\bvolume{238}(\bissue{2}),
\bfpage{489}--\blpage{507}
(\byear{2021})
\end{barticle}
\endbibitem

%%% 110
\bibitem[\protect\citeauthoryear{Ba and Matin}{2022}]{ba2022framework}
\begin{barticle}
\bauthor{\bsnm{Ba}, \binits{Q.}},
\bauthor{\bsnm{Matin}, \binits{J.}}:
\batitle{A framework for building robust deep-learning models against out-of-focus artifact in whole-slide images}.
\bjtitle{Journal of Pathology Informatics}
\bvolume{13},
\bfpage{100085}
(\byear{2022})
\end{barticle}
\endbibitem

%%% 111
\bibitem[\protect\citeauthoryear{Hanna et~al.}{2020}]{hanna2020whole}
\begin{barticle}
\bauthor{\bsnm{Hanna}, \binits{M.G.}},
\bauthor{\bsnm{Parwani}, \binits{A.}},
\bauthor{\bsnm{Sirintrapun}, \binits{S.J.}}:
\batitle{Whole slide imaging: technology and applications}.
\bjtitle{Advances in Anatomic Pathology}
\bvolume{27}(\bissue{4}),
\bfpage{251}--\blpage{259}
(\byear{2020})
\end{barticle}
\endbibitem

%%% 112
\bibitem[\protect\citeauthoryear{Weiss et~al.}{2021}]{weiss2021tutorial}
\begin{barticle}
\bauthor{\bsnm{Weiss}, \binits{K.R.}},
\bauthor{\bsnm{Voigt}, \binits{F.F.}},
\bauthor{\bsnm{Shepherd}, \binits{D.P.}},
\bauthor{\bsnm{Huisken}, \binits{J.}}:
\batitle{Tutorial: practical considerations for tissue clearing and imaging}.
\bjtitle{Nature protocols}
\bvolume{16}(\bissue{6}),
\bfpage{2732}--\blpage{2748}
(\byear{2021})
\end{barticle}
\endbibitem

%%% 113
\bibitem[\protect\citeauthoryear{Zhou et~al.}{2021}]{zhou2021review}
\begin{barticle}
\bauthor{\bsnm{Zhou}, \binits{S.K.}},
\bauthor{\bsnm{Greenspan}, \binits{H.}},
\bauthor{\bsnm{Davatzikos}, \binits{C.}},
\bauthor{\bsnm{Duncan}, \binits{J.S.}},
\bauthor{\bsnm{Van~Ginneken}, \binits{B.}},
\bauthor{\bsnm{Madabhushi}, \binits{A.}},
\bauthor{\bsnm{Prince}, \binits{J.L.}},
\bauthor{\bsnm{Rueckert}, \binits{D.}},
\bauthor{\bsnm{Summers}, \binits{R.M.}}:
\batitle{A review of deep learning in medical imaging: Imaging traits, technology trends, case studies with progress highlights, and future promises}.
\bjtitle{Proceedings of the IEEE}
\bvolume{109}(\bissue{5}),
\bfpage{820}--\blpage{838}
(\byear{2021})
\end{barticle}
\endbibitem

%%% 114
\bibitem[\protect\citeauthoryear{Roy et~al.}{2022}]{roy2022demystifying}
\begin{barticle}
\bauthor{\bsnm{Roy}, \binits{S.}},
\bauthor{\bsnm{Meena}, \binits{T.}},
\bauthor{\bsnm{Lim}, \binits{S.-J.}}:
\batitle{Demystifying supervised learning in healthcare 4.0: A new reality of transforming diagnostic medicine}.
\bjtitle{Diagnostics}
\bvolume{12}(\bissue{10}),
\bfpage{2549}
(\byear{2022})
\end{barticle}
\endbibitem

%%% 115
\bibitem[\protect\citeauthoryear{Lu et~al.}{2022}]{lu2022federated}
\begin{barticle}
\bauthor{\bsnm{Lu}, \binits{M.Y.}},
\bauthor{\bsnm{Chen}, \binits{R.J.}},
\bauthor{\bsnm{Kong}, \binits{D.}},
\bauthor{\bsnm{Lipkova}, \binits{J.}},
\bauthor{\bsnm{Singh}, \binits{R.}},
\bauthor{\bsnm{Williamson}, \binits{D.F.}},
\bauthor{\bsnm{Chen}, \binits{T.Y.}},
\bauthor{\bsnm{Mahmood}, \binits{F.}}:
\batitle{Federated learning for computational pathology on gigapixel whole slide images}.
\bjtitle{Medical image analysis}
\bvolume{76},
\bfpage{102298}
(\byear{2022})
\end{barticle}
\endbibitem

%%% 116
\bibitem[\protect\citeauthoryear{Hill and Hawkes}{2000}]{hill2000across}
\begin{botherref}
\oauthor{\bsnm{Hill}, \binits{D.L.}},
\oauthor{\bsnm{Hawkes}, \binits{D.J.}}:
Across-modality registration using intensity-based cost functions.
Handbook of Medical Imaging: Processing and Analysis,
537--553
(2000)
\end{botherref}
\endbibitem

%%% 117
\bibitem[\protect\citeauthoryear{Varadhan et~al.}{2013}]{varadhan2013framework}
\begin{barticle}
\bauthor{\bsnm{Varadhan}, \binits{R.}},
\bauthor{\bsnm{Karangelis}, \binits{G.}},
\bauthor{\bsnm{Krishnan}, \binits{K.}},
\bauthor{\bsnm{Hui}, \binits{S.}}:
\batitle{A framework for deformable image registration validation in radiotherapy clinical applications}.
\bjtitle{Journal of applied clinical medical physics}
\bvolume{14}(\bissue{1}),
\bfpage{192}--\blpage{213}
(\byear{2013})
\end{barticle}
\endbibitem

%%% 118
\bibitem[\protect\citeauthoryear{Levy et~al.}{2020}]{levy2020topological}
\begin{bchapter}
\bauthor{\bsnm{Levy}, \binits{J.}},
\bauthor{\bsnm{Haudenschild}, \binits{C.}},
\bauthor{\bsnm{Barwick}, \binits{C.}},
\bauthor{\bsnm{Christensen}, \binits{B.}},
\bauthor{\bsnm{Vaickus}, \binits{L.}}:
\bctitle{Topological feature extraction and visualization of whole slide images using graph neural networks}.
In: \bbtitle{BIOCOMPUTING 2021: Proceedings of the Pacific Symposium},
pp. \bfpage{285}--\blpage{296}
(\byear{2020}).
\bcomment{World Scientific}
\end{bchapter}
\endbibitem

%%% 119
\bibitem[\protect\citeauthoryear{Bressan et~al.}{2023}]{bressan2023dawn}
\begin{barticle}
\bauthor{\bsnm{Bressan}, \binits{D.}},
\bauthor{\bsnm{Battistoni}, \binits{G.}},
\bauthor{\bsnm{Hannon}, \binits{G.J.}}:
\batitle{The dawn of spatial omics}.
\bjtitle{Science}
\bvolume{381}(\bissue{6657}),
\bfpage{4964}
(\byear{2023})
\end{barticle}
\endbibitem

%%% 120
\bibitem[\protect\citeauthoryear{Farrokh et~al.}{2019}]{farrokh2019accuracy}
\begin{barticle}
\bauthor{\bsnm{Farrokh}, \binits{A.}},
\bauthor{\bsnm{Maass}, \binits{N.}},
\bauthor{\bsnm{Treu}, \binits{L.}},
\bauthor{\bsnm{Heilmann}, \binits{T.}},
\bauthor{\bsnm{Sch{\"a}fer}, \binits{F.K.}}:
\batitle{Accuracy of tumor size measurement: comparison of b-mode ultrasound, strain elastography, and 2d and 3d shear wave elastography with histopathological lesion size}.
\bjtitle{Acta Radiologica}
\bvolume{60}(\bissue{4}),
\bfpage{451}--\blpage{458}
(\byear{2019})
\end{barticle}
\endbibitem

%%% 121
\bibitem[\protect\citeauthoryear{Akossi et~al.}{2021}]{akossi2021image}
\begin{bchapter}
\bauthor{\bsnm{Akossi}, \binits{A.}},
\bauthor{\bsnm{Wang}, \binits{F.}},
\bauthor{\bsnm{Teodoro}, \binits{G.}},
\bauthor{\bsnm{Kong}, \binits{J.}}:
\bctitle{Image registration with optimal regularization parameter selection by learned auto encoder features}.
In: \bbtitle{2021 IEEE 18th International Symposium on Biomedical Imaging (ISBI)},
pp. \bfpage{702}--\blpage{705}
(\byear{2021}).
\bcomment{IEEE}
\end{bchapter}
\endbibitem

%%% 122
\bibitem[\protect\citeauthoryear{Jaume et~al.}{2024}]{jaume2024multistain}
\begin{botherref}
\oauthor{\bsnm{Jaume}, \binits{G.}},
\oauthor{\bsnm{Vaidya}, \binits{A.}},
\oauthor{\bsnm{Zhang}, \binits{A.}},
\oauthor{\bsnm{Song}, \binits{A.H.}},
\oauthor{\bsnm{Chen}, \binits{R.J.}},
\oauthor{\bsnm{Sahai}, \binits{S.}},
\oauthor{\bsnm{Mo}, \binits{D.}},
\oauthor{\bsnm{Madrigal}, \binits{E.}},
\oauthor{\bsnm{Le}, \binits{L.P.}},
\oauthor{\bsnm{Mahmood}, \binits{F.}}:
Multistain pretraining for slide representation learning in pathology.
arXiv preprint arXiv:2408.02859
(2024)
\end{botherref}
\endbibitem

%%% 123
\bibitem[\protect\citeauthoryear{Bermingham et~al.}{2015}]{bermingham2015application}
\begin{barticle}
\bauthor{\bsnm{Bermingham}, \binits{M.L.}},
\bauthor{\bsnm{Pong-Wong}, \binits{R.}},
\bauthor{\bsnm{Spiliopoulou}, \binits{A.}},
\bauthor{\bsnm{Hayward}, \binits{C.}},
\bauthor{\bsnm{Rudan}, \binits{I.}},
\bauthor{\bsnm{Campbell}, \binits{H.}},
\bauthor{\bsnm{Wright}, \binits{A.F.}},
\bauthor{\bsnm{Wilson}, \binits{J.F.}},
\bauthor{\bsnm{Agakov}, \binits{F.}},
\bauthor{\bsnm{Navarro}, \binits{P.}}, \betal:
\batitle{Application of high-dimensional feature selection: evaluation for genomic prediction in man}.
\bjtitle{Scientific reports}
\bvolume{5}(\bissue{1}),
\bfpage{10312}
(\byear{2015})
\end{barticle}
\endbibitem

%%% 124
\bibitem[\protect\citeauthoryear{Schouten et~al.}{2024}]{schouten2024full}
\begin{barticle}
\bauthor{\bsnm{Schouten}, \binits{D.}},
\bauthor{\bsnm{Laak}, \binits{J.}},
\bauthor{\bsnm{Ginneken}, \binits{B.}},
\bauthor{\bsnm{Litjens}, \binits{G.}}:
\batitle{Full resolution reconstruction of whole-mount sections from digitized individual tissue fragments}.
\bjtitle{Scientific Reports}
\bvolume{14}(\bissue{1}),
\bfpage{1497}
(\byear{2024})
\end{barticle}
\endbibitem

%%% 125
\bibitem[\protect\citeauthoryear{Beyer et~al.}{2022}]{beyer2022survey}
\begin{bchapter}
\bauthor{\bsnm{Beyer}, \binits{J.}},
\bauthor{\bsnm{Troidl}, \binits{J.}},
\bauthor{\bsnm{Boorboor}, \binits{S.}},
\bauthor{\bsnm{Hadwiger}, \binits{M.}},
\bauthor{\bsnm{Kaufman}, \binits{A.}},
\bauthor{\bsnm{Pfister}, \binits{H.}}:
\bctitle{A survey of visualization and analysis in high-resolution connectomics}.
In: \bbtitle{Computer Graphics Forum},
vol. \bseriesno{41},
pp. \bfpage{573}--\blpage{607}
(\byear{2022}).
\bcomment{Wiley Online Library}
\end{bchapter}
\endbibitem

%%% 126
\bibitem[\protect\citeauthoryear{Kashyap et~al.}{2021}]{kashyap2021survey}
\begin{barticle}
\bauthor{\bsnm{Kashyap}, \binits{S.}},
\bauthor{\bsnm{Morse}, \binits{K.E.}},
\bauthor{\bsnm{Patel}, \binits{B.}},
\bauthor{\bsnm{Shah}, \binits{N.H.}}:
\batitle{A survey of extant organizational and computational setups for deploying predictive models in health systems}.
\bjtitle{Journal of the American Medical Informatics Association}
\bvolume{28}(\bissue{11}),
\bfpage{2445}--\blpage{2450}
(\byear{2021})
\end{barticle}
\endbibitem

%%% 127
\bibitem[\protect\citeauthoryear{Saalfeld et~al.}{2010}]{saalfeld2010rigid}
\begin{barticle}
\bauthor{\bsnm{Saalfeld}, \binits{S.}},
\bauthor{\bsnm{Cardona}, \binits{A.}},
\bauthor{\bsnm{Hartenstein}, \binits{V.}},
\bauthor{\bsnm{Toman{\v{c}}{\'a}k}, \binits{P.}}:
\batitle{As-rigid-as-possible mosaicking and serial section registration of large sstem datasets}.
\bjtitle{Bioinformatics}
\bvolume{26}(\bissue{12}),
\bfpage{57}--\blpage{63}
(\byear{2010})
\end{barticle}
\endbibitem

%%% 128
\bibitem[\protect\citeauthoryear{Chen et~al.}{2021}]{chen2021hierarchical}
\begin{barticle}
\bauthor{\bsnm{Chen}, \binits{Z.}},
\bauthor{\bsnm{Zhao}, \binits{S.}},
\bauthor{\bsnm{Hu}, \binits{K.}},
\bauthor{\bsnm{Han}, \binits{J.}},
\bauthor{\bsnm{Ji}, \binits{Y.}},
\bauthor{\bsnm{Ling}, \binits{S.}},
\bauthor{\bsnm{Gao}, \binits{X.}}:
\batitle{A hierarchical and multi-view registration of serial histopathological images}.
\bjtitle{Pattern Recognition Letters}
\bvolume{152},
\bfpage{210}--\blpage{217}
(\byear{2021})
\end{barticle}
\endbibitem

%%% 129
\bibitem[\protect\citeauthoryear{Levy et~al.}{2020}]{levy2020pathflow}
\begin{botherref}
\oauthor{\bsnm{Levy}, \binits{J.J.}},
\oauthor{\bsnm{Jackson}, \binits{C.R.}},
\oauthor{\bsnm{Haudenschild}, \binits{C.C.}},
\oauthor{\bsnm{Christensen}, \binits{B.C.}},
\oauthor{\bsnm{Vaickus}, \binits{L.J.}}:
Pathflow-mixmatch for whole slide image registration: An investigation of a segment-based scalable image registration method.
bioRxiv,
2020--03
(2020)
\end{botherref}
\endbibitem

%%% 130
\bibitem[\protect\citeauthoryear{Lotz et~al.}{2023}]{lotz2023comparison}
\begin{barticle}
\bauthor{\bsnm{Lotz}, \binits{J.}},
\bauthor{\bsnm{Weiss}, \binits{N.}},
\bauthor{\bsnm{Laak}, \binits{J.}},
\bauthor{\bsnm{Heldmann}, \binits{S.}}:
\batitle{Comparison of consecutive and restained sections for image registration in histopathology}.
\bjtitle{Journal of Medical Imaging}
\bvolume{10}(\bissue{6}),
\bfpage{067501}--\blpage{067501}
(\byear{2023})
\end{barticle}
\endbibitem

%%% 131
\bibitem[\protect\citeauthoryear{Zhang et~al.}{2021}]{zhang2021region}
\begin{bchapter}
\bauthor{\bsnm{Zhang}, \binits{J.}},
\bauthor{\bsnm{Li}, \binits{Z.}},
\bauthor{\bsnm{Su}, \binits{A.}}:
\bctitle{Region-aware registration for multi-stained histology images}.
In: \bbtitle{Proceedings of the 2021 4th International Conference on Image and Graphics Processing},
pp. \bfpage{125}--\blpage{130}
(\byear{2021})
\end{bchapter}
\endbibitem

%%% 132
\bibitem[\protect\citeauthoryear{Wodzinski and M{\"u}ller}{2021}]{wodzinski2021deephistreg}
\begin{barticle}
\bauthor{\bsnm{Wodzinski}, \binits{M.}},
\bauthor{\bsnm{M{\"u}ller}, \binits{H.}}:
\batitle{Deephistreg: Unsupervised deep learning registration framework for differently stained histology samples}.
\bjtitle{Computer methods and programs in biomedicine}
\bvolume{198},
\bfpage{105799}
(\byear{2021})
\end{barticle}
\endbibitem

%%% 133
\bibitem[\protect\citeauthoryear{Jones et~al.}{2023}]{jones2023alignment}
\begin{barticle}
\bauthor{\bsnm{Jones}, \binits{A.}},
\bauthor{\bsnm{Townes}, \binits{F.W.}},
\bauthor{\bsnm{Li}, \binits{D.}},
\bauthor{\bsnm{Engelhardt}, \binits{B.E.}}:
\batitle{Alignment of spatial genomics data using deep gaussian processes}.
\bjtitle{Nature Methods}
\bvolume{20}(\bissue{9}),
\bfpage{1379}--\blpage{1387}
(\byear{2023})
\end{barticle}
\endbibitem

%%% 134
\bibitem[\protect\citeauthoryear{Ma and Zhou}{2024}]{ma2024accurate}
\begin{botherref}
\oauthor{\bsnm{Ma}, \binits{Y.}},
\oauthor{\bsnm{Zhou}, \binits{X.}}:
Accurate and efficient integrative reference-informed spatial domain detection for spatial transcriptomics.
Nature Methods,
1--14
(2024)
\end{botherref}
\endbibitem

%%% 135
\bibitem[\protect\citeauthoryear{Shen et~al.}{2022}]{shen2022integrative}
\begin{barticle}
\bauthor{\bsnm{Shen}, \binits{A.}},
\bauthor{\bsnm{Wang}, \binits{F.}},
\bauthor{\bsnm{Paul}, \binits{S.}},
\bauthor{\bsnm{Bhuvanapalli}, \binits{D.}},
\bauthor{\bsnm{Alayof}, \binits{J.}},
\bauthor{\bsnm{Farris}, \binits{A.B.}},
\bauthor{\bsnm{Teodoro}, \binits{G.}},
\bauthor{\bsnm{Brat}, \binits{D.J.}},
\bauthor{\bsnm{Kong}, \binits{J.}}:
\batitle{An integrative web-based software tool for multi-dimensional pathology whole-slide image analytics}.
\bjtitle{Physics in Medicine \& Biology}
\bvolume{67}(\bissue{22}),
\bfpage{224001}
(\byear{2022})
\end{barticle}
\endbibitem

%%% 136
\bibitem[\protect\citeauthoryear{Fanous and Popescu}{2022}]{fanous2022ganscan}
\begin{barticle}
\bauthor{\bsnm{Fanous}, \binits{M.J.}},
\bauthor{\bsnm{Popescu}, \binits{G.}}:
\batitle{Ganscan: continuous scanning microscopy using deep learning deblurring}.
\bjtitle{Light: Science \& Applications}
\bvolume{11}(\bissue{1}),
\bfpage{265}
(\byear{2022})
\end{barticle}
\endbibitem

%%% 137
\bibitem[\protect\citeauthoryear{Liu et~al.}{2023}]{liu2023paste2}
\begin{botherref}
\oauthor{\bsnm{Liu}, \binits{X.}},
\oauthor{\bsnm{Zeira}, \binits{R.}},
\oauthor{\bsnm{Raphael}, \binits{B.J.}}:
Paste2: partial alignment of multi-slice spatially resolved transcriptomics data.
bioRxiv
(2023)
\end{botherref}
\endbibitem

%%% 138
\bibitem[\protect\citeauthoryear{Borovec et~al.}{2013}]{borovec2013registration}
\begin{bchapter}
\bauthor{\bsnm{Borovec}, \binits{J.}},
\bauthor{\bsnm{Kybic}, \binits{J.}},
\bauthor{\bsnm{Bu{\v{s}}ta}, \binits{M.}},
\bauthor{\bsnm{Ortiz-de-Sol{\'o}rzano}, \binits{C.}},
\bauthor{\bsnm{Munoz-Barrutia}, \binits{A.}}:
\bctitle{Registration of multiple stained histological sections}.
In: \bbtitle{2013 IEEE 10th International Symposium on Biomedical Imaging},
pp. \bfpage{1034}--\blpage{1037}
(\byear{2013}).
\bcomment{IEEE}
\end{bchapter}
\endbibitem

%%% 139
\bibitem[\protect\citeauthoryear{Greenwald et~al.}{2022}]{greenwald2022whole}
\begin{barticle}
\bauthor{\bsnm{Greenwald}, \binits{N.F.}},
\bauthor{\bsnm{Miller}, \binits{G.}},
\bauthor{\bsnm{Moen}, \binits{E.}},
\bauthor{\bsnm{Kong}, \binits{A.}},
\bauthor{\bsnm{Kagel}, \binits{A.}},
\bauthor{\bsnm{Dougherty}, \binits{T.}},
\bauthor{\bsnm{Fullaway}, \binits{C.C.}},
\bauthor{\bsnm{McIntosh}, \binits{B.J.}},
\bauthor{\bsnm{Leow}, \binits{K.X.}},
\bauthor{\bsnm{Schwartz}, \binits{M.S.}}, \betal:
\batitle{Whole-cell segmentation of tissue images with human-level performance using large-scale data annotation and deep learning}.
\bjtitle{Nature biotechnology}
\bvolume{40}(\bissue{4}),
\bfpage{555}--\blpage{565}
(\byear{2022})
\end{barticle}
\endbibitem

%%% 140
\bibitem[\protect\citeauthoryear{Theelke et~al.}{2021}]{theelke2021iterative}
\begin{bchapter}
\bauthor{\bsnm{Theelke}, \binits{L.}},
\bauthor{\bsnm{Wilm}, \binits{F.}},
\bauthor{\bsnm{Marzahl}, \binits{C.}},
\bauthor{\bsnm{Bertram}, \binits{C.A.}},
\bauthor{\bsnm{Klopfleisch}, \binits{R.}},
\bauthor{\bsnm{Maier}, \binits{A.}},
\bauthor{\bsnm{Aubreville}, \binits{M.}},
\bauthor{\bsnm{Breininger}, \binits{K.}}:
\bctitle{Iterative cross-scanner registration for whole slide images}.
In: \bbtitle{Proceedings of the IEEE/CVF International Conference on Computer Vision},
pp. \bfpage{582}--\blpage{590}
(\byear{2021})
\end{bchapter}
\endbibitem

%%% 141
\bibitem[\protect\citeauthoryear{Gao et~al.}{2023}]{gao2023comprehensive}
\begin{barticle}
\bauthor{\bsnm{Gao}, \binits{G.}},
\bauthor{\bsnm{Miyasato}, \binits{D.}},
\bauthor{\bsnm{Barner}, \binits{L.A.}},
\bauthor{\bsnm{Serafin}, \binits{R.}},
\bauthor{\bsnm{Bishop}, \binits{K.W.}},
\bauthor{\bsnm{Xie}, \binits{W.}},
\bauthor{\bsnm{Glaser}, \binits{A.K.}},
\bauthor{\bsnm{Rosenthal}, \binits{E.L.}},
\bauthor{\bsnm{True}, \binits{L.D.}},
\bauthor{\bsnm{Liu}, \binits{J.T.}}:
\batitle{Comprehensive surface histology of fresh resection margins with rapid open-top light-sheet (otls) microscopy}.
\bjtitle{IEEE Transactions on Biomedical Engineering}
\bvolume{70}(\bissue{7}),
\bfpage{2160}--\blpage{2171}
(\byear{2023})
\end{barticle}
\endbibitem

%%% 142
\bibitem[\protect\citeauthoryear{Sun et~al.}{2023}]{sun2023bi}
\begin{bchapter}
\bauthor{\bsnm{Sun}, \binits{K.}},
\bauthor{\bsnm{Chen}, \binits{Z.}},
\bauthor{\bsnm{Wang}, \binits{G.}},
\bauthor{\bsnm{Liu}, \binits{J.}},
\bauthor{\bsnm{Ye}, \binits{X.}},
\bauthor{\bsnm{Jiang}, \binits{Y.-G.}}:
\bctitle{Bi-directional feature fusion generative adversarial network for ultra-high resolution pathological image virtual re-staining}.
In: \bbtitle{Proceedings of the IEEE/CVF Conference on Computer Vision and Pattern Recognition},
pp. \bfpage{3904}--\blpage{3913}
(\byear{2023})
\end{bchapter}
\endbibitem

%%% 143
\bibitem[\protect\citeauthoryear{Jiang et~al.}{2019}]{jiang2019robust}
\begin{barticle}
\bauthor{\bsnm{Jiang}, \binits{J.}},
\bauthor{\bsnm{Larson}, \binits{N.B.}},
\bauthor{\bsnm{Prodduturi}, \binits{N.}},
\bauthor{\bsnm{Flotte}, \binits{T.J.}},
\bauthor{\bsnm{Hart}, \binits{S.N.}}:
\batitle{Robust hierarchical density estimation and regression for re-stained histological whole slide image co-registration}.
\bjtitle{Plos one}
\bvolume{14}(\bissue{7}),
\bfpage{0220074}
(\byear{2019})
\end{barticle}
\endbibitem

%%% 144
\bibitem[\protect\citeauthoryear{Liu et~al.}{2021}]{liu2021unpaired}
\begin{barticle}
\bauthor{\bsnm{Liu}, \binits{S.}},
\bauthor{\bsnm{Zhang}, \binits{B.}},
\bauthor{\bsnm{Liu}, \binits{Y.}},
\bauthor{\bsnm{Han}, \binits{A.}},
\bauthor{\bsnm{Shi}, \binits{H.}},
\bauthor{\bsnm{Guan}, \binits{T.}},
\bauthor{\bsnm{He}, \binits{Y.}}:
\batitle{Unpaired stain transfer using pathology-consistent constrained generative adversarial networks}.
\bjtitle{IEEE transactions on medical imaging}
\bvolume{40}(\bissue{8}),
\bfpage{1977}--\blpage{1989}
(\byear{2021})
\end{barticle}
\endbibitem

%%% 145
\bibitem[\protect\citeauthoryear{Di et~al.}{2022}]{di2022generating}
\begin{barticle}
\bauthor{\bsnm{Di}, \binits{D.}},
\bauthor{\bsnm{Zou}, \binits{C.}},
\bauthor{\bsnm{Feng}, \binits{Y.}},
\bauthor{\bsnm{Zhou}, \binits{H.}},
\bauthor{\bsnm{Ji}, \binits{R.}},
\bauthor{\bsnm{Dai}, \binits{Q.}},
\bauthor{\bsnm{Gao}, \binits{Y.}}:
\batitle{Generating hypergraph-based high-order representations of whole-slide histopathological images for survival prediction}.
\bjtitle{IEEE Transactions on Pattern Analysis and Machine Intelligence}
\bvolume{45}(\bissue{5}),
\bfpage{5800}--\blpage{5815}
(\byear{2022})
\end{barticle}
\endbibitem

%%% 146
\bibitem[\protect\citeauthoryear{Hu et~al.}{2024}]{hu2024development}
\begin{barticle}
\bauthor{\bsnm{Hu}, \binits{Q.}},
\bauthor{\bsnm{Rizvi}, \binits{A.A.}},
\bauthor{\bsnm{Schau}, \binits{G.}},
\bauthor{\bsnm{Ingale}, \binits{K.}},
\bauthor{\bsnm{Muller}, \binits{Y.}},
\bauthor{\bsnm{Baits}, \binits{R.}},
\bauthor{\bsnm{Pretzer}, \binits{S.}},
\bauthor{\bsnm{BenTaieb}, \binits{A.}},
\bauthor{\bsnm{Gordhamer}, \binits{A.}},
\bauthor{\bsnm{Nussenzveig}, \binits{R.}}, \betal:
\batitle{Development and validation of a deep learning-based microsatellite instability predictor from prostate cancer whole-slide images}.
\bjtitle{NPJ Precision Oncology}
\bvolume{8}(\bissue{1}),
\bfpage{88}
(\byear{2024})
\end{barticle}
\endbibitem

%%% 147
\bibitem[\protect\citeauthoryear{Ojala}{2023}]{ojala2023differently}
\begin{botherref}
\oauthor{\bsnm{Ojala}, \binits{A.}}:
Differently stained whole slide image registration technique with landmark validation.
Master's thesis,
A. Ojala
(2023)
\end{botherref}
\endbibitem

%%% 148
\bibitem[\protect\citeauthoryear{Li et~al.}{2024}]{li2024navigating}
\begin{barticle}
\bauthor{\bsnm{Li}, \binits{R.}},
\bauthor{\bsnm{Chen}, \binits{X.}},
\bauthor{\bsnm{Yang}, \binits{X.}}:
\batitle{Navigating the landscapes of spatial transcriptomics: How computational methods guide the way}.
\bjtitle{Wiley Interdisciplinary Reviews: RNA}
\bvolume{15}(\bissue{2}),
\bfpage{1839}
(\byear{2024})
\end{barticle}
\endbibitem

\end{thebibliography}
\noindent\textbf{Figure Legends}\\

% \begin{figure}[h!]
%     \centering
%     \includegraphics[width=\textwidth]{methods-2.pdf}
%     % \caption{}
%     \label{fig2}
% \end{figure}
% \usepackage{multirow}

\begin{figure}[h!]
    \centering
    \includegraphics[width=\textwidth]{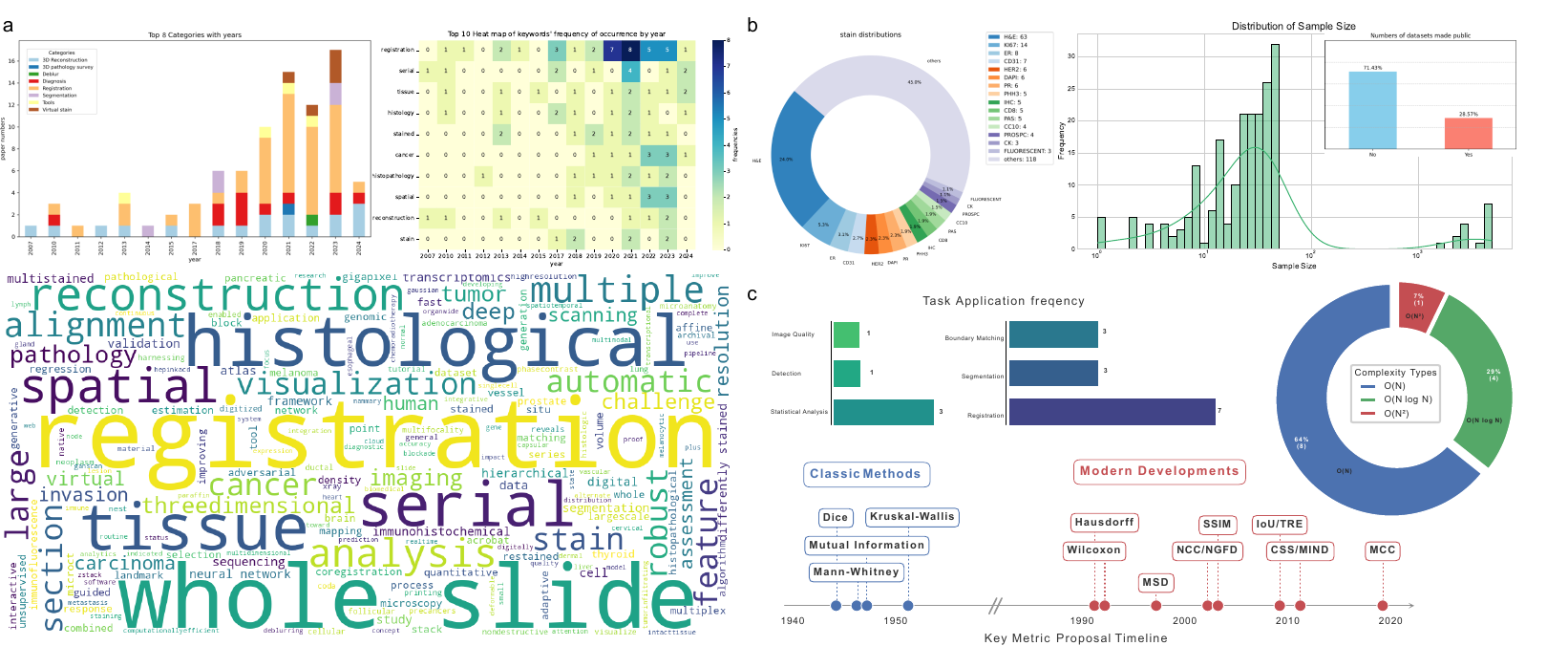}
    \caption{\textbf{a.} Temporal trends of top categories and keywords in the surveyed papers and word cloud visualization of article title phrases. \textbf{b.} Distribution of staining modalities, dataset sizes, and data accessibility in the surveyed datasets. \textbf{c.} Task frequency, complexity distribution, and timeline of proposed evaluation metrics.}
    \label{data-desc}
\end{figure}

\begin{figure}[h!]
    \centering
    \includegraphics[width=\textwidth]{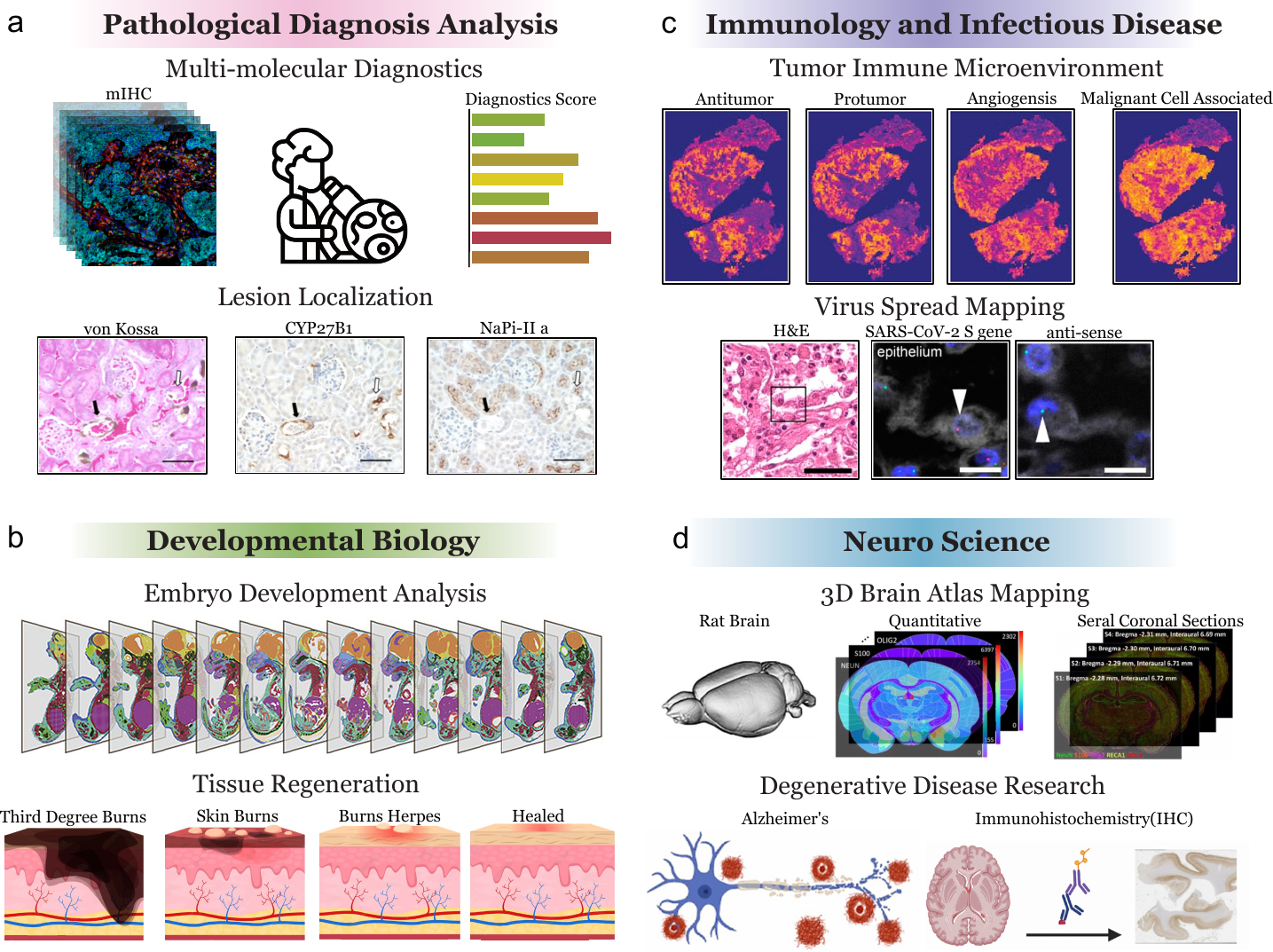}
    \caption{\textbf{a.} Pathological Diagnosis Analysis. Multiplex immunohistochemistry (mIHC) detects co-expression patterns (e.g., Von Kossa, CYP27B1, NaPi-IIa ) within tissue, revealing spatial pathology in mineralization \& metabolic disorders \textbf{b.} Immunology and Infectious Disease. Registers serial sections, combining immunofluorescence/H\&E to map immune entities (Antitumor/Protumor/Angiogenesis/Malignant-associated) in tumors. Applies cross-modality registration (Gene-FISH-H\&E) to track viral dissemination (e.g., COVID-19/influenza) in alveoli/vessels. \textbf{c.} Spatio-Temporal Reconstruction of Development \& Healing 3D reconstruction of serial sections reveals cellular differentiation and organogenesis (e.g., mouse heart via IF/spatial transcriptomics registration). Tracks biomarker dynamics in wound healing to map tissue remodeling. \textbf{d.} Neuroscience. Registers sequential sections + spatial transcriptomics for brain mapping (e.g., Allen Brain Atlas in mouse/rat/human). Integrates IHC/FISH to map pathology spread (plaques/tangles). Enables neural connectivity analysis for learning, memory \& neurodegenerative disorders (e.g., AD).}
    \label{fig3}
\end{figure}

\begin{figure}[h!]
    \centering
    \includegraphics[width=\textwidth]{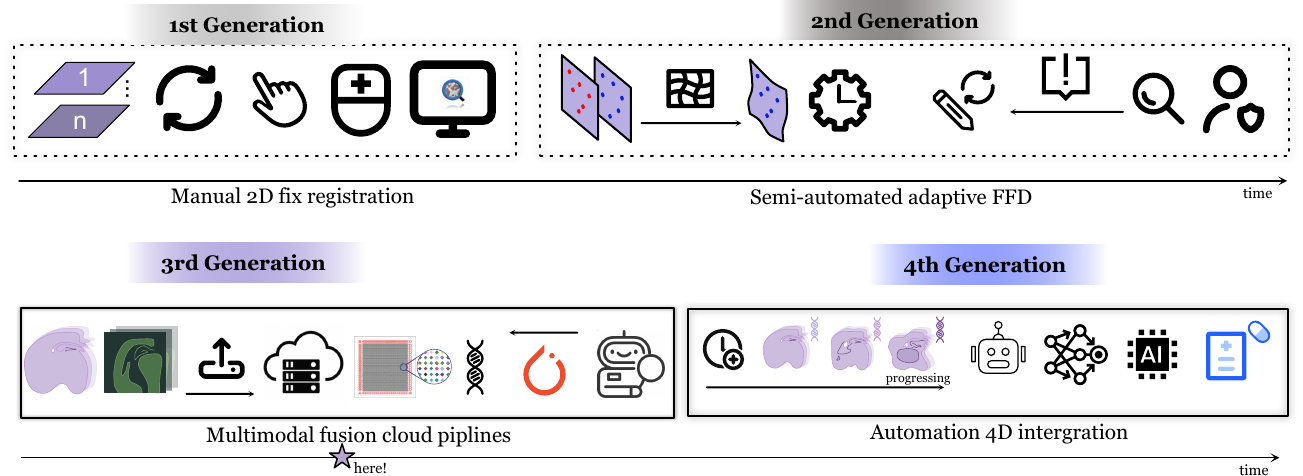}
    \caption{Evolutionary Milestones Illustrated:
    1$^{\text{st}}$ Generation: Manual 2D registration with fixed regularization.
    2$^{\text{nd}}$ Generation: Semi-automated FFD with adaptive parameterization.
    3$^{\text{rd}}$ Generation: AI-driven multimodal sections fusion + cloud-based pipelines. (Currently in this stage)
    4$^{\text{th}}$ Generation : Automation 4D (space + time) integration.
    This framework systematically maps the technological speciation from isolated 2D analysis to integrated 4D (space + time) histopathological ecosystems, fulfilling the promise of evolutionary precision diagnostics.}
    \label{fig4}
\end{figure}

~\\

%%%%%%%%%%%%%%%%%%%%%%%%

% 这张metrics表格的图和论文/数据的图一起优化成统计大图。
%%%%%%%%%%%%%%%%%%%%%%%%%%

% \noindent Table 2 : Summary of the application of the methods. \S3.1 presents the slices registration, mainly divided into rigid and deformable registration and hybrid methods. \S3.2 reviewed from the perspective of Serial sections and the 3D. \S3.3 investigated the application of serial sections in multiple staining. \S3.4 overview of cross-modal applications of serial sections from the perspectives of multimodal fusion, heterogeneous alignment, and morphological integration.

~\\
% \noindent Table 3: Table of research on the dataset of pathological serial sections. Mainly focus on the modality, scale and public situation. The dataset is relatively small in scale due to the large size and rarity of the data. 

\begin{sidewaystable}[h!]
\centering
\renewcommand{\arraystretch}{1.5}
\resizebox{\linewidth}{!}{
\begin{tabular}{ccc|c|cccccc}
\hline
\multicolumn{3}{c|}{\multirow{3}{*}{\textbf{Method}}}                                                                                                 & \multirow{3}{*}{\textbf{Contribution}}                                                                                    & \multicolumn{6}{c}{\textbf{Modality}}                                                                                                                     \\ \cline{5-10} 
\multicolumn{3}{c|}{}                                                                                                                        &                                                                                                                  & \multicolumn{3}{c|}{\textbf{Stain}}                                                             & \multicolumn{3}{c}{\textbf{Multiplexed}}                         \\ \cline{5-10} 
\multicolumn{3}{c|}{}                                                                                                                        &                                                                                                                  & \multicolumn{1}{c|}{Chemical} & \multicolumn{1}{c|}{IHC} & \multicolumn{1}{c|}{Volume} & \multicolumn{1}{c|}{IMC} & \multicolumn{1}{c|}{IF} & ST \\ \hline
\multicolumn{1}{c|}{\multirow{15}{*}{Registration}}  & \multicolumn{1}{c|}{\multirow{6}{*}{Rigid \& Deformable}}    & Elastix \cite{klein2009elastix}                & Integrate lots of registration algorithms applicable to global transformations                                   & \checkmark                             &                          & \checkmark                           &                          &                         &    \\
\hline
\end{tabular}
}
\caption{Summary of the application of the methods. \S3.1 presents the slices registration, mainly divided into rigid and deformable registration and hybrid methods. \S3.2 reviewed from the perspective of Serial sections and the 3D. \S3.3 investigated the application of serial sections in multiple staining. \S3.4 overview of cross-modal applications of serial sections from the perspectives of multimodal fusion, heterogeneous alignment, and morphological integration.}
\end{sidewaystable}

\begin{sidewaystable}[thp]
\centering
\resizebox{\linewidth}{!}{
%
}
\caption{Research on the dataset of pathological serial sections. Mainly focus on the modality, scale and public situation. The dataset is relatively small in scale due to the large size and rarity of the data. }
\end{sidewaystable}

\begin{table}[h!]
\centering
\renewcommand{\arraystretch}{1.5}
\resizebox{\linewidth}{!}{
%
 \\ \bottomrule
\end{tabular}
}
\caption{\textbf{Evaluation Metrics for Histopathological Serial Slice Analysis.} These metrics are pivotal for assessing the efficacy of algorithms across key taskslike  image registration, segmentation, boundary alignment, and classification, which are essential for the precise interpretation of sequential pathological sections. The metrics are selected for their relevance and complexity, ensuring a comprehensive assessment of algorithmic performance.}
\label{tab:metrics}
\end{table}

\end{document}